\documentclass[twocolumn,twocolappendix]{aastex631}

\usepackage{newtxtext,newtxmath,amsmath,float,xspace,threeparttable,bm,amsmath,color}
\usepackage[T1]{fontenc}

\defcitealias{Brok:2022:7}{DB22}

\newcommand{\todo}{\ifmmode \text{\color{red}\Huge{\(\bullet\)}} \else {\color{red}{\Huge$\bullet$}}\fi}
\newcommand{\tido}{\ifmmode {{\color{red}\bullet}} \else {\color{red}$\bullet$}\fi}

\newcommand{\E        }[1]{\ifmmode 10^{#1} \else $10^{#1}$\fi}
\newcommand{\tE        }[1]{\ifmmode \times10^{#1} \else $\times10^{#1}$\fi}
\newcommand{\til}{\ifmmode \sim \else $\sim$\fi}
\renewcommand{\~} {\ifmmode \sim \else $\sim$\fi}

\newcommand{\logNH }{\ifmmode \log (N_{\rm H}/{\rm cm}^{-2}) \else $\log (N_{\rm H}/{\rm cm}^{-2})$\fi}

\newcommand{\Mbh   }{\ifmmode M_{\rm BH} \else $M_{\rm BH}$\fi}

\newcommand{\pc}	{\ifmmode {\rm pc} \else pc\fi}
\newcommand{\ld}	{\ifmmode {\rm l.d.} \else l.d.\fi}
\newcommand{\kms}	{\ifmmode {\rm km\,s}^{-1} \else km\,s$^{-1}$\fi}
\newcommand{\cc}	{\ifmmode {\rm cm}^{-3}    \else cm$^{-3}$\fi}
\newcommand{\cmii}	{\ifmmode {\rm cm}^{-2}    \else cm$^{-2}$\fi}
\newcommand{\ergs}	{\ifmmode {\rm erg\,s}^{-1} \else erg s$^{-1}$\fi}
\newcommand{\ergcms}	{\ifmmode {\rm erg\,cm}^{-2}\,{\rm s}^{-1} \else erg\,cm$^{-2}$\,s$^{-1}$\fi}
\newcommand{\ergcmsA}	{\ifmmode {\rm erg\,cm}^{-2}\,{\rm s}^{-1}\,{\rm\AA}^{-1}
\else erg\,cm$^{-2}$\,s$^{-1}$\,\AA$^{-1}$\fi}
\newcommand{  \ergcmsHz  }{\ifmmode{\rm erg\,cm}^{-2}\,{\rm s}^{-1}\,{\rm Hz}^{-1}
                       \else ergs\,cm$^{-2}$\,s$^{-1}$\,Hz$^{-1}$\fi}
\newcommand{\kev}	{\ifmmode {\rm keV} \else keV\fi}

\newcommand{\mic}	{\ifmmode {\rm \mu m} \else $\mu$m\fi}
\newcommand{\vFWHM}	{\ifmmode v_{\mbox{\tiny FWHM}} \else $v_{\mbox{\tiny FWHM}}$\fi}
\newcommand{\vBLR}	{\ifmmode v_{\mbox{\tiny BLR}} \else $v_{\mbox{\tiny BLR}}$\fi}
\newcommand{\sigBLR}	{\ifmmode \sigma_{\mbox{\tiny BLR}} \else $\sigma_{\mbox{\tiny BLR}}$\fi}
\newcommand{\vNLR}	{\ifmmode v_{\mbox{\tiny NLR}} \else $v_{\mbox{\tiny NLR}}$\fi}
\newcommand{\tauBLR}	{\ifmmode \tau_{\mbox{\tiny BLR}} \else $\tau_{\mbox{\tiny BLR}}$\fi}

\newcommand{\Hubble}	{\ifmmode {\rm km\,s}^{-1}\,{\rm Mpc}^{-1} \else km\,s$^{-1}$\,Mpc$^{-1}$\fi}
\newcommand{\NDunit}	{\ifmmode {\rm Mpc}^{-3} \else Mpc$^{-3}$\fi}
\newcommand{\LFunit}	{\ifmmode {\rm Mpc}^{-3}\,{\rm mag}^{-1} \else Mpc$^{-3}$\,mag$^{-1}$\fi}
\newcommand{\MFunit}	{\ifmmode {\rm Mpc}^{-3}\,{\rm dex}^{-1} \else Mpc$^{-3}$\,dex$^{-1}$\fi}

\newcommand{\Msun}{\ifmmode M_{\odot} \else $M_{\odot}$\fi}
\newcommand{\Lsun}{\ifmmode L_{\odot} \else $L_{\odot}$\fi}
\newcommand{\Zsun}{\ifmmode Z_{\odot} \else $Z_{\odot}$\fi}
\newcommand{\mpyr}{\ifmmode \Msun\,{\rm yr}^{-1} \else $\Msun\,{\rm yr}^{-1}$\fi}

\newcommand{\qnote}{\ifmmode q_{0} \else $q_{0}$\fi}
\newcommand{\Hnote}{\ifmmode H_{0} \else $H_{0}$\fi}
\newcommand{\hnote}{\ifmmode h_{0} \else $h_{0}$\fi}
\newcommand{\anote}{\ifmmode a_{0} \else $a_{0}$\fi}

\newcommand{  \Halpha   }{\ifmmode {\rm H}\alpha \else H$\alpha$\fi}
\newcommand{  \ha   	}{\ifmmode {\rm H}\alpha \else H$\alpha$\fi}
\newcommand{  \Hbeta    }{\ifmmode {\rm H}\beta \else H$\beta$\fi}
\newcommand{  \hb    	}{\ifmmode {\rm H}\beta \else H$\beta$\fi}
\newcommand{  \Hgamma   }{\ifmmode {\rm H}\gamma \else H$\gamma$\fi}
\newcommand{  \Hdelta   }{\ifmmode {\rm H}\delta \else H$\delta$\fi}
\newcommand{  \Lya      }{\ifmmode {\rm Ly}\alpha \else Ly$\alpha$\fi}
\newcommand{  \Lyb      }{\ifmmode {\rm Ly}\beta \else Ly$\beta$\fi}
\newcommand{  \Pa       }{\ifmmode {\rm P}\alpha \else P$\alpha$\fi}
\newcommand{  \Pb       }{\ifmmode {\rm P}\beta \else P$\beta$\fi}
\newcommand{  \Bra      }{\ifmmode {\rm Br}\alpha \else Br$\alpha$\fi}
\newcommand{  \Brg      }{\ifmmode {\rm Br}\gamma \else Br$\gamma$\fi}
\newcommand{  \hii      }{\ifmmode {\rm H}\,\textsc{ii} \else H\,\textsc{ii}\fi}
\newcommand{  \hei      }{\ifmmode {\rm He}\,\textsc{i} \else He\,\textsc{i}\fi}
\newcommand{  \heii     }{\ifmmode {\rm He}\,\textsc{ii} \else He\,\textsc{ii}\fi}
\newcommand{  \HeIIuv   }{\ifmmode {\rm He}\,\textsc{ii}\,\lambda1640 \else He\,\textsc{ii}\,$\lambda1640$\fi}
\newcommand{  \HeIIop   }{\ifmmode {\rm He}\,\textsc{ii}\,\lambda4686 \else He\,\textsc{ii}\,$\lambda4686$\fi}
\newcommand{  \cii      }{\ifmmode {\rm C}\,\textsc{ii}  \else C\,\textsc{ii}\fi}
\newcommand{  \ciii     }{\ifmmode {\rm C}\,\textsc{iii}\right] \else C\,\textsc{iii}]\fi}
\newcommand{  \CIII     }{\ifmmode {\rm C}\,\textsc{iii}\right]\,\lambda1909 \else C\,\textsc{iii}]\,$\lambda1909$\fi}
\newcommand{  \civ      }{\ifmmode {\rm C}\,\textsc{iv}  \else C\,\textsc{iv}\fi}
\newcommand{  \CIV      }{\ifmmode {\rm C}\,\textsc{iv}\,\lambda1549 \else C\,\textsc{iv}\,$\lambda1549$\fi}
\newcommand{  \nii      }{\ifmmode [{\rm N}\,\textsc{ii}]  \else [N\,\textsc{ii}]\fi}
\newcommand{  \niii     }{\ifmmode {\rm N}\,\textsc{iii} \else N\,\textsc{iii}\fi}
\newcommand{  \niv      }{\ifmmode {\rm N}\,\textsc{iv}  \else N\,\textsc{iv}\fi}
\newcommand{  \NIVuv    }{\ifmmode {\rm N}\,\textsc{iv}\,\lambda1486 \else N\,\textsc{iv}\,$\lambda1486$\fi}
\newcommand{  \nv       }{\ifmmode {\rm N}\,\textsc{v}   \else N\,\textsc{v}\fi}
\newcommand{\oi}{\ifmmode \left[{\rm O}\,\textsc{i}\right] \else [O\,{\sc i}]\fi}
\newcommand{\OI}{\ifmmode \left[{\rm O}\,\textsc{i}\right]\,\lambda6300 \else [O\,{\sc i}]$\,\lambda6300$\fi}

\newcommand{\oii}{\ifmmode \left[{\rm O}\,\textsc{ii}\right] \else [O\,{\sc ii}]\fi}
\newcommand{\OII}{\ifmmode \left[{\rm O}\,\textsc{ii}\right]\,\lambda3727 \else [O\,{\sc ii}]\,$\lambda3727$\fi}
\newcommand{\oiii}{\ifmmode \left[{\rm O}\,\textsc{iii}\right] \else [O\,{\sc iii}]\fi}
\newcommand{\OIII}{\ifmmode \left[{\rm O}\,\textsc{iii}\right]\,\lambda5007 \else [O\,{\sc iii}]\,$\lambda5007$\fi}

\newcommand{\NII}{\ifmmode \left[{\rm N}\,\textsc{ii}\right]\,\lambda6583 \else [N\,{\sc ii}]$\,\lambda6583$\fi}

\newcommand{\NeIII}{\ifmmode \left[{\rm Ne}\,\textsc{iii}\right]\,\lambda3968 \else [Ne\,{\sc iii}]$\,\lambda3968$\fi}

\newcommand{\NeV}{\ifmmode \left[{\rm Ne}\,\textsc{v}\right]\,\lambda3426 \else [Ne\,{\sc v}]$\,\lambda3426$\fi}

\newcommand{\HeII}{\ifmmode {\rm He}\,\textsc{ii}\,\lambda4686 \else He\,{\sc ii}$\,\lambda4686$\fi}

\newcommand{\sii}{\ifmmode \left[{\rm S}\,\textsc{ii}\right] \else [S\,{\sc ii}]\fi}

\newcommand{\SII}{\ifmmode \left[{\rm S}\,\textsc{ii}\right]\,\lambda6717,6731 \else [S\,{\sc ii}]$\,\lambda6717,6731$\fi}

\newcommand{  \OIIIuv   }{\ifmmode {\rm O}\,\textsc{iii}\,\lambda1663 \else O\,\textsc{iii}\,$\lambda1663$\fi}
\newcommand{  \oiv      }{\ifmmode {\rm O}\,\textsc{iv}  \else O\,\textsc{iv}\fi}
\newcommand{  \OIVuv    }{\ifmmode {\rm O}\,\textsc{iv}\,\lambda1402  \else O\,\textsc{iv}\,$\lambda1402$\fi}
\newcommand{  \OIVIR    }{\ifmmode {\rm O}\,\textsc{iv}\,25.9\,\mu {\rm m} \else O\,\textsc{iv}\,$25.9\,\mu$m\fi}
\newcommand{  \ovi      }{\ifmmode {\rm O}\,\textsc{vi}   \else O\,\textsc{vi}\fi}
\newcommand{  \Ovi      }{\ifmmode {\rm O}\,\textsc{vi}\,\lambda1035 \else O\,\textsc{vi}\,$\lambda1035$\fi}
\newcommand{  \nei      }{\ifmmode {\rm Ne}\,\textsc{i}   \else Ne\,\textsc{i}\fi}
\newcommand{  \neii     }{\ifmmode {\rm Ne}\,\textsc{ii}  \else Ne\,\textsc{ii}\fi}
\newcommand{  \NeiiIR   }{\ifmmode {\rm Ne}\,\textsc{ii}\,12.8\,\mu {\rm m} \else Ne\,\textsc{ii}\,$12.8\,\mu$m\fi}
\newcommand{  \neiii    }{\ifmmode {\rm Ne}\,\textsc{iii} \else Ne\,\textsc{iii}\fi}
\newcommand{  \neiv     }{\ifmmode {\rm Ne}\,\textsc{iv}  \else Ne\,\textsc{iv}\fi}
\newcommand{  \NevIR    }{\ifmmode {\rm Ne}\,\textsc{v}\,24.3\,\mu {\rm m} \else Ne\,\textsc{v}\,$24.3\,\mu$m\fi}
\newcommand{  \nevi     }{\ifmmode {\rm Ne}\,\textsc{vi}  \else Ne\,\textsc{vi}\fi}
\newcommand{  \mgi      }{\ifmmode {\rm Mg}\,\textsc{i}   \else Mg\,\textsc{i}\fi}
\newcommand{  \mgii     }{\ifmmode {\rm Mg}\,\textsc{ii}  \else Mg\,\textsc{ii}\fi}
\newcommand{  \MgII     }{\ifmmode {\rm Mg}\,\textsc{ii}\,\lambda2798 \else Mg\,\textsc{ii}\,$\lambda2798$\fi}
\newcommand{  \siv      }{\ifmmode {\rm S}\,\textsc{iv}  \else S\,\textsc{iv}\fi}
\newcommand{  \sili     }{\ifmmode {\rm Si}\,\textsc{i}   \else Si\,\textsc{i}\fi}
\newcommand{  \silii    }{\ifmmode {\rm Si}\,\textsc{ii}  \else Si\,\textsc{ii}\fi}
\newcommand{  \Siliv    }{\ifmmode {\rm Si}\,\textsc{iv}  \else Si\,\textsc{iv}\fi}
\newcommand{  \SilIVuv  }{\ifmmode {\rm Si}\,\textsc{iv}\,\lambda1400  \else Si\,\textsc{iv}\,$\lambda1400$\fi}

\newcommand{  \caii     }{\ifmmode {\rm Ca}\,\textsc{ii}   \else Ca\,\textsc{ii}\fi}
 \newcommand{\Mgb}{\ifmmode \left{\rm Mg}\,\textsc{i}\right\,\lambda5175 \else Mg\,{\sc i}\,$\lambda5175$\fi}
\newcommand{\Cahk}{\ifmmode \left[{\rm Ca H+K}\,\textsc{ii}\right\,\lambda3935,3968 \else Ca H+K$\,\lambda3935,3968$\fi}
\newcommand{ \Lhb   }{\ifmmode L\left(\hb\right) \else $L\left(\hb\right)$\fi}
\newcommand{ \fwhb  }{\ifmmode {\rm FWHM}\left(\hb\right) \else FWHM(\hb)\fi}
\newcommand{ \Lha   }{\ifmmode L\left(\ha\right) \else $L\left(\ha\right)$\fi}
\newcommand{ \fwha  }{\ifmmode {\rm FWHM}\left(\ha\right) \else FWHM(\ha)\fi}
\newcommand{ \Lmg   }{\ifmmode L\left(\mgii\right) \else $L\left(\mgii\right)$\fi}
\newcommand{ \fwmg  }{\ifmmode {\rm FWHM}\left(\mgii\right) \else FWHM(\mgii)\fi}
\newcommand{ \Lciv  }{\ifmmode L\left(\civ\right) \else $L\left(\civ\right)$\fi}
\newcommand{ \fwciv }{\ifmmode {\rm FWHM}\left(\civ\right) \else FWHM(\civ)\fi}
\newcommand{ \fwhm  }{\ifmmode {\rm FWHM} \else FWHM\fi} 
\newcommand{ \voff  }{\ifmmode v_{\rm off} \else $v_{\rm off}$\fi} 

\newcommand{ \mumg  }{\ifmmode \mu\left(\mgii\right) \else $\mu\left(\mgii\right)$\fi}
\newcommand{ \fmg   }{\ifmmode f\left(\mgii\right) \else $f\left(\mgii\right)$\fi}
\newcommand{ \muciv }{\ifmmode \mu\left(\civ\right) \else $\mu\left(\civ\right)$\fi}
\newcommand{ \fciv  }{\ifmmode f\left(\civ\right) \else $f\left(\civ\right)$\fi}
\newcommand{  \auvo     }{\ifmmode \alpha_{\nu,{\rm UVO}} \else $\alpha_{\nu,{\rm UVO}}$\fi}
\newcommand{  \Ledd     }{\ifmmode L_{\rm Edd} \else $L_{\rm Edd}$\fi}
\newcommand{  \lamLlam  }{\ifmmode \lambda L_{\lambda} \else $\lambda L_{\lambda}$\fi}
\newcommand{  \lLl      }{\ifmmode \lambda L_{\lambda} \else $\lambda L_{\lambda}$\fi}
\newcommand{  \nuLnu    }{\ifmmode \nu L_{\nu} \else $\nu L_{\nu}$\fi}
\newcommand{  \nLn      }{\ifmmode \nu L_{\nu} \else $\nu L_{\nu}$\fi}
\newcommand{  \Luv      }{\ifmmode L_{1450} \else $L_{1450}$\fi}
\newcommand{  \Lop      }{\ifmmode L_{5100} \else $L_{5100}$\fi}
\newcommand{  \lLop     }{\ifmmode \log\left(\Lop/\ergs\right) \else $\log\left(\Lop/\ergs\right)$\fi}
\newcommand{  \Lthree   }{\ifmmode L_{3000} \else $L_{3000}$\fi}
\newcommand{  \lLthree  }{\ifmmode \log\left(\Lthree/\ergs\right) \else $\log\left(\Lthree/\ergs\right)$\fi}

\newcommand{\Fthree}{\ifmmode F_{3000} \else $F_{3000}$\fi}
\newcommand{\fuv}{\ifmmode f_{\lambda}\left(1450{\rm \AA}\right) \else $f_{\lambda}\left(1450 {\rm \AA}\right)$\fi}
\newcommand{\fthree}{\ifmmode f_{\lambda}\left(3000{\rm \AA}\right) \else $f_{\lambda}\left(3000{\rm \AA}\right)$\fi}
\newcommand{\fH}{\ifmmode f_{\lambda}\left(1.65\micron\right) \else
$f_{\lambda}\left(1.65\micron\right)$\fi}

\newcommand{\fbol}{\ifmmode f_{\rm bol} \else $f_{\rm bol}$\fi}
\newcommand{\fbolwv}{\ifmmode f_{\rm bol}\left(\lambda\right) \else $f_{\rm bol}\left(\lambda\right)$\fi}
\newcommand{\fbolopt}{\ifmmode f_{\rm bol}\left(5100{\rm \AA}\right) \else $f_{\rm bol}\left(5100{\rm \AA}\right)$\fi}
\newcommand{\fbolthree}{\ifmmode f_{\rm bol}\left(3000{\rm \AA}\right) \else $f_{\rm bol}\left(3000{\rm \AA}\right)$\fi}
\newcommand{\fboluv}{\ifmmode f_{\rm bol}\left(1450{\rm \AA}\right) \else $f_{\rm bol}\left(1450{\rm \AA}\right)$\fi}

\newcommand{  \mbh      }{\ifmmode M_{\rm BH} \else $M_{\rm BH}$\fi}
\newcommand{  \lmbh     }{\ifmmode \log\left(\mbh/\Msun\right) \else $\log\left(\mbh/\Msun\right)$\fi} 
\newcommand{  \lledd    }{\ifmmode L_{\rm bol}/L_{\rm Edd} \else $L_{\rm bol}/L_{\rm Edd}$\fi}
\newcommand{  \Lbol     }{\ifmmode L_{\rm bol} \else $L_{\rm bol}$\fi}
\newcommand{  \lbol     }{\ifmmode L_{\rm bol} \else $L_{\rm bol}$\fi}
\newcommand{  \lLbol    }{\ifmmode \log\left(\Lbol/\ergs\right) \else $\log\left(\Lbol/\ergs\right)$\fi} 
\newcommand{  \Lagn     }{\ifmmode L_{\rm AGN} \else $L_{\rm AGN}$\fi}
\newcommand{  \lagn     }{\ifmmode L_{\rm AGN} \else $L_{\rm AGN}$\fi}

\newcommand{  \tgrow     }{\ifmmode t_{\rm growth} \else $t_{\rm growth}$\fi}
\newcommand{  \tUni      }{\ifmmode t_{\rm Universe} \else $t_{\rm Universe}$\fi}

\newcommand{  \Mindot	}{\ifmmode \dot{M}_{\rm infall} \else $\dot{M}_{\rm infall}$\fi}
\newcommand{  \Mbhdot	}{\ifmmode \dot{M}_{\rm BH} \else $\dot{M}_{\rm BH}$\fi}
\newcommand{  \Maddot	}{\ifmmode \dot{M}_{\rm AD} \else $\dot{M}_{\rm AD}$\fi}

\newcommand{  \as	}{\ifmmode a_{\rm *} 		\else $a_{\rm *}$\fi}
\newcommand{  \avec	}{\ifmmode \vec{a}_{\rm *} 	\else $\vec{a}_{\rm *}$\fi}
\newcommand{  \re	}{\ifmmode \eta      	\else $\eta$\fi}

\newcommand{  \mseed    }{\ifmmode M_{\rm seed} \else $M_{\rm seed}$\fi}
\newcommand{  \mbul     }{\ifmmode M_{\rm Bulge} \else $M_{\rm Bulge}$\fi} 
\newcommand{  \mstar    }{\ifmmode M_{*} \else $M_{*}$\fi} 
\newcommand{  \mgal     }{\ifmmode M_{*} \else $M_{*}$\fi} 
\newcommand{  \mhost    }{\ifmmode M_{\rm Host} \else $M_{\rm Host}$\fi}
\newcommand{  \mm       }{\ifmmode M_{*}/M_{\rm BH} \else $M_{*}/M_{\rm BH}$\fi}
\newcommand{  \mmsmall  }{\ifmmode M_{\rm BH}/M_{*} \else $M_{\rm BH}/M_{*}$\fi}
\newcommand{  \mmlarge  }{\ifmmode M_{*}/M_{\rm BH} \else $M_{*}/M_{\rm BH}$\fi}
\newcommand{  \mmwp     }{\ifmmode \left(M_{*}/M_{\rm BH}\right) \else $\left(M_{*}/M_{\rm BH}\right)$\fi}
\newcommand{  \ml       }{\ifmmode M_{*}/L_{*} \else $M_{*}/L_{*}$\fi}
\newcommand{  \mlwp     }{\ifmmode \left(M_{*}/L\right) \else $\left(M_{*}/L\right)$\fi}
\newcommand{  \mlk      }{\ifmmode \left(M_{*}/L_{K}\right) \else $\left(M_{*}/L_{K}\right)$\fi}
\newcommand{  \sigs     }{\ifmmode \sigma_{*} \else $\sigma_{*}$\fi}
\newcommand{  \Reff     }{\ifmmode R_{\rm e} \else $R_{\rm e}$\fi}

\newcommand{\bj}{\ifmmode b_{\rm J} \else $b_{\rm J}$\fi}

\newcommand{\iab}{\ifmmode i_{\rm AB} \else $i_{\rm AB}$\fi}

\newcommand{\jab}{\ifmmode J_{\rm AB} \else $J_{\rm AB}$\fi}
\newcommand{\hab}{\ifmmode H_{\rm AB} \else $H_{\rm AB}$\fi}
\newcommand{\kab}{\ifmmode K_{\rm AB} \else $K_{\rm AB}$\fi}

\newcommand{\jveg}{\ifmmode J_{\rm Vega} \else $J_{\rm Vega}$\fi}
\newcommand{\hveg}{\ifmmode H_{\rm Vega} \else $H_{\rm Vega}$\fi}
\newcommand{\kveg}{\ifmmode K_{\rm Vega} \else $K_{\rm Vega}$\fi}

\newcommand{  \Chisq    }{\ifmmode \chi^{2} \else $\chi^{2}$}
\newcommand{  \nelec    }{\ifmmode n_{e} \else $n_{e}$\fi}     % electron density
\newcommand{  \nh       }{\ifmmode n_{\rm H} \else $n_{\rm H}$\fi}     % hydrogen density
\newcommand{  \Ncol     }{\ifmmode N_{col} \else $N_{col}$\fi} % column density
\newcommand{  \NH       }{\ifmmode N_{\rm H} \else $N_{\rm H}$\fi}     % column density

\def\sun{\hbox{$\odot$}}

\def\ion#1#2{#1$\;${\small\rm\@Roman{#2}}\relax}

\newcommand{\OIIIa}{\ifmmode \left[{\rm O}\,\textsc{iii}\right]\,\lambda4959 \else [O\,{\sc iii}]\,$\lambda4959$\fi}
\newcommand{\NIIa}{\ifmmode \left[{\rm N}\,\textsc{ii}\right]\,\lambda6548 \else [N\,{\sc ii}]\,$\lambda6548$\fi}
\newcommand{\SIIa}{\ifmmode \left[{\rm S}\,\textsc{ii}\right]\,\lambda6716 \else [S\,{\sc ii}]\,$\lambda6716$\fi}
\newcommand{\SIIb}{\ifmmode \left[{\rm S}\,\textsc{ii}\right]\,\lambda6732 \else [S\,{\sc ii}]\,$\lambda6731$\fi}
\newcommand{\NeVa}{\ifmmode \left[{\rm Ne}\,\textsc{v}\right]\,\lambda3346 \else [Ne\,{\sc v}]\,$\lambda3346$\fi}
\newcommand{\NeVb}{\ifmmode \left[{\rm Ne}\,\textsc{v}\right]\,\lambda3426 \else [Ne\,{\sc v}]\,$\lambda3426$\fi}
\newcommand{\NeIIIa}{\ifmmode \left[{\rm Ne}\,\textsc{iii}\right]\,\lambda3869 \else [Ne\,{\sc iii}]\,$\lambda3869$\fi}
\newcommand{\NeIIIb}{\ifmmode \left[{\rm Ne}\,\textsc{iii}\right]\,\lambda3968 \else [Ne\,{\sc iii}]\,$\lambda3968$\fi}

\newcommand{\mgb}{\ifmmode \left{\rm Mg}\,\textsc{i}\right \else Mg\,{\sc i}\fi}

\def\cm{{\rm\thinspace cm}}

\def\g{{\rm\thinspace g}}

\def\K{{\rm\thinspace K}}

\def\Lsun{\hbox{$\rm\thinspace L_{\odot}$}}
\def\m{{\rm\thinspace m}}

\def\pc{{\rm\thinspace pc}}

\def\s{{\rm\thinspace s}}

\newcommand{\HeIIir}{\ifmmode {\rm He}\,\textsc{ii}\,\lambda8237 \else He\,{\sc ii}$\,\lambda8237$\fi}
\newcommand{\HeIir}{\ifmmode {\rm He}\,\textsc{i}\,\lambda10830 \else He\,{\sc i}$\,\lambda10830$\fi}

\newcommand{\SIII}{\ifmmode \left[{\rm S}\,\textsc{iii}\right]\,\lambda9531 \else [S\,\textsc{ii}]\,$\lambda9531$\fi}

\newcommand {\Lsoftint} {\ifmmode L^{\rm in}_{\mathrm{2-10\ keV}} \else $L^{\rm in}_{\mathrm{2-10\ keV}}$\fi}
\newcommand {\ergpersec} {\ifmmode {\rm erg~s}^{-1} \else erg~s$^{-1}$ \fi}

\def\micron{{\mbox{$\mu{\rm m}$}}}

\def\cm{{\rm\thinspace cm}}

\def\g{{\rm\thinspace g}}

\def\K{{\rm\thinspace K}}

\def\Lsun{\hbox{$\rm\thinspace L_{\odot}$}}
\def\m{{\rm\thinspace m}}

\def\pc{{\rm\thinspace pc}}

\def\s{{\rm\thinspace s}}

\def\apjl{\hbox{ApJL}}
\def\apjs{\hbox{ApJS}}

\def\aap{\hbox{AAP}}
\def\micron{{\mbox{$\mu{\rm m}$}}}

\newcommand {\molecfit}{\texttt{molecfit}}

\newcommand{\nuvr}{\ifmmode {\rm NUV}-r \else NUV-$r$\fi}
\newcommand{\mh}{\ifmmode M_{\rm H_2} \else $M_{\rm H_2}$\fi}
\newcommand{\mhi}{\ifmmode M_{\rm HI} \else $M_{\rm HI}$\fi}

\newcommand{\must}{\ifmmode \mu_{\ast} \else $\mu_{\ast}$\fi}
\newcommand{\hmol}{\ifmmode H_2 \else $H_2$\fi}
\newcommand{\rmol}{\ifmmode R_{\rm mol} \else $R_{\rm mol}$\fi}
\newcommand{\tdep}{\ifmmode t_{\rm dep}({\rm H_2}) \else $t_{\rm dep}({\rm H_2})$\fi}
\newcommand{\tdepHI}{\ifmmode t_{\rm dep}({\rm HI}) \else $t_{\rm dep}({\rm HI})$\fi}
\newcommand{\fgas}{\ifmmode f_{\rm H_2} \else $f_{\rm H_2}$\fi}
\newcommand{\fhi}{\ifmmode f_{\rm HI} \else $f_{\rm HI}$\fi}
\newcommand{\xco}{\ifmmode \alpha_{\rm CO} \else $\alpha_{\rm CO}$\fi}

\newcommand{\SiX}{\ifmmode \left[{\rm Si}\,\textsc{x}\right]\,\lambda14300 \else [Si\,{\sc x}]\,$\lambda14300$\fi}
\newcommand{\SiVI}{\ifmmode \left[{\rm Si}\,\textsc{vi}\right]\,\lambda19640 \else [Si\,{\sc vi}]\,$\lambda19640$\fi}
\newcommand{\SXI}{\ifmmode \left[{\rm S}\,\textsc{xi}\right]\,\lambda19196 \else [S\,{\sc xi}]\,$\lambda19196$\fi}
\newcommand{\SVIII}{\ifmmode \left[{\rm S}\,\textsc{viii}\right]\,\lambda9915 \else [S\,{\sc viii}]\,$\lambda9915$\fi}
\newcommand{\SIX}{\ifmmode \left[{\rm S}\,\textsc{ix}\right]\,\lambda12520 \else [S\,{\sc ix}]\,$\lambda12520$\fi}
\newcommand{\FeXIII}{\ifmmode \left[{\rm Fe}\,\textsc{xiii}\right]\,\lambda10747 \else [Fe\,{\sc xiii}]\,$\lambda10747$\fi}
\newcommand{\SiXI}{\ifmmode \left[{\rm Si}\,\textsc{xi}\right]\,\lambda19320 \else [Si\,{\sc xi}]\,$\lambda19320$\fi}

\newcommand{\cmN}{cm$^{-2}$}

\newcommand{\mum}{$\mu$m}
\newcommand{\lam}{$\lambda$}

\newcommand{\ebv}{\mbox{$E(B$$-$$V)$}}
\newcommand{\brd}{\mbox{Br$\delta$}}
\newcommand{\brg}{\mbox{Br$\gamma$}}
\newcommand{\paa}{\mbox{Pa$\alpha$}}
\newcommand{\pab}{\mbox{Pa$\beta$}}
\newcommand{\pag}{\mbox{Pa$\gamma$}}
\newcommand{\pad}{\mbox{Pa$\delta$}}
\newcommand{\pae}{\mbox{Pa$\epsilon$}}
\newcommand{\alix}{\mbox{[Al\,{\sc ix}]}}
\newcommand{\cai}{\mbox{Ca\,{\sc i}}}
\newcommand{\cav}{\mbox{[Ca\,{\sc v}]}}
\newcommand{\caviii}{\mbox{[Ca\,{\sc viii}]}}

\newcommand{\ci}{\mbox{C\,{\sc i}}}

\newcommand{\siii}{\mbox{[Si\,{\sc ii}]}}
\newcommand{\sivi}{\mbox{[Si\,{\sc vi}]}}
\newcommand{\six}{\mbox{[Si\,{\sc x}]}}
\newcommand{\sixi}{\mbox{[Si\,{\sc xi}]}}
\newcommand{\sfiii}{\mbox{[S\,{\sc iii}]}}

\newcommand{\sfviii}{\mbox{[S\,{\sc viii}]}}

\newcommand{\sfix}{\mbox{S\,{\sc ix}}}
\newcommand{\sfxi}{\mbox{[S\,{\sc xi}]}}

\newcommand{\feii}{\mbox{[Fe\,{\sc ii}]}}

\newcommand{\fevi}{\mbox{[Fe\,{\sc vi}]}}
\newcommand{\fevii}{\mbox{[Fe\,{\sc vii}]}}
\newcommand{\fex}{\mbox{[Fe\,{\sc x}]}}
\newcommand{\fexiii}{\mbox{Fe\,{\sc xiii}}}

\newcommand{\pii}{\mbox{P\,{\sc ii}}}

\newcommand{\nev}{\mbox{[Ne\,{\sc viii}]}}

\newcommand{\nai}{\mbox{Na\,{\sc i}}}

\newcommand{\hto}{\mbox{H$_2$O}}
\newcommand{\cdo}{\mbox{CO$_2$}}
\newcommand{\chf}{\mbox{CH$_4$}}
\newcommand{\scikitsurv}{\textsc{scikit-survival}}
\newcommand{\rpear}{$R_{\text{pear}}$}
\newcommand{\ppear}{$P_{\text{pear}}$}

\shorttitle{BASS NIR Atlas}
\shortauthors{}

\graphicspath{{./}{./figures/}}

\begin{document}

\title{BASS.L. Near-infrared Data Release 3: A Spectral Atlas and Characterization of AGN}

\correspondingauthor{Jarred Gillette}
\email{jgill016@ucr.edu}

\author[0000-0001-8528-4839]{Jarred Gillette}
\affiliation{Eureka Scientific, 2452 Delmer Street, Suite 100, Oakland, CA 94602-3017, USA}
\affiliation{Department of Physics \& Astronomy, University of California Riverside, 900 University Avenue, Riverside CA 92521, USA
}
\author[0000-0002-7998-9581]{Michael J. Koss}
\affiliation{Eureka Scientific, 2452 Delmer Street, Suite 100, Oakland, CA 94602-3017, USA}
\affiliation{Space Science Institute, 4750 Walnut Street, Suite 205, Boulder, CO 80301, USA}

\author[0000-0002-2603-2639]{Darshan Kakkad}
\affiliation{Centre for Astrophysics Research, University of Hertfordshire, Hatfield, AL10 9AB, UK}

\author[0000-0001-5742-5980]{Federica Ricci}
\affiliation{Department of Mathematics and Physics, Roma Tre University, Via della Vasca Navale 84, 00146 Rome, Italy}
\affiliation{INAF-Osservatorio Astronomico di Roma, Via Frascati 33, 00078 Monteporzio Catone, Italy}

\author[0000-0003-3336-5498]{Isabella Lamperti}
\affiliation{Dipartimento di Fisica e Astronomia, Università di Firenze, Via G. Sansone 1, 50019, Sesto F.no (Firenze), Italy}
\affiliation{INAF - Osservatorio Astrofisico di Arcetri, largo E. Fermi 5, 50127 Firenze, Italy}

\author[0000-0002-5037-951X]{Kyuseok Oh}
\affiliation{Korea Astronomy \& Space Science Institute, 776, Daedeokdae-ro, Yuseong-gu, Daejeon 34055, Republic of Korea}

\author[0000-0003-0006-8681]{Alejandra Rojas}
\affiliation{Departamento de F\'isica, Universidad T\'ecnica Federico Santa Mar\'ia, Vicu\~{n}a Mackenna 3939, San Joaqu\'in, Santiago de Chile, Chile}

\author[0000-0002-8604-1158]{Yaherlyn Diaz}
\affiliation{Instituto de Estudios Astrofísicos, Facultad de Ingeniería y Ciencias, Universidad Diego Portales, Av. Ejército Libertador 441, Santiago, Chile}

\author[0000-0002-9144-2255]{Turgay Caglar}
\affiliation{George P. and Cynthia Woods Mitchell Institute for Fundamental Physics and Astronomy, Texas A\&M University, College Station, TX, 77845, USA}
\affiliation{Leiden Observatory, PO Box 9513, 2300 RA Leiden, The Netherlands}

\author[0000-0002-4377-903X]{Kohei Ichikawa}
\affiliation{Global Center for Science and Engineering, Faculty of Science and Engineering, Waseda University, 3-4-1, Okubo, Shinjuku, Tokyo 169-8555, Japan}
\affiliation{Department of Physics, School of Advanced Science and Engineering, Faculty of Science and Engineering, Waseda University, 3-4-1,
Okubo, Shinjuku, Tokyo 169-8555, Japan}

\author[0000-0001-8931-1152]{Ignacio del Moral-Castro} 
\affiliation{Instituto de Astrof{\'i}sica, Facultad de F{\'i}sica, Pontificia Universidad Cat{\'o}lica de Chile, Casilla 306, Santiago 22, Chile}

\author[0000-0001-5231-2645]{Claudio Ricci}
\affiliation{N\'ucleo de Astronom\'ia de la Facultad de Ingenier\'ia, Universidad Diego Portales, Av. Ej\'ercito Libertador 441, Santiago 22, Chile}
\affiliation{Kavli Institute for Astronomy and Astrophysics, Peking University, Beijing 100871, People's Republic of China}
% \affiliation{George Mason University, Department of Physics \& Astronomy, MS 3F3, 4400 University Drive, Fairfax, VA 22030, USA}
 
\author[0000-0001-7568-6412]{Ezequiel Treister}
\affiliation{Instituto de Alta Investigaci{\'{o}}n, Universidad de Tarapac{\'{a}}, Casilla 7D, Arica, Chile}

\author[0000-0002-8686-8737]{Franz E. Bauer}
\affiliation{Instituto de Alta Investigaci{\'{o}}n, Universidad de Tarapac{\'{a}}, Casilla 7D, Arica, Chile}

\author[0000-0002-7962-5446]{Richard Mushotzky}
\affiliation{Department of Astronomy, University of Maryland, College Park, MD 20742, USA}
\affiliation{Joint Space-Science Institute, University of Maryland, College Park, MD 20742, USA}

\author[0000-0003-0476-6647]{Mislav Balokovi\'c}
\affiliation{Yale Center for Astronomy \& Astrophysics, 52 Hillhouse Avenue, New Haven, CT 06511, USA}
\affiliation{Department of Physics, Yale University, P.O. Box 2018120, New Haven, CT 06520, USA}

\author[0000-0002-8760-6157]{Jakob S. den Brok}
\affiliation{Institute for Particle Physics and Astrophysics, ETH Z{\"u}rich, Wolfgang-Pauli-Strasse 27, CH-8093 Z{\"u}rich, Switzerland}
\affiliation{Argelander Institute for Astronomy, Auf dem H{\"u}gel 71, 53231, Bonn, Germany}

\author[0000-0002-3683-7297]{Benny Trakhtenbrot}
\affiliation{School of Physics and Astronomy, Tel Aviv University, Tel Aviv 69978, Israel}

\author[0000-0002-0745-9792]{C. Megan Urry}
\affiliation{Yale Center for Astronomy \& Astrophysics and Department of Physics, Yale University, P.O. Box 208120, New Haven, CT 06520-8120, USA}

\author[0000-0002-4226-8959]{Fiona Harrison}
\affiliation{Cahill Center for Astronomy and Astrophysics, California Institute of Technology, Pasadena, CA 91125, USA}

\author[0000-0003-2686-9241]{Daniel Stern}
\affiliation{Jet Propulsion Laboratory, California Institute of Technology, 4800 Oak Grove Drive, MS 169-224, Pasadena, CA 91109, USA}

\begin{abstract}
We present an analysis of near-infrared (NIR) emission-line properties, AGN diagnostics, and circumnuclear gas dynamics for 453 hard X-ray selected (14$-$195~keV) AGN from the BAT AGN Spectroscopic Survey (BASS) NIR Data Release 3 (DR3; $\langle z \rangle = 0.036$, $z < 1.0$). This dataset is the largest compilation of rest-frame NIR spectroscopic observations of hard-X-ray-selected AGN and includes the full DR2 sample. Observations were obtained with VLT X-shooter, a multiwavelength (0.3$-$2.5~\mum) spectrograph ($R$~=~4,000$-$18,000), using a $\geq$~2$\sigma$ detection threshold, enabling broad analysis of emission features. We find that NIR coronal lines, particularly \sivi~$\lambda1.964$, are more reliable tracers of AGN luminosity than optical \oiii, showing a tighter correlation with hard X-ray luminosity ($\sigma$~=~0.25~dex) than \oiii~$\lambda5007$ ($\sigma$~=~0.55~dex). Broad Paschen lines (\paa\ and \pab) are detected in 12\% of Seyfert~2 and 57\% of Seyfert~1.9, consistent with previous hidden BLR studies. We introduce a refined NIR diagnostic diagram (\feii~$\lambda1.257$~\mum/\pab\ and H$_2$~$\lambda~2.122$~\mum/\brg) that effectively distinguishes AGN, star-forming, and composite sources even when contamination limits individual diagnostics or only upper limits are available. Additionally, we find a moderate correlation ($p \approx 7.4 \times 10^{-3}$) between hot molecular gas mass (traced by H$_2$~2.121~\micron) and X-ray luminosity, while its relation with Eddington ratio is weaker. The hot-to-cold gas mass ratio spans four orders of magnitude, averaging $\sim 3 \times 10^{-7}$, indicating diverse molecular gas excitation processes likely driven by star formation and AGN feedback. Our results underscore the value of NIR spectroscopy in probing AGN activity, obscured BLRs, and the complex interactions between AGN and their circumnuclear environments.
\end{abstract}

\section{Introduction} 
\label{sec:intro}

Active galactic nuclei (AGN) are among the most luminous sources in the Universe, powered by accretion onto supermassive black holes (SMBHs). Their diverse observational properties and classification arise from a combination of intrinsic physical conditions and orientation-dependent obscuration, often denoted as unification models \citep{Antonucci:1993:473, Urry:1995:803}. While orientation and obscuration are central to unification, evolutionary scenarios have also been proposed \citep[e.g.,][]{Ballantyne:2006:1070,Tozzi:2024:141}. The study of obscured AGN has gained renewed interest with advancements in near-infrared (NIR) spectroscopy, which penetrates dust more effectively than optical wavelengths by up to a factor 10 \citep{Goodrich:1994, Veilleux:1997:631, Veilleux:2002:315}, offering a unique window into (optically) hidden broad-line regions (BLRs; e.g., Seyfert 1.9 and 2\footnote{Seyfert 1.9 galaxies are defined by optical emission that shows only narrow \hb, \& broad \ha,  whereas Seyfert 2 galaxies exhibit narrow \hb\ \& \ha.}; \citealt{Oh:15:1}). In particular, NIR emission lines such as Brackett and Paschen series transitions are less affected by dust due to their longer wavelengths and have become crucial for identifying broad-line components obscured in optical surveys \citep{Marinucci:2016:L94,RamosAlmeida:2017:679,Lamperti:2017:540,Onori:2017:1783O,Caglar:2020:A114,Ricci:2022:8}. For example, \citet[][hereafter DB22]{Brok:2022:7} highlighted that NIR diagnostics are particularly effective for Seyfert 1.9 galaxies, where optical signatures of the BLR may be strongly biased or suppressed. 

NIR spectroscopy has provided robust methods to infer black hole masses (\Mbh) using emission-line widths and luminosities \citep{Kim:2010:386,Landt:2013,LaFranca:2015:1526,Ricci:598A:51R}, complementing or even surpassing optical-based techniques in heavily obscured AGN. Additionally, high-resolution interferometric observations have spatially resolved the innermost hot dust continuum and broad-line regions in nearby Seyfert 1 AGN \citep{Amorim:23:14,Amorim:24:167}, providing independent constraints on BLR sizes and \Mbh. Mid-infrared (MIR) diagnostics (color–color selection and spectroscopy) have proven especially powerful at revealing heavily obscured AGN by tracing the hot dust and power-law continuum that penetrate large columns of host dust \citep{Lacy:2004:166,Petric:2011:28}. Recent JWST/MIRI and large-area WISE studies have confirmed that MIR methods reveal substantial populations of heavily obscured AGN missed by X-ray/optical surveys, while also emphasizing important caveats for star-formation contamination and selection limits \citep[e.g.,][]{Yang:2023:L5,Omaira:2025:2158}. These findings underscore the importance of multiwavelength approaches in AGN studies to mitigate the effects of dust obscuration and refine our understanding of AGN unification and black hole growth across diverse AGN populations and cosmic epochs.

In addition to BLRs, the AGN's circumnuclear regions host a variety of emission processes, including coronal lines (CLs; ionization potential $>100$~eV; \citealt{Oliva:97:288,Mazzalay:10:1315}) and molecular gas emission, both of which serve as diagnostics and proxies of the bolometric luminosity of the AGN. Coronal lines are highly ionized, and trace the energetic influence of the central SMBH, while molecular hydrogen (e.g., H$_2$ 1-0 S(1) at 2.121 \mum) traces the hot ($\sim$2000\,K) gas phase, influenced by UV fluorescence, shocks, or X-ray heating \citep{Black_Dishoeck:1987, Hollenbach_McKee:1989, Maloney:1996, Rodriguez-Ardila:2004:457, Rodriguez-Ardila:2005, Riffel:2009:273, Riffel:2013:2002}. Cold molecular hydrogen can serve to supply accretion material for the SMBH, and/or become involved in outflows at low speeds but high mass outflow rates \citep[e.g., traced by CO emission,][]{Cicone:2014:562A,Feruglio:2015:583A}. The relationship between molecular gas phases, AGN luminosity, and accretion efficiency remains a topic of significant interest, as it sheds light on the mutual evolution of AGNs and their host galaxies. 

We present NIR spectroscopy for an AGN sample selected at hard X-ray (14$-$195 keV) from the Swift/Burst Alert Telescope (BAT) survey, \citep{Baumgartner:2013:19} as part of the BAT AGN Spectroscopic Survey \citep[BASS,][]{Koss:2022:1}, which is nearly unbiased to obscuration up to Compton-thick column density (\NH$>10^{24} \text{cm}^{-2}$). Optical spectra and properties are detailed in \citealt{Koss:2017:74,Koss:2022:2,Oh:2022:4,Mejia-Restrepo:2022:5}, and \NH\ obscuration measurements in \citealt{Ricci:2017:17}. Building on a solid foundation from previous BASS investigations, our work leverages novel NIR spectroscopy to advance our understanding of AGN properties. In this context, \citetalias{Brok:2022:7} presented a comprehensive analysis of 168 AGN that revealed a substantial fraction of both Seyfert 1 and Seyfert 2 galaxies exhibit high-ionization coronal lines, with a tight correlation between the \sivi\ line and X-ray emission, and identified systematic biases in \Mbh\ measurements in obscured systems. Similarly, \citet{Lamperti:2017:540} conducted a detailed census of NIR spectroscopic features of 102 AGN, showing that while traditional diagnostics may underperform in distinguishing AGN from star-forming (SF) galaxies, they remain crucial for probing obscured nuclei and refining black hole mass estimates. By consistently applying these established methodologies and significantly expanding the sample size by several hundred objects, our study enhances the statistical robustness of NIR-based diagnostics and extends their applicability to a broader, more diverse AGN population.

More specifically, this work analyzes BASS AGN obscuration and emission properties in the NIR, focusing on detecting hidden broad lines and their implications for \Mbh\ estimation. We explore correlations among hard X-ray luminosities, coronal line fluxes, and the distribution of hot molecular gas masses, building on earlier findings \citep{Koss:2017:74,Lamperti:2017:540,Brok:2022:7} while incorporating an expanded dataset that includes higher-redshift sources. Our results highlight the significance of NIR diagnostics in addressing the limitations of optical surveys and in advancing our understanding of AGN physics. For distance calculations in this work, we use the concordance cosmological model with $\Omega_\text{M} =0.3$, $\Omega_{\Lambda} =0.7$, and H$_0 =70$ \kms\ Mpc$^{-1}$.

\section{Sample, Observations, and Reductions} 

\subsection{Sample Selection}
\label{sec:sample}

Our work is a contribution to the collaborative effort of the BASS project, to broadly characterize 14$-$195 keV selected AGN at low to intermediate redshift \citep{Koss:2017:74, Ricci:2017:17}. This hard X-ray selection allows us to obtain a nearly unbiased sample up to Compton-thick AGN \citep{Ricci:2015:L13, Koss:2016:85} and faint AGN due to high X-ray flux sensitivity \citep{Barthelmy:2005:143}. 

New data for this sample come from observations from Very Large Telescope (VLT) X-shooter, a multiwavelength (0.3$-$2.5 \mum) echelle spectrograph with medium spectral resolution $R=4$,000$-$18,000 \citep{D'Odorico:2006, Vernet:2011:A105}. This work includes 249 new spectra obtained with X-shooter. Its three spectroscopic arms provide efficient simultaneous and overlapping coverage of ultraviolet, visible, and NIR passbands. Utilizing these unique features of X-shooter's wide wavelength coverage, we selected AGN at $z < 0.9$ to have coverage of the most prominent NIR features (e.g., \pab), excluding beamed AGN \citep{Paliya:2019:154}. 

We add to this all X-shooter sources previously measured and analyzed by \citetalias{Brok:2022:7}, 82 of which have spectral coverage limited to the J and H-bands (0.994$-$2.101~\mum), encompassing sources with detectable \pab\ emission at $z < 0.6$. \citetalias{Brok:2022:7} did not measure and analyze their remaining sources in the K-band. We remeasure and analyze the remaining X-shooter sources from \citetalias{Brok:2022:7} with K-band coverage (2.101$-$2.5~\mum) to extend the analysis to the full NIR range. Incorporating all additional X-shooter observations through December 12, 2023, the final long-slit sample consists of 417 AGN. An additional 27 sources observed with X-shooter are classified as blazars or gravitationally lensed sources. These sources are not included in any analysis, but their spectral properties are measured and included in our final table. We then incorporate 19 AGN from X-shooter-IFU observations \citep[0.994$-$2.479~\mum]{Davies:2015:127,Burtscher:2021}, all of which have \pab\ coverage ($z \lesssim 0.9$). We include 17 unique sources observed with Magellan using the Folded-port InfraRed Echellette (FIRE, 0.8$-$2.5~\mum) that were not observed with X-shooter \citep{Ricci:2022:8}, and are of comparable spectral resolution, $R=6,000$. All FIRE spectra are at $z \lesssim 0.2$, and yields a total sample of 453 AGN, with a sample median of $z \approx 0.036$.

Our final sample totals 453 unbeamed AGN (Table~\ref{tab:sample_tab}), of which 223/453 (49\%) are Seyfert 2, 76/453 (17\%) are Seyfert 1.9, and 154/453 (34\%) are Seyfert 1–1.8 type AGN with broad \hb. The sample is dominated by Seyfert 2 AGN, followed by Seyfert 1, and Seyfert 1.9, similar to the parent sample of BAT-detected AGN \citep{Koss:2017:74}, as well as previous BASS NIR data releases DR1 \& DR2 (\citealt{Lamperti:2017:540}, \citealt{Brok:2022:7}, respectively).

\begin{figure}
\centering
\includegraphics[width=\columnwidth]{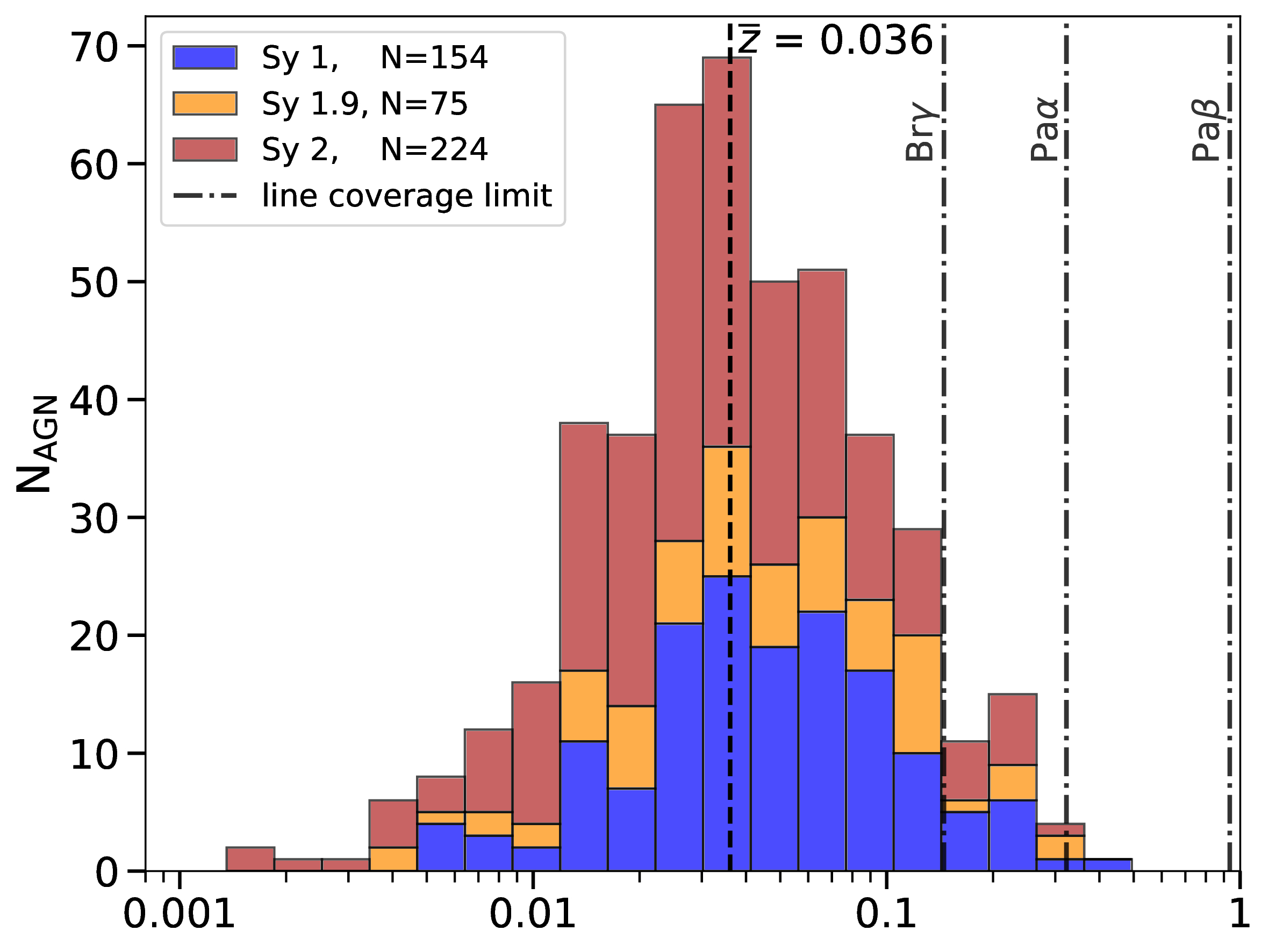}
\includegraphics[width=\columnwidth]{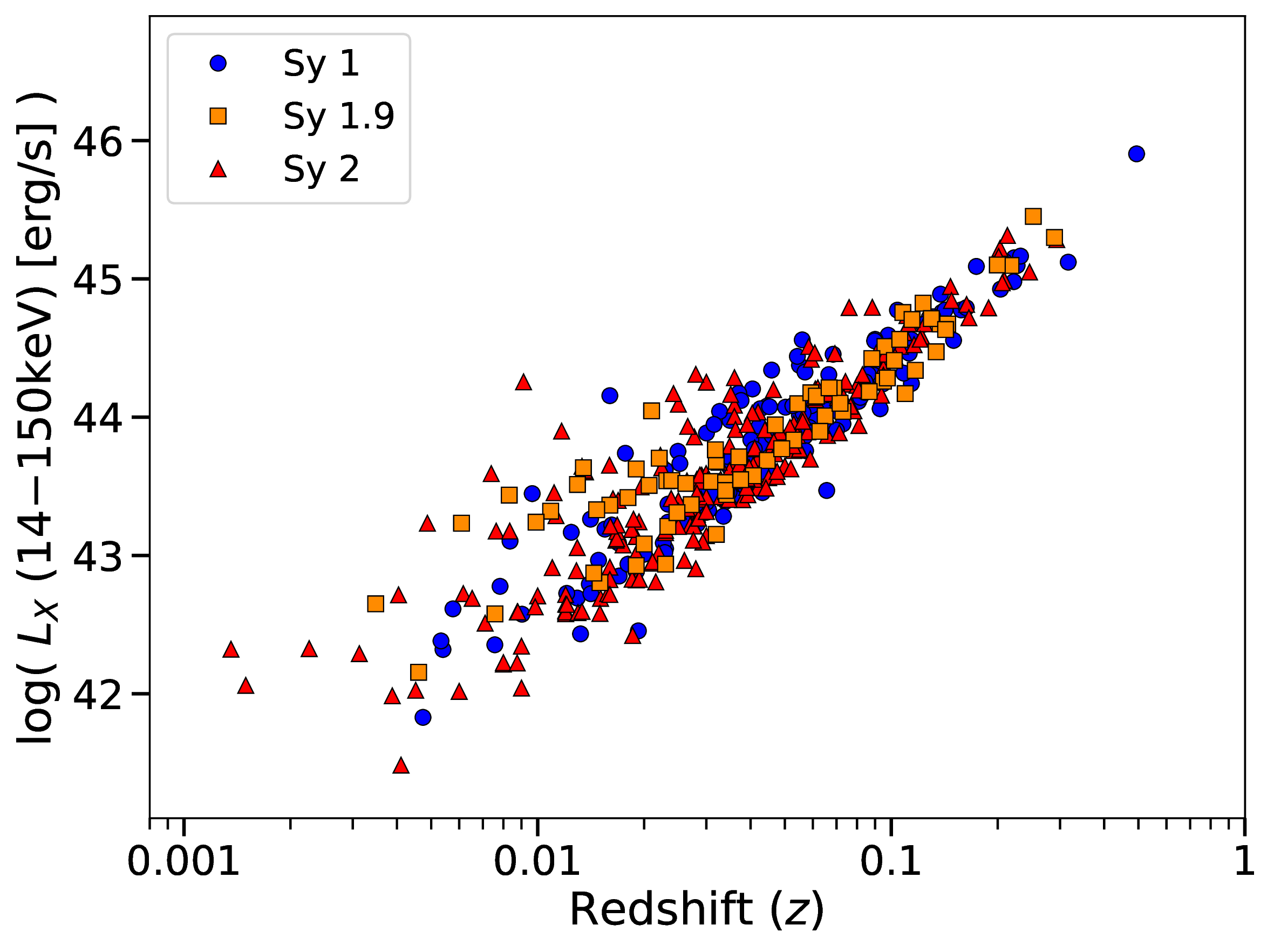}
\caption{\textit{Top}: Redshift histogram (binned for uniform width) for the full sample of 453 AGN, with median redshift $z\approx0.036$, and a vertical dash-dotted line indicating maximum redshift for emission line coverage. \textit{Bottom}: Distribution of intrinsic $L_{\text{X}}$(14$-$150 keV) vs redshift. Blue dots indicate AGN with Seyfert 1 optical classification, orange squares for Seyfert 1.9, and red triangles for Seyfert 2. }
\label{fig:fig_z_hist_xlum}
\end{figure}

Figure~\ref{fig:fig_z_hist_xlum} presents the distribution of redshifts for the sample, with median $z\approx0.036$, and the distribution of redshift against intrinsic X-ray luminosity $L_{\text{X}}$(14$-$150 keV). This sample of NIR data observed up to December 2023 is a component of the data release 3 (DR3) of BASS, and the additional reduced spectra will be made public on the BASS survey website\footnote{\url{https://www.bass-survey.com/}}.  The optical spectra for these sources will be presented in a separate publication (Koss et al., in prep.). 

\subsection{Observations} 
\label{sec:observations}

Previous BASS NIR data release 2 \citepalias[DR2;][]{Brok:2022:7} included X-shooter observations carried out in service mode between 2016 and 2019 (under the ESO run IDs 98.A-0635, 99.A-0403, 100.B-0672, 101.A-0765, 102.A-0433, 103. A-0521, and 104.A-0353). Our latest X-shooter observations were conducted after 2018 and most recently up to 2023 (105.20DA.001, 106.21B1.001, 108.229H.001, 109.22YE.001, 110.241X.001, and 112.25LU.001). These observations focus on Seyfert 1.9/2 or newly identified AGN, and follow a similar configuration and strategy as previous X-shooter observations. They are configured for full NIR spectral coverage (0.994$-$2.479 \mum), and used uniform slit widths of 0.9" ($R \approx 5400$) and exposure times ranging from 500 to 2000s (median of 1000s). The observations are summarized in Table~\ref{appendix:tab_info}. This work focuses on the NIR observations (J, H, and K-bands).

\subsection{Data Reduction} 
\label{sec:sec_reductions}
Spectra from X-shooter were reduced using the standard ESO reflex pipeline \citep{Freudling:2013:A96}, following the same procedures as \citetalias{Brok:2022:7}. The ESO Pipeline v2.9.3 was used with the default parameters for creating the calibration frames. Science and flux-standard frames are transformed into flat-fielded, rectified, and wavelength-calibrated 2D-order spectra using the \textsc{xsh\_scired\_slit\_nod} recipe. A standard 4$''$ extraction on a spectrophotometric standard star observed during the same night was used to derive the nightly instrument response function. Thus, typical calibrated science spectra have relative flux accuracy of a few percent, and typical absolute flux calibration uncertainty $\lesssim10\%$ \citep{Schonebeck:2014:A13,Sana:2024:A104}. 

Spectra obtained with FIRE were reduced using the \textsc{IDL} pipeline \textsc{FireHose} package \citep[v2;][]{Gagne:2015}, which performs 2D sky subtraction and extracts an optimally weighted 1D spectrum. A0V stars were observed to derive relative flux calibrations. Further details of these observations are presented in \citet{Ricci:2022:8}.

We removed atmospheric absorption effects that contaminated the spectra (e.g., H2O, O2, and CO2) using the software \molecfit\ (v1.5.9; \citealt{Kausch:2015:A78, Smette:2015:A77}), following the method described in \citetalias{Brok:2022:7}. \molecfit\ uses a radiative transfer code to simulate the atmosphere by adopting atmospheric parameters recorded during observations, including ambient temperature, pressure, mirror temperature, and outside humidity. For best results with \molecfit, the telluric features should not be saturated, and AGN absorption/emission features should be avoided. It can also not correct telluric absorption properly in low SN conditions (SN~$>5$ per resolution element). \molecfit\ does not require telluric standard star observations and can simulate atmospheric change at shorter timescales than standard star methods, making it account better for rapid changes from water, with smaller scatter \citep{Ulmer:2019}.

\begin{deluxetable}{lcr}
\tablecaption{Instruments composing our full sample, their wavelength coverage, and total contribution.\label{tab:sample_tab}}
\tablewidth{0pt}
\tablehead{
   \colhead{Instrument name} & \colhead{Wavelength coverage} & \colhead{Number} \\
   \colhead{} & \colhead{(\(\mu\)m)} & \colhead{}
}
\startdata
VLT/XShooter         & 0.994$-$2.101 & 82 \\
VLT/XShooter         & 0.994$-$2.479 & 335 \\
VLT/XShooter-IFU     & 0.994$-$2.479 & 19 \\
Magellan/FIRE        & 0.8$-$2.5     & 17 \\
Total                & ---           & 453 \\
\enddata
\end{deluxetable}

\section{Spectral Measurements} \label{sec:measurements}

This work expands and complements the NIR analysis done for the BASS sample in DR2 with additional analysis to the K-band (2$-$2.5~\mum) that was not performed. Extending the spectral coverage to 2.5~\mum\ allows us to observe the \brg\ emission line in sources ideally up to $z \approx$~$0.145$. Fitting is performed with the spectroscopic toolkit \textsc{PySpecKit} (v0.1.20; \citealt{Ginsburg:2011:1109.001}), and following similar procedures as \citet{Lamperti:2017:540} and \citetalias{Brok:2022:7}. We first correct Galactic extinction for every spectrum by using the built-in deredden function, which considers the \ebv\ value (values from \citealt{Schlegel:1998:525}). Following \citetalias{Brok:2022:7}, we subtract the instrumental dispersion from the measured line FWHM in quadrature. For both FIRE and X-shooter, the instrumental dispersion is $\sim$50 \kms, which is negligible compared to our results.

NIR spectra are subdivided into regions to best fit the local continuum near the strong and common emission lines before line fitting. These regions are: \pae\ (0.94$-$0.98~\mum), \sfviii\ (0.97$-$1.0~\mum), \pag\ (1.0$-$1.15~\mum), \pab\ (1.15$-$1.35~\mum), Br10 (1.4$-$1.5~\mum), \feii\ (1.6$-$1.7~\mum), \paa\ (1.8$-$2.02~\mum), and \brg\ (2.02$-$2.35~\mum). \sfviii\ is near a spectral cut at 1~\mum\ because, depending on redshift, part of the rest 0.94$-$1.0~\mum\ region is in the observed NIR arm and part in the VIS arm. By separating \sfviii\ and \pae\ into two fitting regions, flux calibrations, and continuum modeling issues are minimized.

\begin{figure*}
\centering
\includegraphics[width=\textwidth]{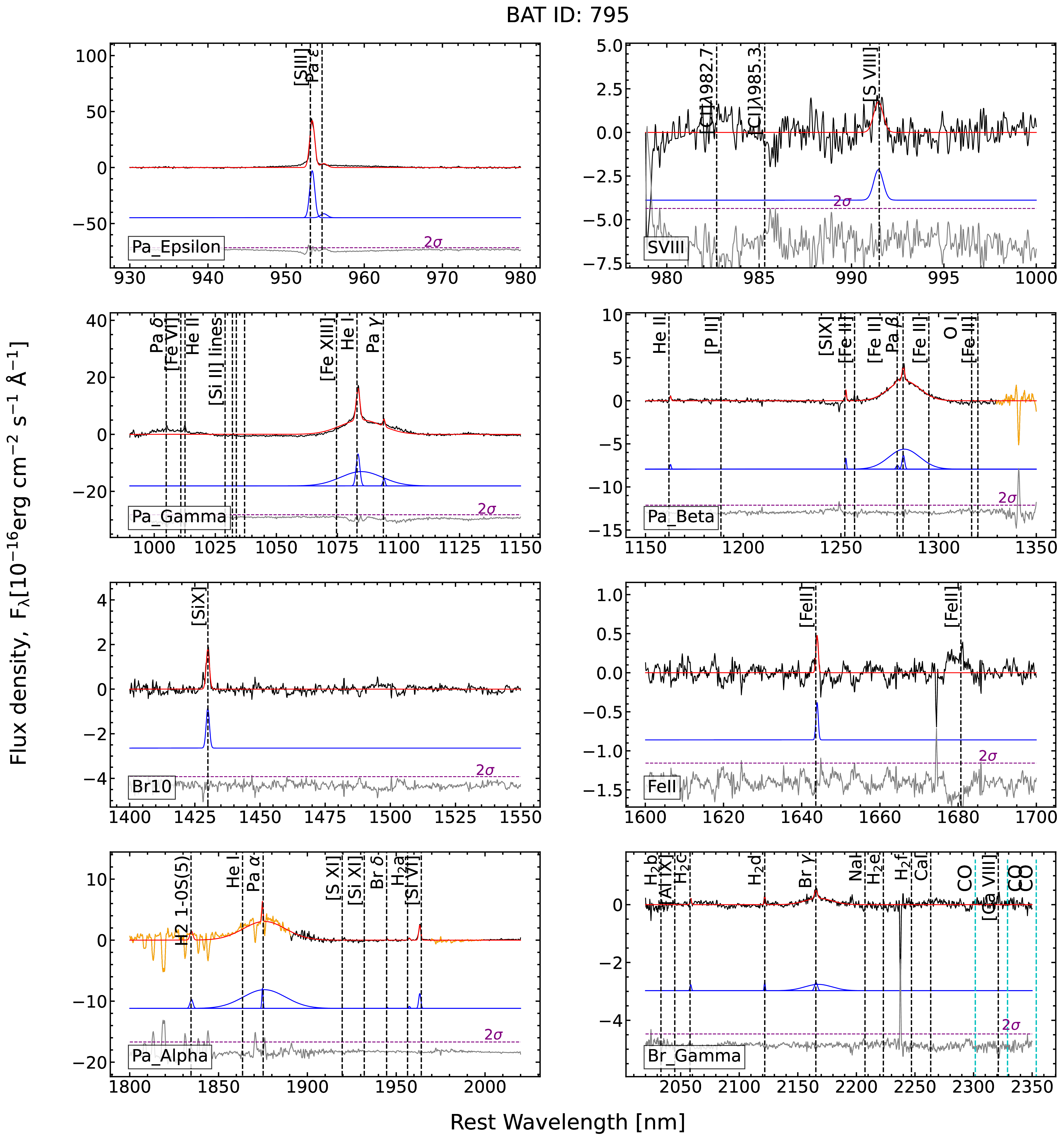}
\caption{Example of simultaneous emission-line fits for galaxy LEDA 2793282 (BAT ID 795), observed with VLT/X-shooter. The eight panels show the fitting for each region, labeled in the lower left of each panel (from the upper left to the bottom right: \pae, \sfviii, \pag, \pab, Br10, \feii, \paa, and \brg). In each panel, spectral data are plotted in black, and the best fit is in red. Regions of intense telluric absorption are in orange. Below the spectrum is the emission-line model fit in blue, below the model is the spectrum residual in grey, and a 2$\sigma$ threshold in dashed magenta. Cyan indicates CO absorption regions, which may complicate fitting. In this example spectrum, Br10 is absent. \\(The complete figure set will be available.) }
\label{fig:fig_bat_795}
\end{figure*}

We first fit the local continuum in each spectral region similarly using a fourth-order polynomial as \citetalias{Brok:2022:7}, and we take care to mask emission lines and strong telluric regions (using regions $\approx$~0.015\mum\ to both the blue and red sides adjacent of the line location). AGN continuum modeling using a fourth-order polynomial has been done in several previous studies \citep[e.g.,][]{Krajnovic:2007, Raimundo:2013, Zeimann:2015, Husemann:2020}. An alternative spline-fit function is applied to some fitting regions, such as ``\sfviii'' and ``\feii'' as described in \citetalias{Brok:2022:7}. Additional regions are applied to 82/453 (18\%) of the spectra based on careful visual inspection of the continuum fit. Spline-fit is applied to better correct for intrinsically unusual continuum shapes, or to cases with strong telluric residuals, which allow for a more accurate estimate of the continuum level. An example fit is given in Figure~\ref{fig:fig_bat_795}.

Our wide passband allows for the detection of several strong hydrogen recombination lines (e.g., \paa\ and \pab), as well as \hei, \siii\ \lam 9531, \feii\, and \brg. All emission lines are initially modeled with a single Gaussian profile. For lines that exceed our defined detection threshold (see details below), we include a second Gaussian component to better capture complex line profiles, ensuring consistency with previous studies (e.g., \citetalias{Brok:2022:7}).\footnote{While Bayesian Information Criterion (BIC) can guide model selection, we adopt a simpler approach based on amplitude significance and noise, ensuring consistency with prior work and emphasizing physical detectability.} First, we fit the \pab\ region to constrain the broad line widths and velocity offsets for fitting other regions (\pag\ is used if \pab\ is not detected). We fit Gaussian profiles to narrow emission lines (FWHM $<$1200 \kms) and broad lines (FWHM $>$1200 \kms), which is consistent with the previous data releases \citet{Lamperti:2017:540} \& \citetalias{Brok:2022:7}. Relative velocity centers of the narrow lines are also tied together, and the width of the most prominent narrow line is utilized to limit the widths of other narrow lines in velocity space, permitting a variation of up to 200 \kms\ for narrow lines and 500 \kms\ for the broad lines. Many AGN emission line studies have empirically shown velocity shifts in \hb\ from the systemic redshift by up to a thousand \kms, with an average shift of $\sim$100 \kms\ \citep{Shen:2016}. It has been shown that the velocity shifts can be more extreme in high-ionization lines \citep[e.g.,][]{Gillette:2023:2578}. For these spectra, the high-ionization lines can be shifted up to 400 \kms, and this flexibility is adequate in previous data released for fitting these lines. We tie the line width to the FWHM of \pab, a prominent line, or \pag\ if \pab\ is not found.

To be considered as detected, the line amplitude of the fit Gaussian must exceed a signal-to-noise threshold, which we define as 2$\sigma$, where $\sigma$ is the noise level in the surrounding continuum. To estimate $\sigma$, we compute the root-mean-square (RMS) in a $\approx$~30\,nm window 15\,nm to both the blue and red sides adjacent to the expected line position to avoid the line itself. For broad lines this is increased to a $\approx$~150\,nm window 75\,nm to the blue and red sides. The integration width is set to the FWHM of other, more prominent emission lines. As a result, our sample is limited by equivalent width (EW) rather than flux. For emission lines that are not detected, we present conservative upper limits on the flux using a Gaussian amplitude F$_\text{UL}=3\sigma$, and assume a line width of FWHM $=1200$\kms. We estimate errors in fitted parameters using the same procedure as \citetalias{Brok:2022:7}, performing Monte Carlo simulations drawn from a normal distribution with a standard deviation equal to the noise level in the spectrum. 

\subsection{Virial Black Hole Masses} \label{sec:bhm}

Wide wavelength coverage (0.94$-$2.4\mum) allows observation of several NIR broad lines, and we compute estimates of \Mbh\ from multiple indicators (assuming that the broad emission is virialized). In this work, we include many sources with limited or no optical broad-line detections (e.g., Seyfert 1.9/2). Thus, we use the broad-line detection of \pab, or alternatively \paa,  to estimate \Mbh. 

Throughout our analysis, we adopt a common virial factor of one for \Mbh\ estimates based on the FWHM of broad emission lines by scaling the masses by $-0.13$~dex from the prescription in \citet{Kim:2010:386}. We acknowledge that the systematic uncertainty in the virial factor is one of the largest sources of scatter in virial mass estimates \citep[with literature values typically spanning $\approx$~0.7 to 1.1; e.g.,][and references therein]{Greene:2005:122,LaFranca:2015:1526,Woo:2015,Yong:2016:009,Mejia-Restrepo:2018:63}. This uncertainty is sub-dominant compared to other sources of error (e.g., uncertainties in line width, luminosity measurements, and calibration of scaling relations). Our choice is consistent with previous BASS/DR2 analyses, as well as previous NIR studies \citep[e.g.,][]{Greene:2005:122, Woo:2015, Mejia-Restrepo:2018:63}, and facilitates a direct comparison between \Mbh\ estimates derived from different emission lines. Moreover, as our analysis primarily focuses on relative differences and trends, the systematic uncertainty in virial factors does not qualitatively affect our main conclusions.

For sources with Paschen $\alpha$ or $\beta$ broad-line detections, the virial black hole mass ($M_\text{BH,vir}$) is defined by: 

\begin{align*}\label{eq_vbhm}
\centering
\log \left( \frac{ M_\text{BH,\paa}}{M_{\odot}} \right) = 7.16 &+ 1.92 \, \log \left( \frac{ \text{FWHM}_{\paa}}{10^{3} \text{ km/s } } \right) \\ &+ 0.43 \, \log \left( \frac{ L_{\paa}}{ 10^{42} \text{ erg/s} } \right),\\
\text{or,} \\
\log \left( \frac{ M_\text{BH,\pab}}{M_{\odot}} \right) = 7.20 &+ 1.69 \, \log \left( \frac{ \text{FWHM}_{\pab}}{10^{3} \text{ km/s } } \right) \\ &+ 0.45 \, \log \left( \frac{ L_{\pab}}{ 10^{42} \text{ erg/s} } \right),
\end{align*}

where $\text{FWHM}$ and $L$ are the full width half maximum and line luminosity of the broad component of the emission line profile, respectively \citep{Kim:2010:386}.

\section{Results}

Our analysis first examines the Paschen line properties (\paa\ and \pab). We then present our findings with the CL measurements and the X-ray emission. In addition, we continuously compare our results against the previous BASS NIR data releases. 

% Results
\subsection{Broad and Narrow Paschen Lines}

We examine the distribution of line properties from the broad \paa\ and \pab\ measurements, and we consider an emission component with FWHM greater than $1200$~\kms\ to be a broad line source, consistent with what is usually done for optical lines \citep[e.g.,][]{Oh:2022:4}. First, among the 442 spectra with redshifts that allow \paa\ coverage and have adequate continuum on either side to estimate spectral noise for upper limits, we detect broad \paa\ in 153/442 (35\%) cases. For broad \pab, we find 162/453 (36\%) detections. We have 211 sources with both or either broad component detections. Both detections correlate with the luminosity of the AGN and other hydrogen lines \citep[e.g.,][]{Landt:11:106}. These total detections are comparable, and we prefer to use \paa\ for analysis because its longer wavelength (1.875 \mum) makes it less affected by possible extinction compared to \pab\ (1.282 \mum). We find the average FWHM of broad \paa\ is $3518 \pm 1582$~\kms, and the average FWHM of broad \pab\ is $3829 \pm 1870$~\kms. Performing a Kolmogorov–Smirnov test (K-S), the p-value $= 0.24$, suggesting that it cannot be excluded that they are from the same distribution.\footnote{A p-value $\leq$ 0.05, suggests statistically significant.} Details of \paa, \pab, and \brg\ detection, and fraction which have broad or narrow measurements, are presented in Table~\ref{tab:tab_paschen_det}.

\begin{deluxetable}{lccccc}
\tablecaption{Detection totals for the \pab, \paa, and \brg\ emission lines. \label{tab:tab_paschen_det}}
\tablewidth{0pt}
\tablehead{
    \colhead{Line} & \colhead{Source} & \colhead{N$_\text{Telluric}$} & \colhead{N$_\text{Det.}$/Total} & \colhead{Broad} & \colhead{Narrow}
}
\startdata
\pab\ & Sy 1   & 145      & 118 / 154 & 105 & 94 \\
      &        & (94.2\%) & (76.6\%)  & (68.2\%) & (61.0\%) \\
      & Sy 1.9 & 69        & 41 / 75   & 31  & 41 \\
      &        & (92.0\%)  & (54.7\%)  & (41.3\%) & (54.7\%) \\
      & Sy 2   & 209       & 85 / 224  & 26  & 114 \\
      &        & (93.3\%)  & (37.9\%)  & (11.6\%) & (50.9\%) \\
\paa\ & Sy 1 & 145      & 132 / 151 & 104   & 106 \\
      &           & (96.0\%)  & (87.4\%)  & (68.9\%) & (70.2\%) \\
      & Sy 1.9   & 72       & 59 / 74   & 26    & 57 \\
      &           & (97.3\%)  & (79.7\%)  & (35.1\%) & (77.0\%) \\
      & Sy 2     & 212       & 150 / 217 & 34    & 157 \\
      &           & (97.7\%)  & (69.1\%)  & (15.7\%) & (72.4\%) \\
\brg\ & Sy 1   & 39       & 35 / 131  & 23    & 23 \\
      &        & (29.8\%)  & (26.7\%)  & (17.6\%) & (17.6\%) \\
      & Sy 1.9 & 19       & 19 / 60   & 12    & 14 \\
      &        & (31.7\%)  & (31.7\%)  & (20.0\%) & (23.3\%) \\
      & Sy 2   & 40       & 46 / 155  & 14    & 37 \\
      &        & (25.8\%)  & (29.7\%)  & (9.0\%)  & (23.9\%)  \\
\enddata
\tablecomments{For each line, we report the total number of detections, as well as the number of sources exhibiting broad and narrow emission components. We also include the total number of sources close enough to the Telluric regions in Table~\ref{tab:tab_telluric} to potentially interfere with emission line detection and continuum fitting.}
\end{deluxetable}

\begin{deluxetable}{cc}
\tablecaption{Primary NIR Telluric Regions.\label{tab:tab_telluric}}
\tablewidth{0pt}
\tablehead{
  \colhead{Molecule} & \colhead{Telluric Region Wavelength} \\
  \colhead{}       & \colhead{(\(\mu\)m)}
}
\startdata
\hto\ & 1.12 $-$ 1.15 \\ 
\hto\ & 1.35 $-$ 1.49 \\ 
\hto\ & 1.79 $-$ 1.98 \\ 
\cdo\ & 2.00 $-$ 2.02 \\ 
\cdo\ & 2.05 $-$ 2.08 \\ 
\chf\ & 2.35 $-$ 2.36 \\ 
\enddata
\end{deluxetable}

\begin{figure}
\centering
\includegraphics[width=\columnwidth]{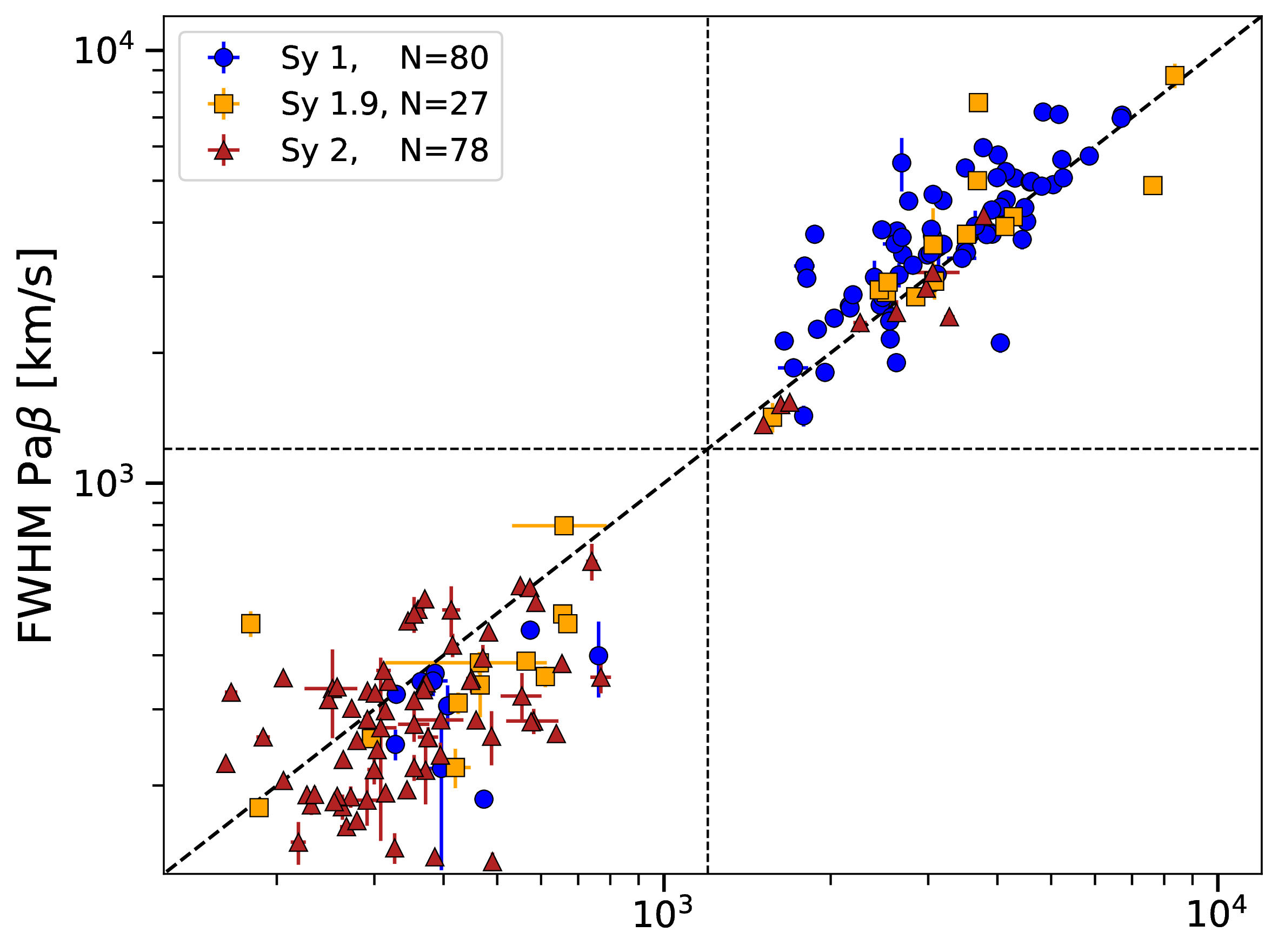}
\includegraphics[width=\columnwidth]{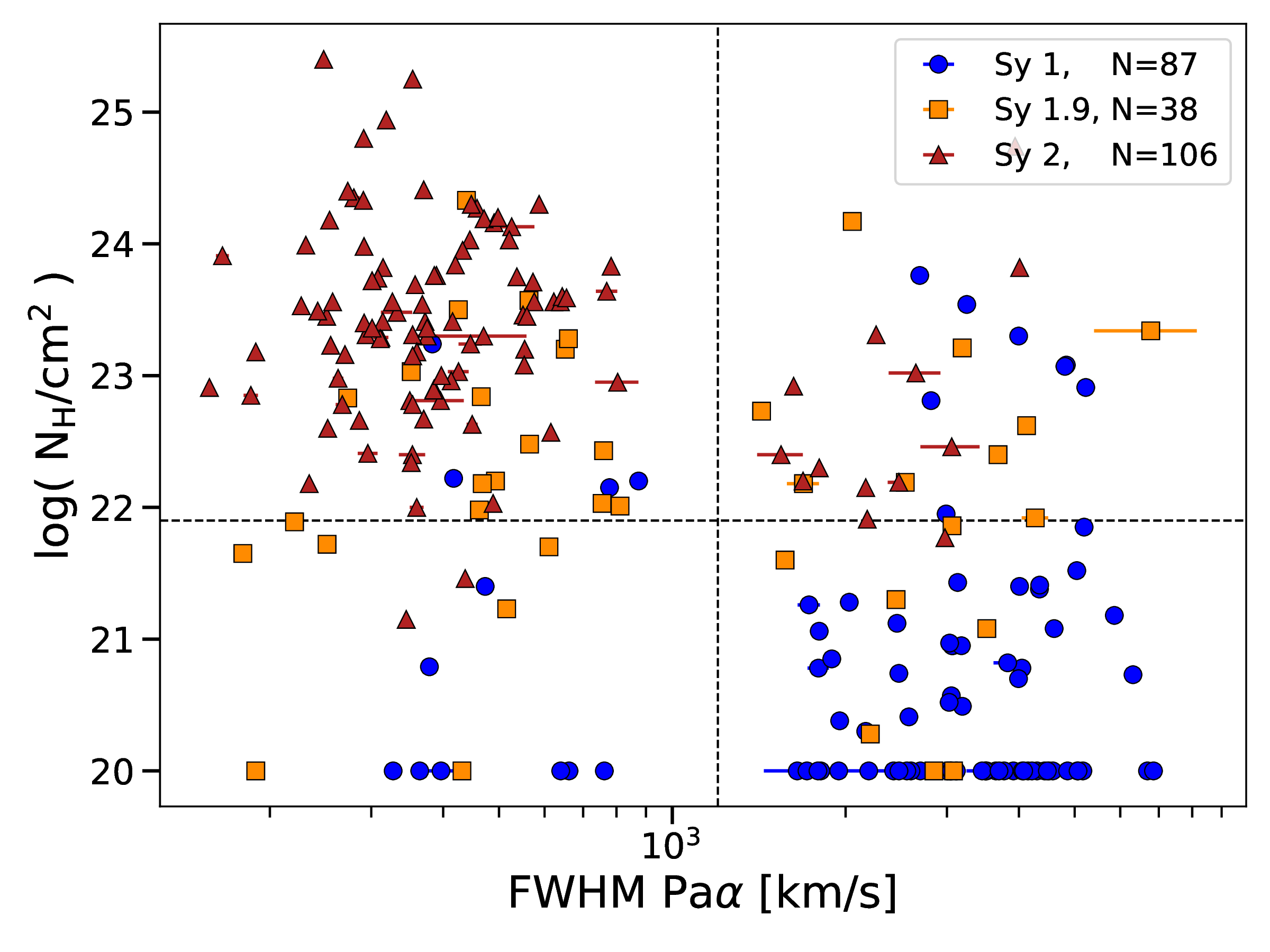}
\caption{\textit{Top}: FWHM \paa\ vs. \pab, separated by optical Seyfert type, and with a diagonal dashed black line indicating a one-to-one ratio of widths, and dashed lines indicating the division between narrow and broad Paschen lines vertically and horizontally. \textit{Bottom}: FWHM \paa\ vs \NH, the vertical dashed line indicates the boundary for broad/narrow \paa. The horizontal dashed line to indicates the threshold \NH\ value that separates optical Seyfert 1 or Seyfert 2 using \hb\ \citep[21.9,][]{Koss:2017:74}. A similar lack of sources in the FWHM 800$-$1400 \kms\ range is also seen in \citetalias{Brok:2022:7}, likely reflecting standard fitting constraints rather than a physical gap. }
\label{fig:fig_fwhm_paa_pab_nh}
\end{figure}

The top panel of Figure~\ref{fig:fig_fwhm_paa_pab_nh} shows the distribution of broad or narrow component FWHM line widths for AGN sources containing both \paa\ and \pab\ detections. Measurement errors are included for plotting; however, it should be noted that these errors can be dominated by systematic factors, particularly from weak emission or telluric absorption. \citet{Lamperti:2017:540} found in their distribution 16 AGN with broad component \paa\ and \pab\ lines, and the p-value derived from the K-S~$ = 0.63$, suggesting no significant difference in the distributions. Line widths of the \pab\ broad component appear larger than \paa\ in our figure, with the average FWHM of broad \pab\ $\approx 300$~\kms\ larger, but does not each a 3$\sigma$ significance level. This trend between the FWHM of the broad component of \paa\ being narrower than other lines, including \pab, is also present in nearly all sources in \citet{Lamperti:2017:540}. 

Comparing the optical Seyfert classification to NIR line properties in the top panel of Figure~\ref{fig:fig_fwhm_paa_pab_nh} reveals a general agreement with expectations. Counting sources with simultaneous \paa\ and \pab\ detections, 71/80 (89\%) of Seyfert 1 sources exhibit a broad component in Paschen lines, and 12/21 (57\%) and 9/75 (12\%) for Seyfert 1.9 and Seyfert 2, respectively. We do not detect both Paschen lines for all sources because one may be low signal-to-noise or extincted by Telluric absorption. Both narrow and broad-line classifications are consistent between \paa\ and \pab, meaning that there are no narrow \pab\ profiles with simultaneous broad component \paa, as what was used to define Seyfert 1.9 with Balmer lines in the optical. 

A linear fit with a slope of one was performed between the broad \paa\ and broad \pab\ emission widths, resulting in a linear offset of ($0.058 \pm 0.009$)~dex, a small difference that is not visible in the figure. This fit offset is smaller than the previous correlation in \citealt{Lamperti:2017:540} ($0.093 \pm 0.005$)~dex. This distribution in broad \paa\ and \pab\ also indicates an agreement between these lines if one is not detected due to systematics, e.g., telluric absorption; in such a case, the detected line may be used to approximate \Mbh.

% Results
\subsection{Hidden Broad Lines}

Our sample has 224/453 (49\%) sources categorized as Seyfert 2 based on their narrow optical Balmer lines (FWHM $<$ 1200\kms). 156/224 (70\%) of these sources have \NH\ measurements, and from those, 8/156 (5\%) have both broad \paa\ and \pab\ lines. 

The bottom panel of Figure~\ref{fig:fig_fwhm_paa_pab_nh} shows the distribution of 232 sources based on their \paa\ line widths, using the broad component when available or the narrow component otherwise, and their hydrogen column density (\NH). \citet{Oh:2022:4} shows a similar trend, which uses the optical emission line \ha. A vertical line in Fig.~\ref{fig:fig_fwhm_paa_pab_nh} indicates the \paa\ broad/narrow line distinction and the horizontal is the X-ray unobscured/obscured boundary \citep[log \NH/\cmN $>$ 21.9;][]{Koss:2017:74}. The line widths follow the established trend, where Seyfert 1 AGNs exhibit lower X‑ray obscuration, while Seyfert 2 show higher obscuration. For Seyfert 2, the average is log(\NH\/\cmN)$= 23.4 \pm 0.8$, well above the assumed lower limit for Seyfert 2 of log(\NH/\cmN)$>$ 21.9. Seyfert 1.9 sources are generally scattered between the two distributions. 

There may be some scatter in these general trends due to systemic errors, where residuals or chi-squared values may not indicate poor fits, but visual inspection of Seyfert 1 with narrow \paa\ and Seyfert 2 with broad \paa\ confirms that the fits are generally good. There are a handful of Seyfert 2 sources (11/153, 7\%) with broad \paa, and Seyfert 1 (12/105, 11\%) having no broad \paa, excluding Seyfert 1.9 (examples shown in Section~\ref{appendix:hidden_lines} in the appendix). Instances of narrow \paa\ in Seyfert 1 may be cases of the broad emission being weakly present. Of these Seyfert 1, 5/12 (42\%) have measured \ha\ EW, and four are greater than or equal to the median (EW~$\geq165$\AA), ruling out the possibility that these are associated with weak \ha\ systems. Cases of broad \paa\ in Seyfert 2 may indicate that these are hidden broad line sources along a classification similar to Seyfert 1.9.

Considering the Seyfert 1.9 sources, there are 16/51 (31\%) with broad \paa\ detected. Among these, 14/16 have available \NH\ measurements (ranging from 20.28 $<$ log(\NH/\cmN)$<$ 24.17), and have an average value of log(\NH/\cmN)$=22.05 \pm 0.96$. There are 20 sources with narrow \paa\ and \NH\ (ranging from 21.23 $<$ log(\NH/\cmN)$<$ 24.33), with an average value of log(\NH/\cmN)$=22.39 \pm 0.76$. Of sources with \NH\ available, the narrow and broad \paa\ Seyfert 1.9 appear to be from the same underlying distribution (K-S p-value~=~0.90). Broad \paa\ shows a similar distribution as \citet{Ricci:2022:8}, with mostly Seyfert 1 at low \NH\ ($<21.9$ \cmN) and gradually becoming replaced by Seyfert 1.9 and 2 with increasing \NH. 

\begin{figure}
\centering
\includegraphics[width=\columnwidth]{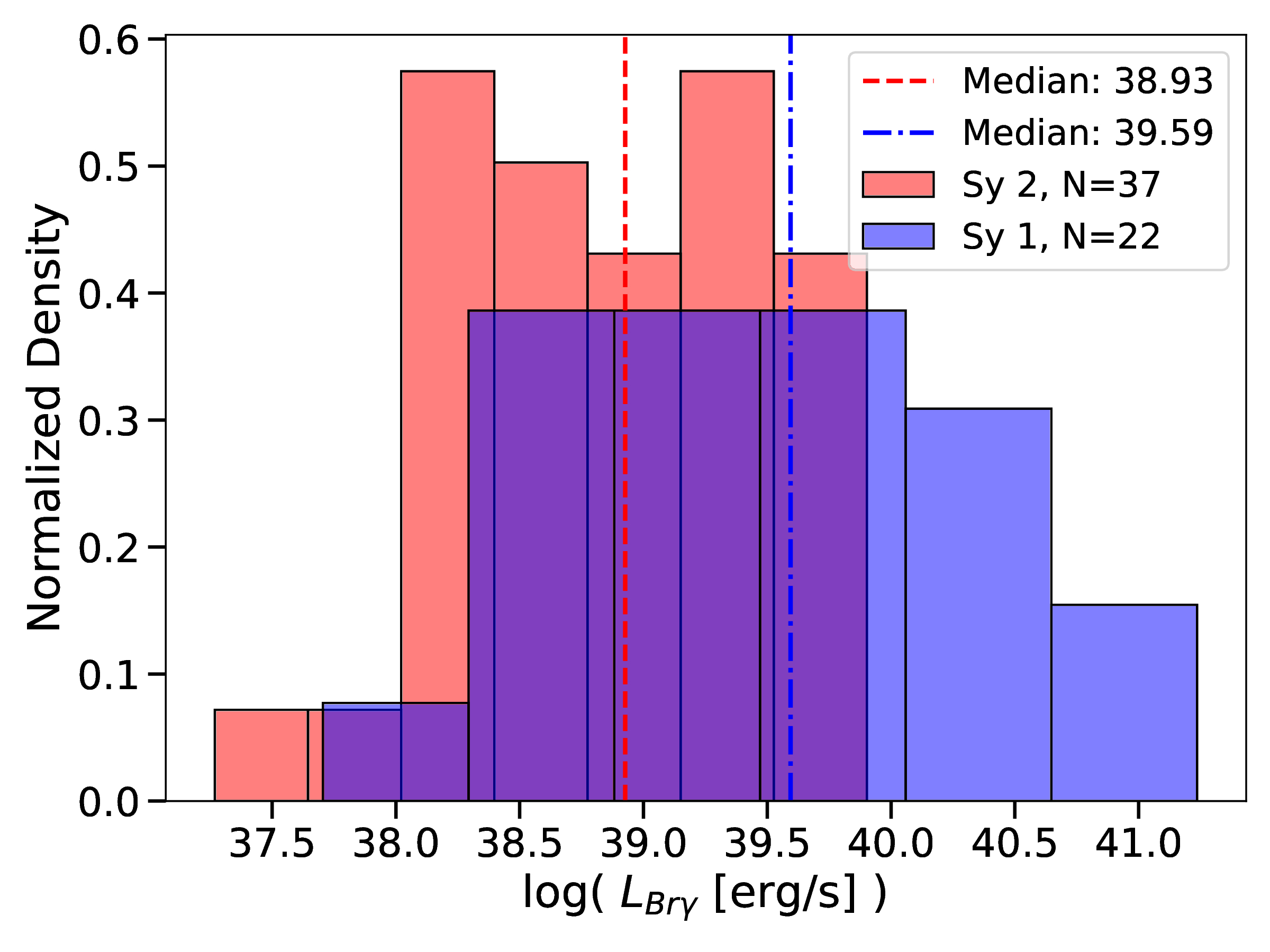}
\caption{Histogram across narrow \brg\ luminosity, categorized by Seyfert 1 or 2, and blue or red lines indicate kernel density estimation of the two distributions. Statistical analysis suggests these detections are from different distributions (p-value~$\approx 2 \times 10^{-3}$).}
\label{fig:fig_hist_brg}
\end{figure}

Extending to yet redder recombination lines, we detect many \brg\ lines. Figure~\ref{fig:fig_hist_brg} shows the distribution of luminosity of the narrow component of \brg, and the distributions of Seyfert 1 (including 1.0$-$1.8) and Seyfert 2 appear to show an offset. For this analysis, we use the \scikitsurv\ package to perform a Kaplan–Meier (KM) survival test between the datasets, similar to the K-S method but allowing the inclusion of non-detections \citep{Polsterl:2019}. This yields a p-value $\approx 1 \times 10^{-4}$, suggesting we can reject that these come from the same distribution. Regarding \brg\ line emission detections, we find: 23/131 (18\%) narrow and 23/131 (18\%) broad \brg\ among Seyfert 1s; 37/155 (27\%) narrow and 14/155 (9.0\%) broad \brg\ among Seyfert 2s; and 14/60 (23\%) narrow and 12/60 (20\%) broad \brg\ among Seyfert 1.9s. 

% Results
\subsection{Black Hole Masses}

\begin{figure}
\centering
\includegraphics[width=\columnwidth]{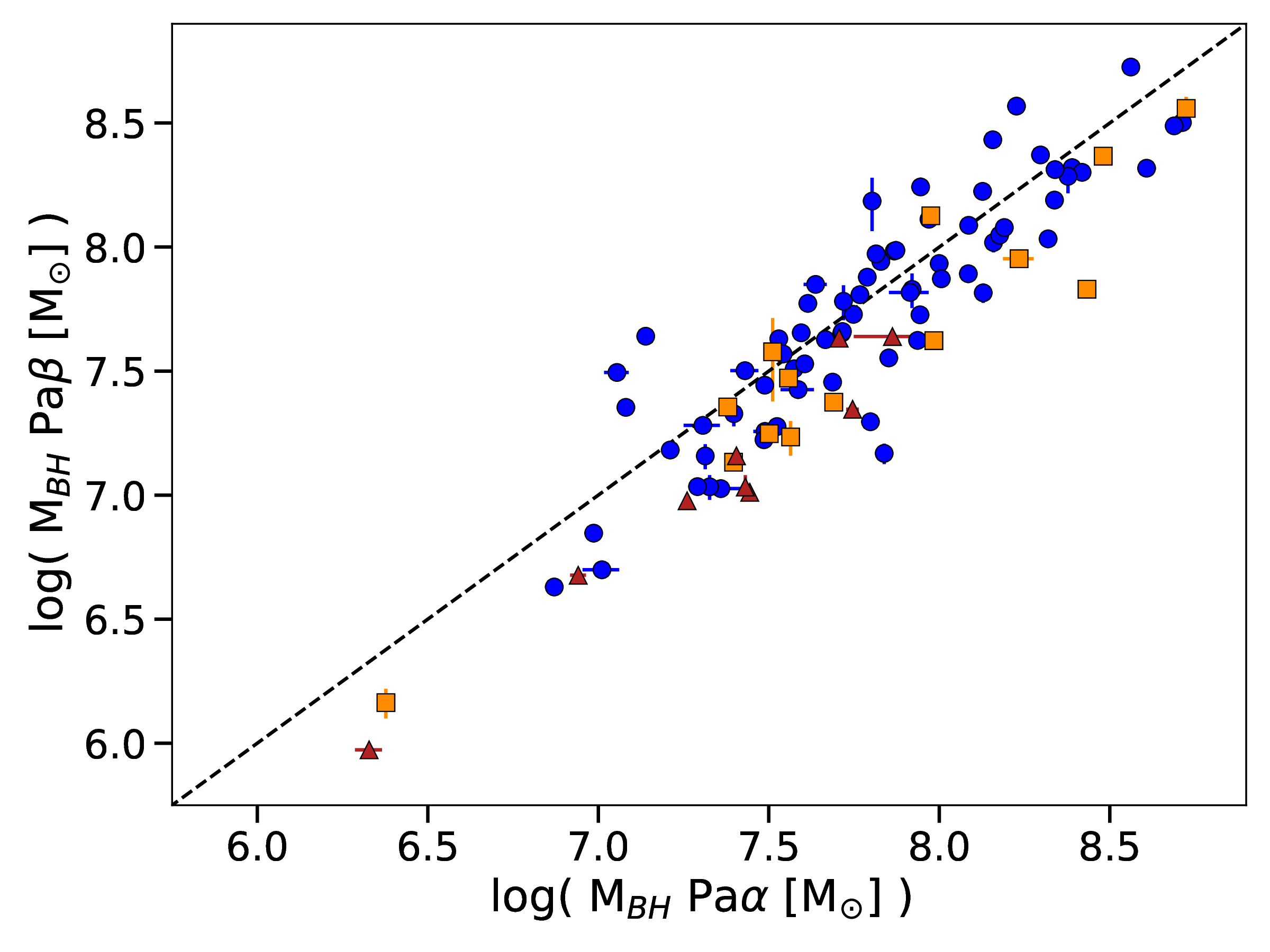}
\caption{\Mbh\ derived from FWHM \paa\ vs. \pab, using identical markers as Fig. \ref{fig:fig_z_hist_xlum}, and with a dashed black line indicating a one-to-one ratio of widths.}
\label{fig:fig_fwhm_paa_pab_bhm}
\end{figure}

Figure~\ref{fig:fig_fwhm_paa_pab_bhm} shows the distribution of virial black hole masses as derived from 96 sources with broad components detected both in \paa\ and \pab, and their respective mass estimates plotted against each other. Their distributions have an offset of $-$0.09~dex and a scatter of 0.23~dex. \paa\ is near a spectral region that is frequently impacted by atmospheric absorption, resulting in NIR studies typically having more \pab\ measurements than \paa. \citet{Lamperti:2017:540} found that \pab‐based \Mbh\ estimates are in good agreement with those derived from reverberation mapping, \hb, or the stellar velocity dispersion, using the M$-\sigma_*$. Therefore, we generally prefer to use \pab\ for \Mbh\ when available. Additionally, \citet{Lamperti:2017:540} observed that \paa\ \Mbh\ estimates also agree with those obtained from \hei, irrespective of Seyfert classification, as long as the broad line luminosity is not suppressed by high extinction (\NH$>$21$-$22~\cmN).

% Results
\subsection{NIR Diagnostics}

Recent JWST studies have used diagnostic diagrams to classify new identified objects \citep[e.g.,][]{Harkiane:23:39,Ubler:23:145,Kocevski:23:4,Killi:24:52}, and NIR diagnostics could further benefit such studies since redder emission suffers less from extinction in obscured sources. We use a NIR diagnostic diagram to distinguish AGN from SF galaxies, using a wide range of NIR J and K-band measurements (\pab, \feii\ \lam1.257, H$_2$ \lam2.121, and \brg; e.g., \citealt{Larkin:98:59}; \citealt{Riffel+06}; \citealt{Lamperti:2017:540}).  

\citet{Lamperti:2017:540} found that SF galaxies appear indistinguishable from AGN based on the \feii/\pab line ratio regions defined by \citet{Riffel:13:2587}, while the H$_2$/\brg\ ratio significantly separates SF galaxies and AGN. We instead use a refined version of the NIR diagnostic discussed in \citet{Riffel:13:2587} to more clearly distinguish AGN from SF galaxies. We shift the emission ratio bounds to define a SF region that excludes all AGN sources, and vis a versa, as well as third region that comprises both sources overlapping.

\begin{figure}
\centering
\includegraphics[width=\columnwidth]{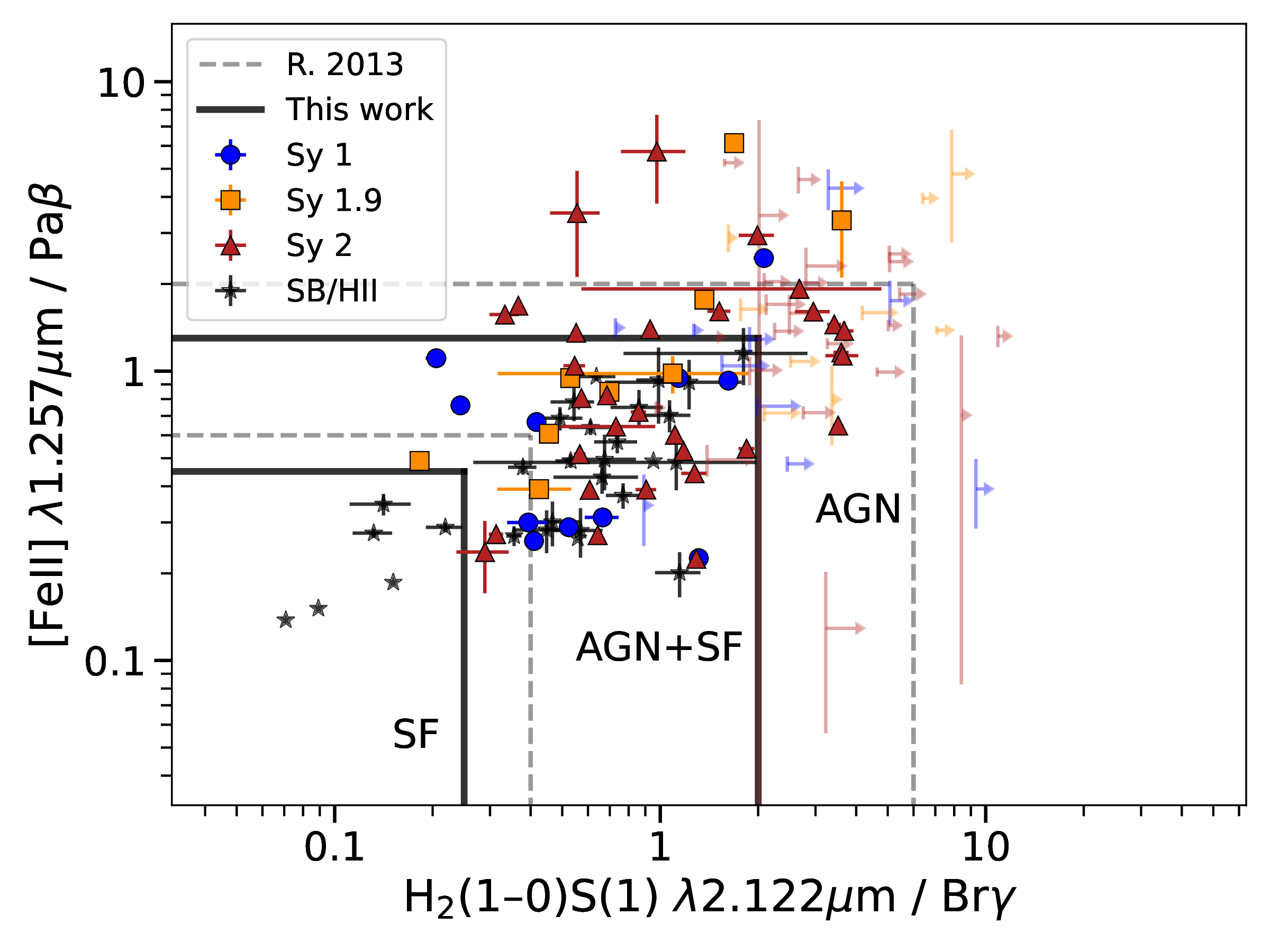}
\caption{NIR diagnostic plot refined from \citet{Riffel:13:2587}, using identical markers as Fig. \ref{fig:fig_z_hist_xlum}, and additional gray stars for starburst (SB) galaxies and \hii\ region sources from \citet{Larkin:98:59,Dale:2004:813,Riffel:2006:61,Martins:2013:1823}. We show 1$\sigma$ error bars for detections, and 3$\sigma$ upper limits. We find that our sample is more effectively classified by our modified regions of H$_2$/\brg\ and \feii/\pab\ ratios denoted by solid lines \citep[compared to the dashed lines from][]{Riffel:13:2587}, which define an overlap region that includes many AGN and SB/\hii\ sources. }
\label{fig:fig_diagnostic_plot}
\end{figure}

Figure~\ref{fig:fig_diagnostic_plot} presents our refined NIR diagnostic diagram, the regions are defined by eye, and are guided by the aim of cleanly separating AGN from SF sources and informed by the distributions seen in our sample. We find 51/372 (14\%, excluding sources without K-band coverage) sources with simultaneous detections of these four emission lines. This low rate is partially due to \brg\ being relatively faint, and often near a Telluric region. Of the sources with four detected lines, none fall in the SF galaxy region, 32/48 (67\%) are in the AGN and SF region, and 16/48 (33\%) are in the AGN region. Nearly all sources with upper limits for H$_2$ (35/42, 83\%) fall within the AGN classification, yielding a total for detection and upper limits of 54/93 (58\%). We do not consider our low-ionization nuclear emission-line sources (LINERS) classification according to the line ratios in \citet{Riffel:13:2587}. It should be noted that \citet{Riffel:13:2587} defined the upper bounds for AGN classification to distinguish sources with LINER characteristics. The BASS sample is almost exclusively AGN, and are not LINERS, sometimes also assigned to shocks \citep{Riffel:20:4857}. We consider trends in the fraction of sources in the AGN region to the composite AGN and SF, binned in X-ray (14$-$150~keV) luminosity. We find marginal difference between the fraction of AGN to composite sources from $\sim7\times 10^{42}$ to $\sim5\times10^{44}$~erg/s, and across Seyfert classification. We do not investigate scaling with permitted or forbidden line luminosity. Totals in each region are summarized in Table~\ref{tab:nir_diagnostic_tab}. We only plot sources with upper limits in \brg, as it is the weakest line in this diagnostic (upper limits are at 3$\sigma$), typically followed by H$_2$. When H$_2$ is undetected, \brg\ is also undetected. In contrast, \feii\ and \pab\ are generally stronger; if these lines are undetected, the other axis in the diagram would also be unconstrained. Additional details of the molecular H$_2$ line detections are presented in Section~\ref{sec:section_h2_emit}.

\begin{deluxetable}{lccc}
\tablecaption{Distribution of sources across NIR diagnostic region, with the percentage of the sources found in each region.\label{tab:nir_diagnostic_tab}}
\tablewidth{0pt}
\tablehead{
  \colhead{Source} & \colhead{SF} & \colhead{SF+AGN} & \colhead{AGN} \\
  \colhead{}     & \colhead{H$_2$/\brg\ $<$ 0.25} & \colhead{} & \colhead{H$_2$/\brg\ $>$ 2} \\
  \colhead{}     & \colhead{and} & \colhead{} & \colhead{and/or} \\
  \colhead{}     & \colhead{\feii/\pab\ $<$ 0.45} & \colhead{} & \colhead{\feii/\pab\ $>$ 1.3}
}
\startdata
SB/\hii\    & 6 (20.7\%)   & 23 (79.3\%)  & 0 \\
Total AGN   & 0            & 39 (43.8\%)  & 50 (56.2\%) \\
Detected    & 0            & 32 (66.7\%)  & 16 (33.3\%) \\
Upper-limit & 0            & 7 (17.1\%)   & 34 (82.9\%)
\enddata
\end{deluxetable}

% Results
\subsection{Coronal Line Measurements}

\begin{figure*}
\centering
\includegraphics[width=\textwidth]{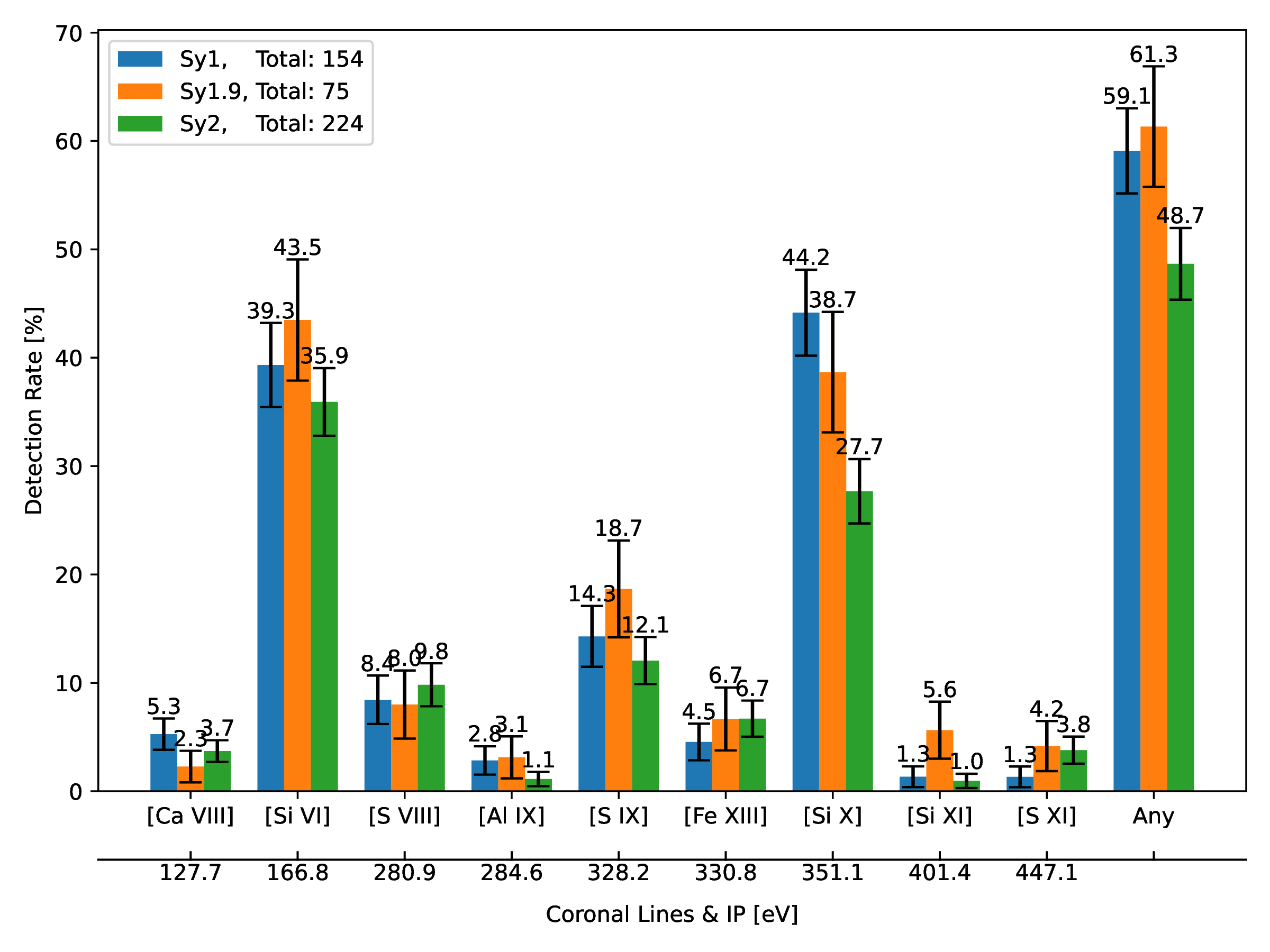}
\caption{Histogram of NIR CLs, ordered left to right in increasing ionization potential and divided by Seyfert classification. The error bars presented here use the 1-$\sigma$ binomial proportion confidence interval.}
\label{fig:fig_CL_histogram}
\end{figure*}

If CLs trace AGN activity, they should be observable in all bright, nearby AGN detected in hard X-ray \citep{Rodriguez-Ardila:2011:100, Lamperti:2017:540, Brok:2022:7, Negus:21:62, Negus:23:127, Bierschenk:24:257B}. Figure~\ref{fig:fig_CL_histogram} shows the distribution of CL detection percentages, with increasing ionization potential from left to right. Detection rates are shown, provided there is adequate wavelength coverage, and lines that have redshifted out of observability are not considered. Two lines stand out with the highest detection rates, \sivi\lam 1.940 with 161/422 (38\%) and \six\\lam 1.4300 with 150/449 (33\%) detections. There is little distinction between Seyfert 1 and Seyfert 2 in detection rates, although \six\ shows more significant detection in Seyfert 1/1.9 than in Seyfert 2, and is consistent with results in \citetalias{Brok:2022:7}.

Generally, we do not see a trend in detection rates with ionization potential (IP). It has been suggested that lines such as Calcium and Iron (e.g., \cav\ $\lambda5309$\AA , \fevii\ $\lambda6087$\AA, and \fex\ $\lambda6374$\AA) are more affected, and suppressed, by the presence of dust \citep[as the gas is depleted into dust;][]{Doan:25:2501}. Photoionization simulations indicate that dust can reduce the strength of optical CLs by up to three orders of magnitude via metal depletion, highlighting that a dust-free environment is essential for their prominence \citep{McKaig:24:130}. \sfviii\ has a lower detection frequency of 39/449 (9\%), which is in general agreement with \citetalias{Brok:2022:7}. This may result from a selection effect because our samples are primarily from X-shooter. X-shooter's NIR arm has an edge that frequently falls near \sfviii, due to our redshift range, and is dominated by noise from spectral fringing. \six\ may also be detected less because that region experiences heavy atmospheric absorption at certain redshifts. \alix, \sixi, and \sfxi\ have the lowest detection rates, with 8/383 (2.0\%), 8/431 (1.8\%), and 13/434 (3.0\%), respectively. 

Among Seyfert galaxies with at least one CL detection, 91/154 (59$\pm$4\%) of Seyfert 1$-$1.8 have a detection, 46/75 (61$\pm$6\%) of Seyfert 1.9, and 109/224 (49$\pm$3\%) of Seyfert 2. We report the uncertainty in these detections using the 1$\sigma$ binomial confidence interval. These are consistent with findings in \citetalias{Brok:2022:7}.

\begin{figure}
\centering
\includegraphics[width=\columnwidth]{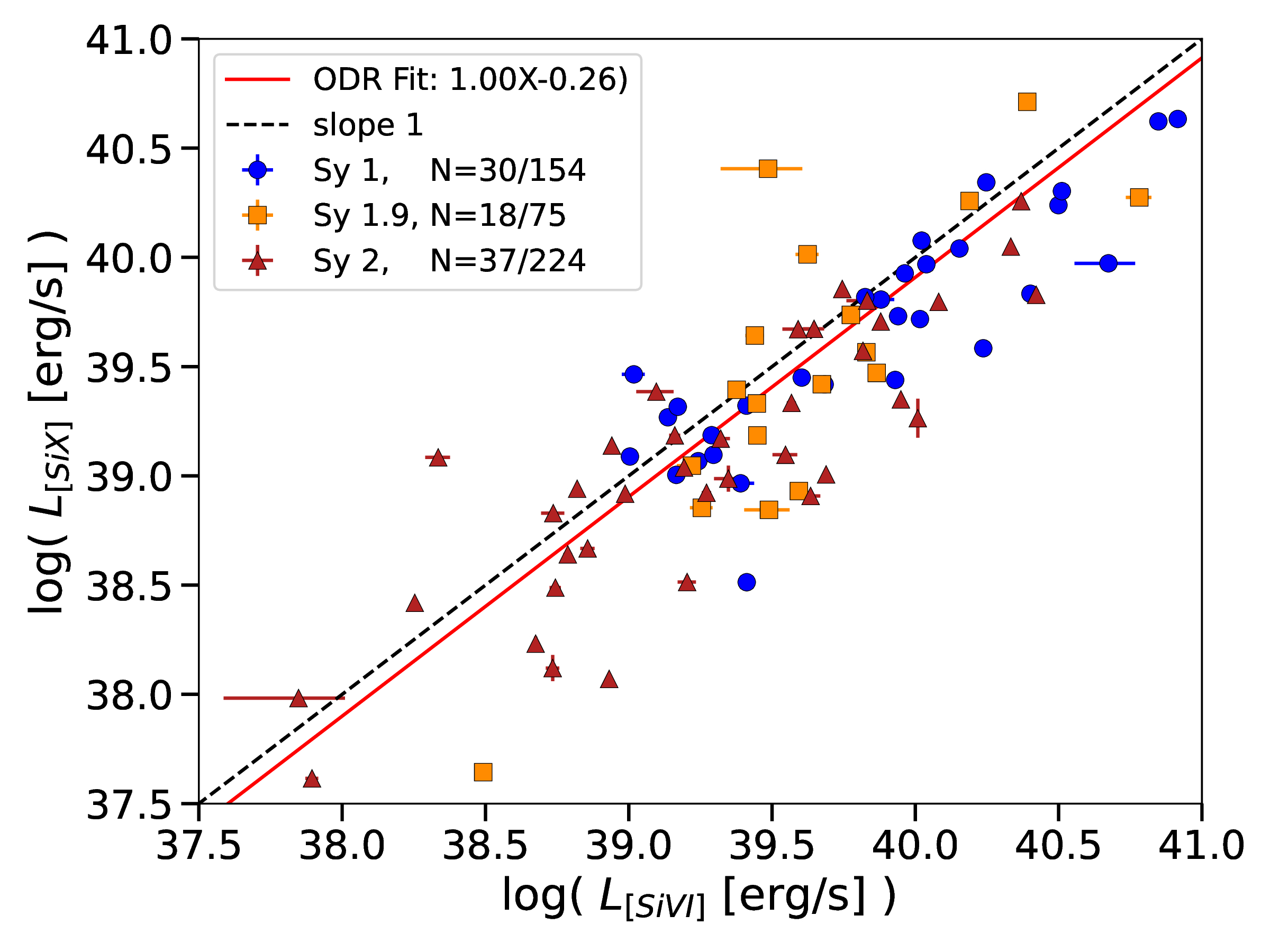}
\caption{Luminosities of \six\ vs \sivi, the dashed black line indicates a one-to-one luminosity ratio, and the red line is an ODR fit to the detected points.}
\label{fig:fig_si10_si6}
\end{figure}

Figure~\ref{fig:fig_si10_si6} shows the distribution of \sivi\ luminosity vs. \six\ luminosity for sources with simultaneous detections. They have a \rpear~$=0.77$ (\ppear~$=~3.4\times10^{-18}$). We include an Orthogonal Distance Regression (ODR) fit, which accounts for errors in both luminosity axes, to give a slope of 1.00 ($\pm~0.05$), an intercept of $-$0.26 ($\pm~2.04$), and a scatter of $\sigma~=~0.47$~dex.

A trend of decreasing CL detection frequency with higher redshift has been seen in previous studies \citep{Rodriguez-Ardila:2011:100, Lamperti:2017:540, Brok:2022:7}, and across Seyfert classification. This decrease has been attributed to emission lines redshifting out of spectral coverage or shifting into telluric absorption, and generally, line fluxes are weaker due to increased distance.

% Results
\subsection{Coronal Lines and X-ray Emission}
\label{sec:sec_discussion_cl}

To first order, CL emission may be thought of as powered by high-energy photons ($>$100~eV) ionizing the various species \citep{Done:2012}. An important step in understanding the interplay between CL and ionizing X-ray emission is to investigate any existing correlations. We use the model-independent Swift-BAT observed X-ray (14$-$195~keV) emission \citep{Tueller:2010:378,Koss:2017:74,Ricci:2017:17}. We also select \sivi~$\lambda$1.940\mum, as it is a bright line used in previous studies (see, \citealt{Lamperti:2017:540}, \citetalias{Brok:2022:7}), and because it is one of the most frequently detected CL in our sample (173 detections, 38\%). The following analysis shows that a positive correlation can be induced by using correlated axes (e.g., luminosities).

\begin{figure}
\centering
\includegraphics[width=\columnwidth]{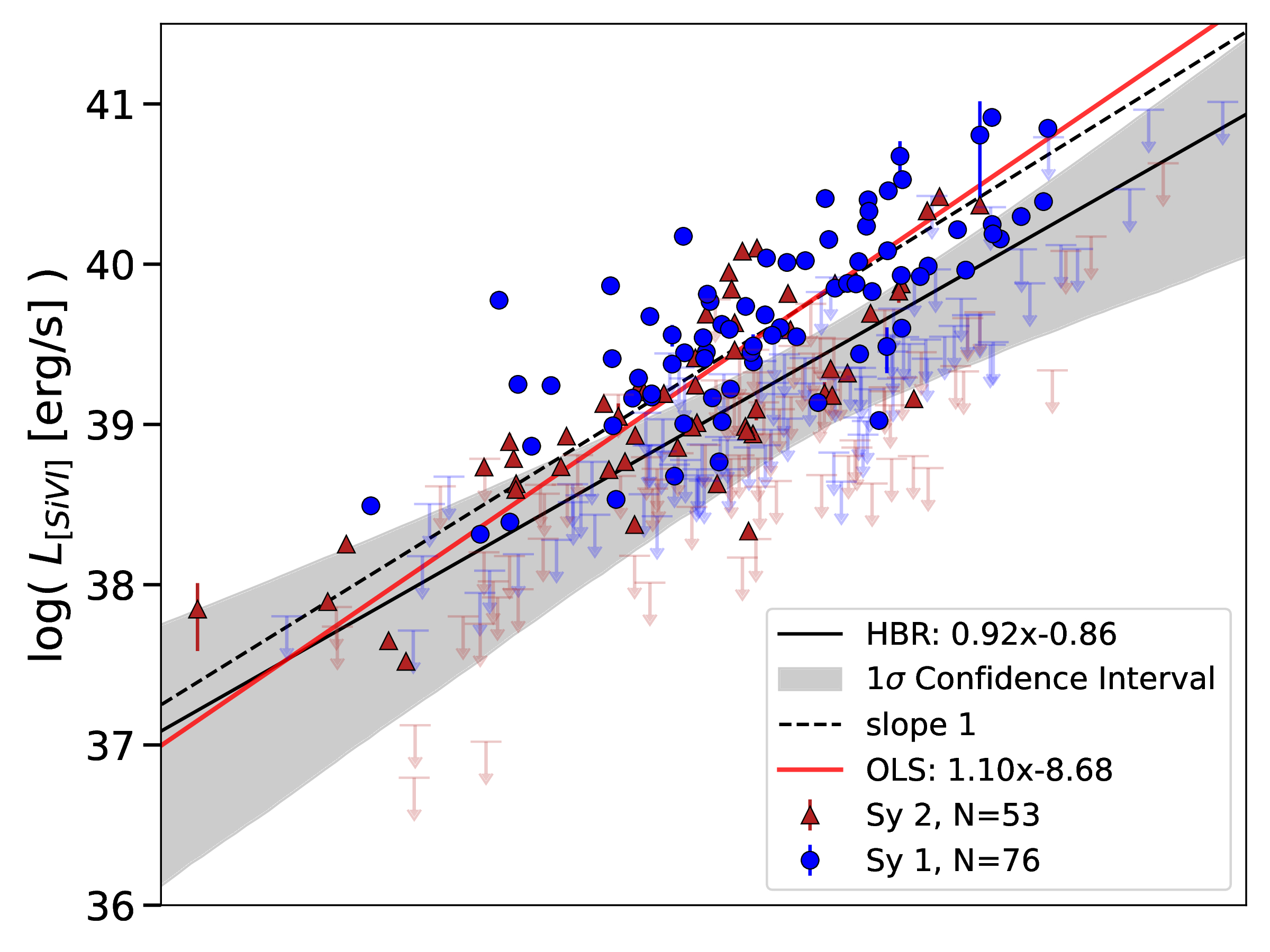}
\includegraphics[width=\columnwidth]{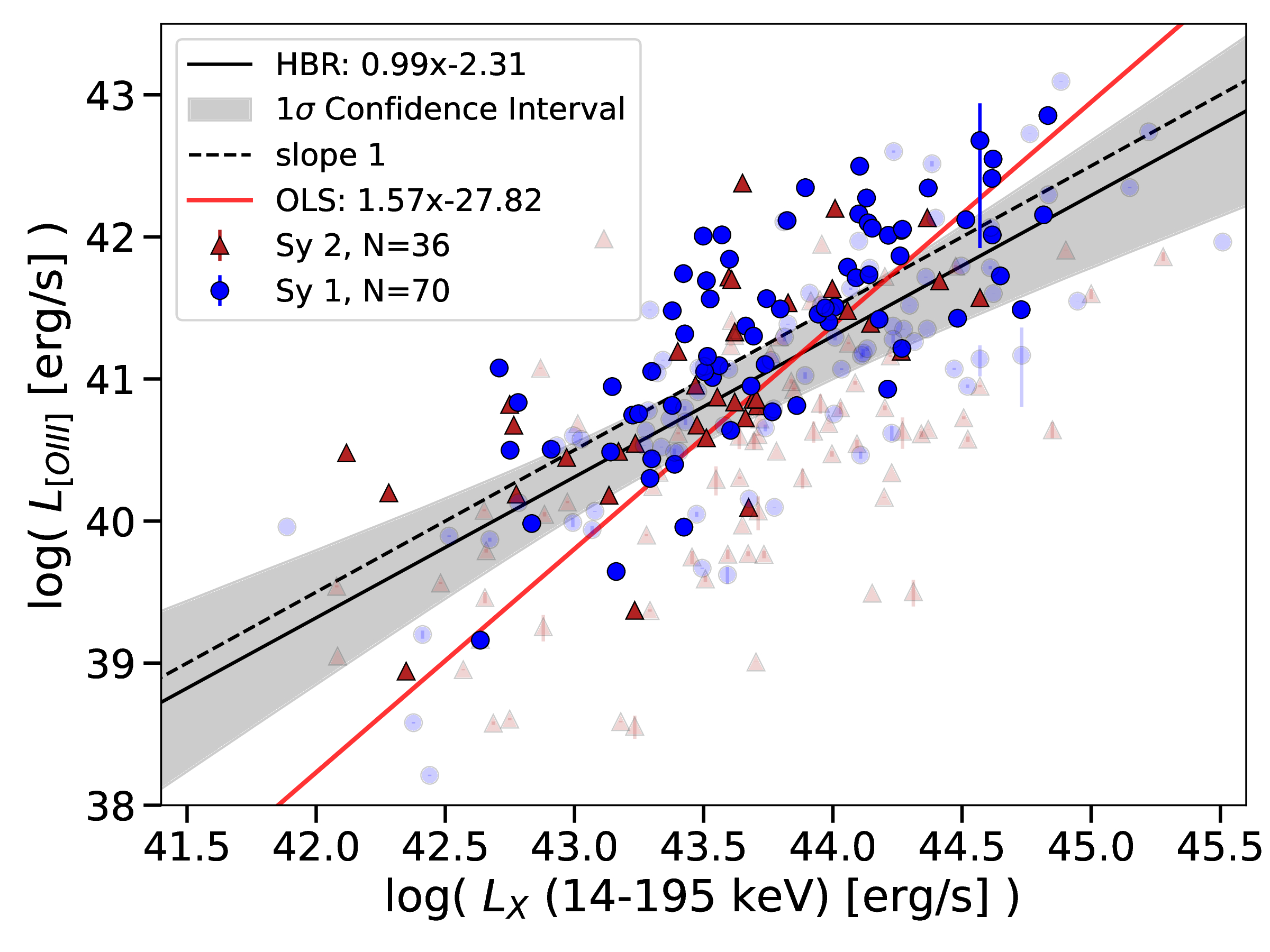}
\caption{\textit{Top}: Luminosities of \sivi\ vs $L_\mathrm{X-ray}$(14$-$195 keV). Blue dots are Seyfert 1/1.9, red squares are Seyfert 2, and downward arrows are upper limits on \sivi. The black dashed line has a slope of 1, scaled to go through the distribution, and the red solid line is the OLS bisector fit to the detections. The black line is the best fit while factoring in upper limits, with a 1-$\sigma$ error bar shaded region. \textit{Bottom}: Luminosities of \oiii vs $L_\mathrm{X-ray}$(14$-$195 keV), with the same color scheme for points and curves. All sources with \sivi\ detection or upper limits have simultaneously detected \oiii. We want to make direct comparison of the detected emission, and we show the distribution of \oiii\ for the sources with both detected. Transparent points are detections in \oiii\ that correspond to upper limits in \sivi. }
\label{fig:fig_lumx_si6_o3}
\end{figure}

The top panel of Figure~\ref{fig:fig_lumx_si6_o3} presents the distribution of luminosities for \sivi\ vs X-ray, and the bottom panel presents \oiii\ vs X-ray. We show an Ordinary Least Squares (OLS) bisector fit to the 129 detections in \sivi\ (slope~$=1.10 \pm 0.05$, and intercept~$=-8.68 \pm 2.29$), with a scatter of $\sigma=0.25$~dex. Since this fit is only to the detected points, the measured scatter is probably smaller than the real intrinsic scatter. Performing a Pearson correlation between the luminosities, we find a strong correlation, \rpear~$=0.70$ (\ppear=$1.6\times10^{-20}$). To include non-detections,  we performed a fit using upper limits as left-censored data (i.e., upper limits) in survival analysis. We performed the analysis using the package \textsc{linmix}\footnote{Software module by Joshua E. Meyers (https://linmix. readthedocs.io) based on the model described in \citet{Kelly:2007:1489}}, which makes hierarchical Bayesian regression (HBR). Including the upper limits yields a smaller intercept value, but the slopes are in agreement with previous analysis, showing that the \sivi\ and X-ray luminosities are strongly correlated (slope~$=0.92^{+0.32}_{-0.29}$, and intercept~$=-0.86^{+12.65}_{-13.94}$). We then perform a partial correlation of \sivi\ and X-ray luminosities, controlling for distance, and found that \sivi\ luminosity is statistically, intrinsically correlated with X-ray luminosity (partial  Spearman $\rho \approx 0.44,\ p \approx 1.9 \times 10^{-7}$), not just because both scale with distance.

The same analysis on 106 \oiii\ detections (from \citealt{Oh:2022:4}, corrected for intrinsic galaxy extinction), which have simultaneous \sivi\ measurements, yields similar results. From the OLS fit (slope~$=1.57 \pm 0.07$, and intercept~$=-27.8 \pm 3.1$), there is a scatter of $\sigma=0.79$~dex. Including the upper limits yields a slope~$=0.99^{+0.25}_{-0.30}$, intercept~$=-2.3^{+12.9}_{-11.1}$. The Pearson correlation coefficient shows a moderate correlation, \rpear~$=0.70$ (\ppear~=~$1.7\times10^{-19}$). While the relationship between \oiii\ and X-ray emission shows more scatter than that between \sivi\ and X-ray emission, the Fisher Z-test (p~$\approx 17.71$) indicates no statistically significant difference between the two correlations. Previous BASS studies focusing on \oiii\ have consistently reported similar relationships between \oiii\ and X-ray emission \citep[e.g.,][]{Ueda:2015:1,Berney:2015:3622}. With a sample size twice that of DR2, our results are comparable to those of \citetalias{Brok:2022:7} when comparing \sivi\ and \oiii\ with $L_\mathrm{X-ray}$(14$-$195 keV). \citetalias{Brok:2022:7} report that for the \sivi\ vs $L_\mathrm{X-ray}$(14$-$195 keV) relation, $\sigma=0.37$~dex with \rpear~$=0.86$, and for the \oiii\ vs $L_\mathrm{X-ray}$(14$-$195 keV) relation, $\sigma=0.71$~dex with \rpear~$=0.68$. We note that these relations do not account for the upper limits on \sivi, which would likely introduce additional scatter into the distribution. However, the consistency between \sivi\ and \oiii\ supports the idea that the detected coronal lines may serve as a better proxy for AGN power than \oiii, as noted in \citetalias{Brok:2022:7}.

\begin{figure}
\centering
\includegraphics[width=\columnwidth]{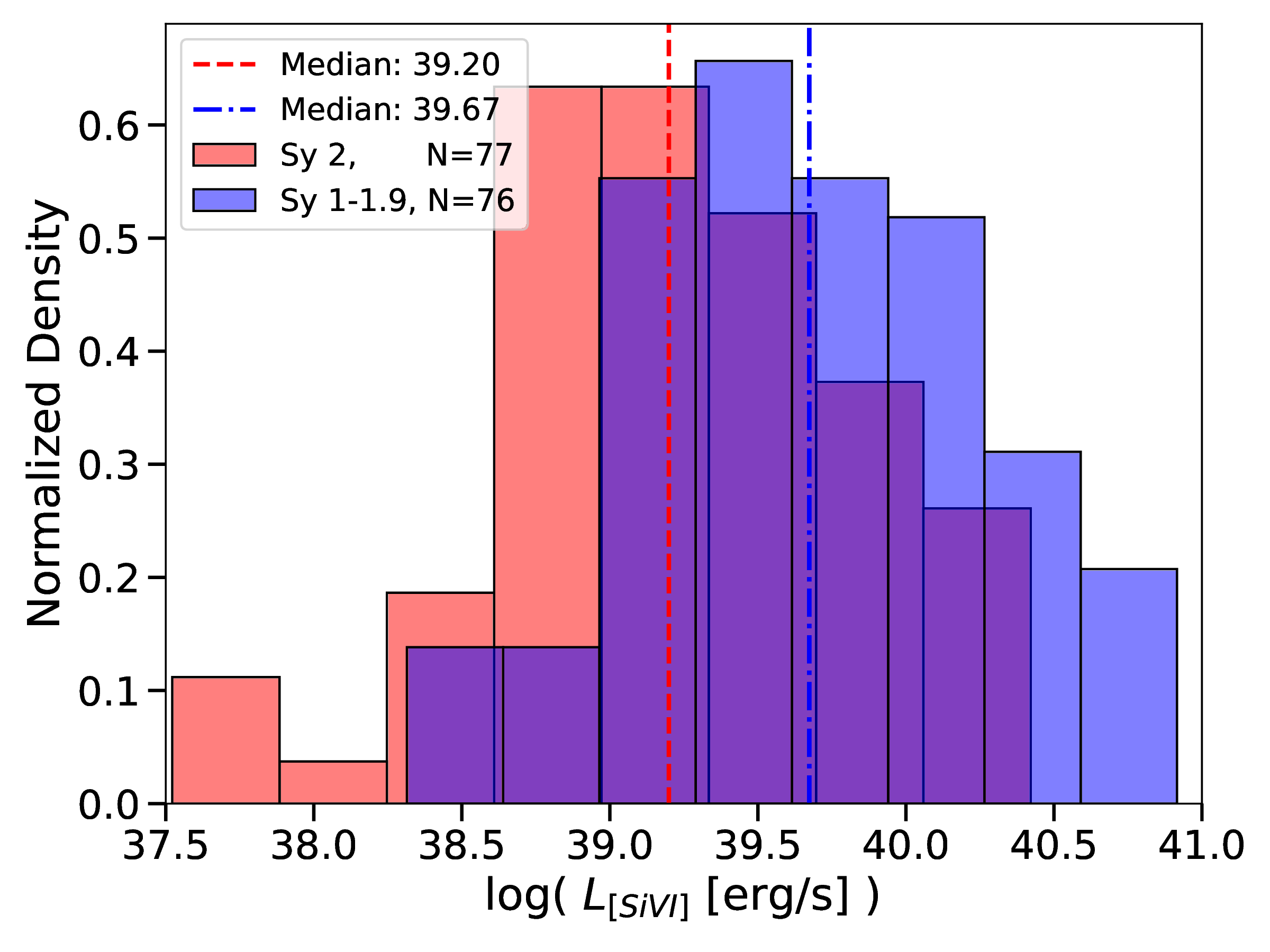}
\caption{Histrogram of \sivi\ luminosity, categorized by Seyfert 1 or 2. The blue and red lines indicate the kernel density estimation of the two distributions. }
\label{fig:fig_hist_si6}
\end{figure}

Figure~\ref{fig:fig_hist_si6} shows the distribution of sources with \sivi\ luminosity separated by Seyfert classification, confirming previous studies that Seyfert 1$-$1.9 sources tend to be more luminous than Seyfert 2 objects (e.g., \citetalias{Brok:2022:7}). Using the KM test to incorporate upper limits, we find a significant difference in narrow \sivi\ luminosity between Seyfert 1 and Seyfert 2 sources (p-value~$= 4.4\times10^{-4}$), with Seyfert 1 sources exhibiting higher luminosities. We note that we have not performed per-source host-extinction corrections in this study. This pattern is consistent with the broader trend we observe across coronal and hydrogen recombination lines, including \brg, where Seyfert 1 sources are more luminous by 0.3 to 0.5dex (Figures~\ref{fig:fig_hist_brg} and \ref{fig:fig_hist_si6}). Typical host extinctions of a few magnitudes would alter NIR line luminosities by $\lesssim$0.2 dex, and to produce a 0.3$-$0.5dex suppression in the NIR would require very large line-of-sight extinction (roughly $A_V \gtrsim 7-10$). Thus, while host-galaxy extinction could contribute for individual objects, it is unlikely to uniformly produce the luminosity offset we observe. Such luminosity offsets, likely linked to differences in Eddington ratio and nuclear covering factor, are well established in the literature \citep[e.g.,][]{Ricci:2017:488,Rojas:2020:5867,Kawamuro:2022:87}.

% Results
\subsection{H$_2$ Molecular Emission}
\label{sec:section_h2_emit}
Hot molecular gas in the NIR provides critical insights into galaxy centers' physical conditions and excitation mechanisms. By studying the emission, particularly in the NIR rotational and vibrational lines, we can probe the excitation mechanisms behind AGN and/or SF feedback processes. These lines may be excited by fluorescent emission from soft-UV photons in both AGN and star-forming regions, or via thermal processes from X-ray or shock heating, offering deeper insights into the interactions between these phenomena and the interstellar medium \citep{Brum:19:691B,Riffel:20:4857,Lai:22:36,Riffel:23:1832R,Bianchin:24:103B,Costa-Souza:24:127}.

Considering the molecular H$_2$~$\lambda$2.121\mum\ line, which is the strongest H$_2$ line in our wavelength range, we find 33/135 (24\%) detections in Seyfert 1, 27/61 (44\%) detections in Seyfert 1.9, and 82/157 (52\%) detections in Seyfert 2. H$_2$~$\lambda$2.247\mum\ is another molecular line commonly used for gas diagnostics, but is less frequently detected than other H$_2$ molecular lines. We find 3/120 (2.5\%) detections in Seyfert 1, 3/50 (6.0\%) detections in Seyfert 1.9, and 9/150 (6.0\%) detections in Seyfert 2. A full list of individual NIR molecular H$_2$ emission line detection statistics is in Table~\ref{appendix:tab_mol_lines}.

\begin{figure}
\centering
\includegraphics[width=\columnwidth]{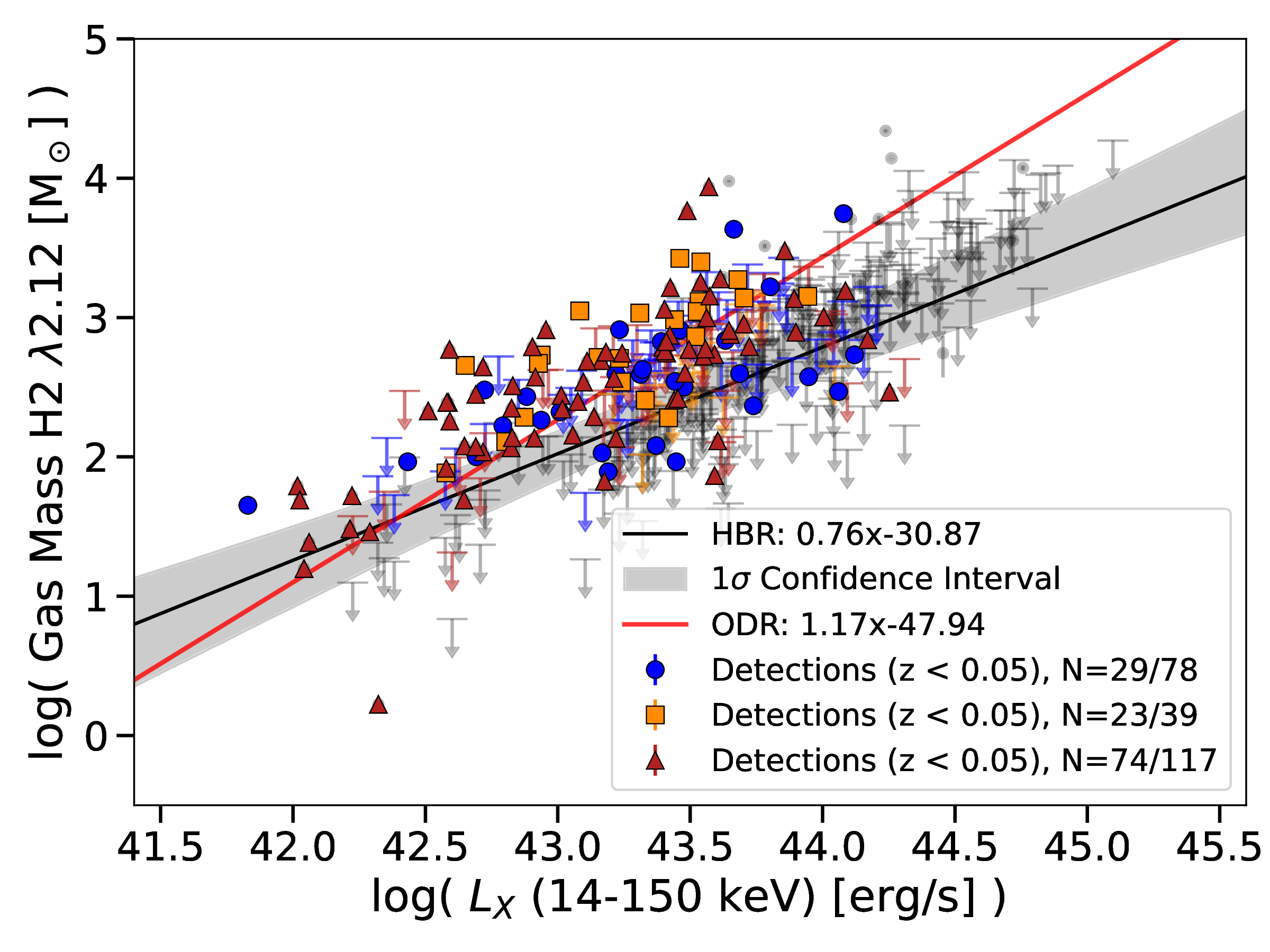}
\includegraphics[width=\columnwidth]{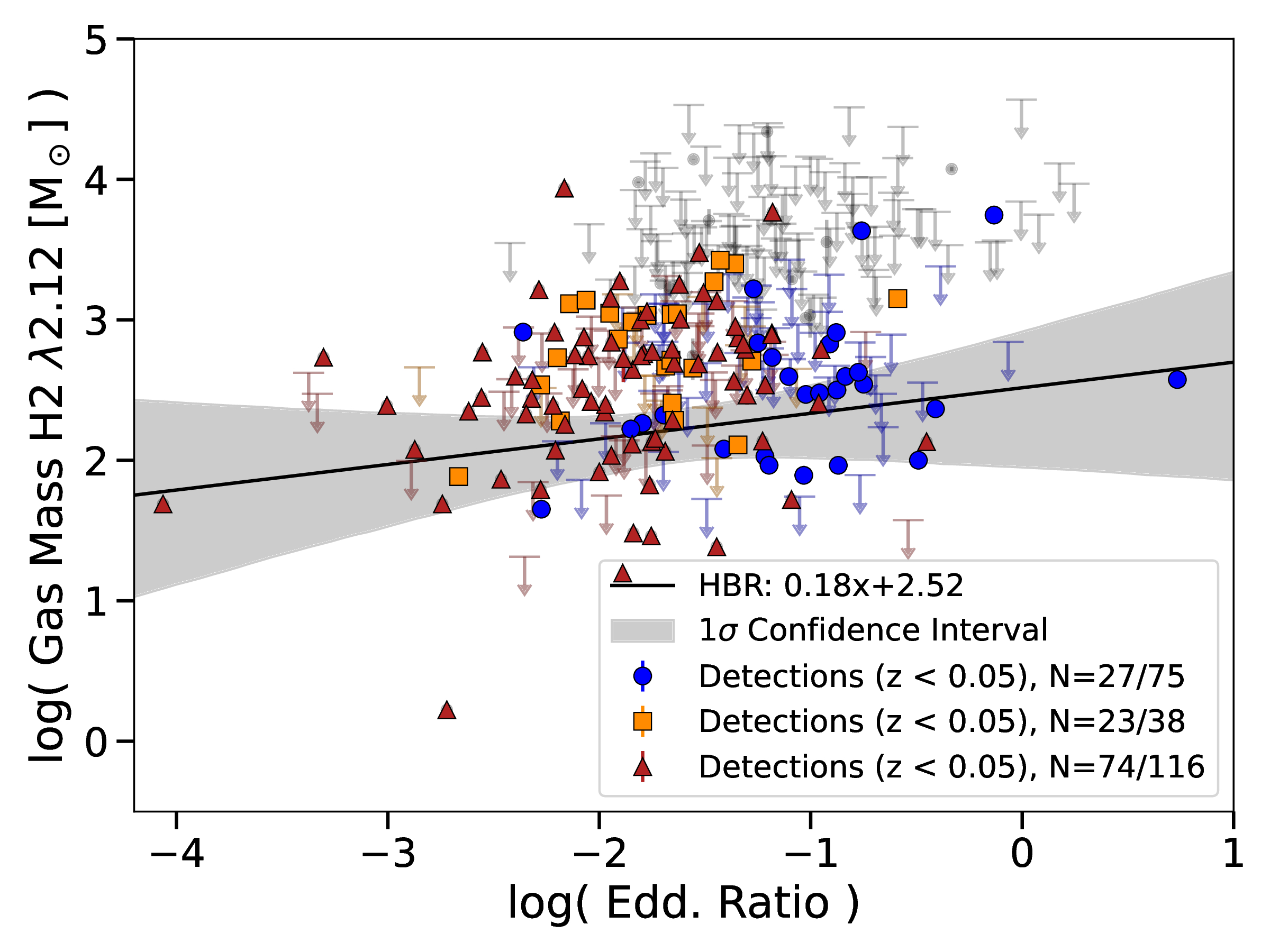}
\caption{Distributions of hot H$_2$ gas mass, markers defined similar to Fig.~\ref{fig:fig_z_hist_xlum}. Colored points and upper limits include the Seyfert classified volume-limited sample ($z<0.05$), and gray points and upper limits indicate sources with $z\geq 0.05$. Fit lines are to the volume-limited data, and follow the same description as Fig.~\ref{fig:fig_si10_si6}~\&~\ref{fig:fig_lumx_si6_o3}. \textit{Top}: Distribution of hot gas mass vs. $L_\text{X-ray}$(14$-$150~keV) with a detection total 137/353 (39\%). \textit{Bottom}: Gas mass vs. Eddington ratio with a detection total 113/334 (34\%). Hot gas mass appears to have a weak correlation with Eddington ratio, but partial correlation analysis does not suggest it is statistically significant. }
\label{fig:fig_h2_gas_mass_xray_eddr}
\end{figure}

Figure~\ref{fig:fig_h2_gas_mass_xray_eddr} illustrates the hot molecular gas mass trends vs X-ray luminosity and Eddington ratio. Gas mass is computed from the luminosity of H$_2$ $\lambda 2.121 $\mum, assuming a temperature $T=2000$~K, using $m_\text{H2} \simeq 5.0875 \times 10^{13} D^{2} I_{1-0S(1)}$ \citep{Reunanen:2002:154}, where $m_\text{H2}$ is in $M_\odot$, $D$ is distance in Mpc, and $I_{1-0S(1)}$ is the observed flux in erg~cm$^{-2}$s$^{-1}$. To analyze these H$_2$ fitting trends, we volume-limit the AGN to redshifts less than 0.05 to reduce bias from luminosity correlations. There is a clear trend in hot $m_\text{H2}$ vs $L_{\text{X}}$, although the trend is much flatter when accounting for non-detections. We show a linear ODR fit to the detected fluxes and an HDR fit based on survival analysis to include the upper limits, to AGN with $z<0.05$. The ODR fit to the detected points follows a slope of 1.17~$\pm0.08$ (intercept~$=-47.94\pm3.49$), while including upper limits yields a slope of $0.76^{+0.21}_{-0.16}$ (intercept~$=-30.87^{+7.06}_{-9.04}$). The Pearson correlation \rpear~$=0.70$ (\ppear$=~1.7\times10^{-19}$) implies a significant linear trend. We perform a partial correlation analysis and find there is a statistically significant corelation ($\rho\ = 0.18$, $p = 7.4\times 10^{-3}$), when controlling for distance. We see no separation by Seyfert classification, with AGN types evenly distributed across the range of X-ray luminosities. However, when comparing the gas mass from H$_2$ $\lambda 2.121 $\mum\ to Eddington ratios we find only a marginal trend. A fit to the detections and upper limits yields a slope of $0.18^{+0.25}_{-0.28}$ (intercept~$=2.52^{+0.40}_{-0.57}$). Controlling for distance using partial correlation analysis shows no statistically significant trend ($\rho = -0.098$,$ p = 0.14$), and the curve is likely dominated by correlations with distance. The distribution of Eddington ratios for Seyfert 1 sources is a magnitude higher (Eddington~$\sim 0.1$) than Seyfert 1.9 and Seyfert 2 sources, consistent with previous BASS studies \citep{Ricci:2017:488,Ananna:2022:9,Koss:2022:1}.

\begin{figure}
\centering
\includegraphics[width=\columnwidth]{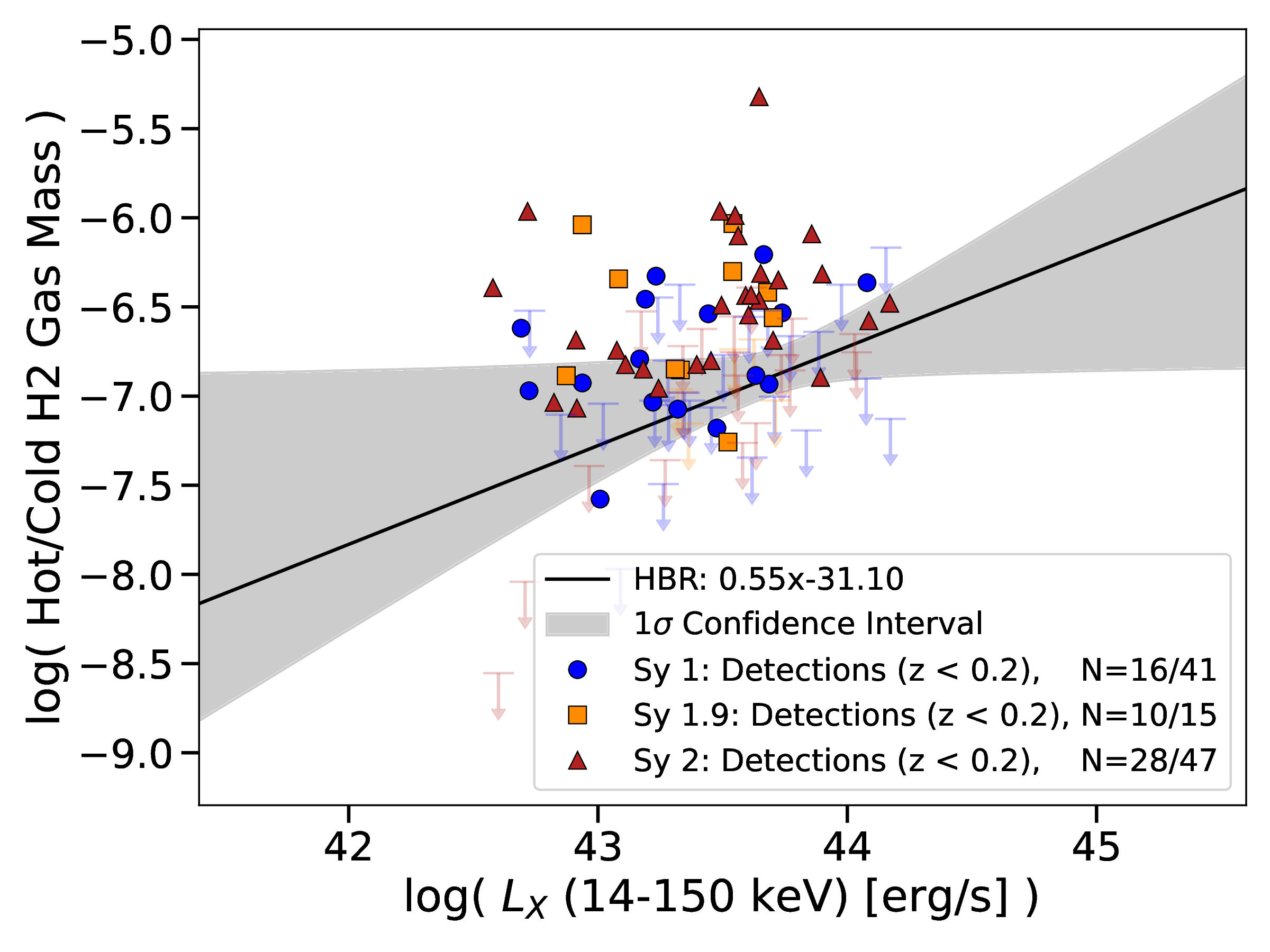}
\includegraphics[width=\columnwidth]{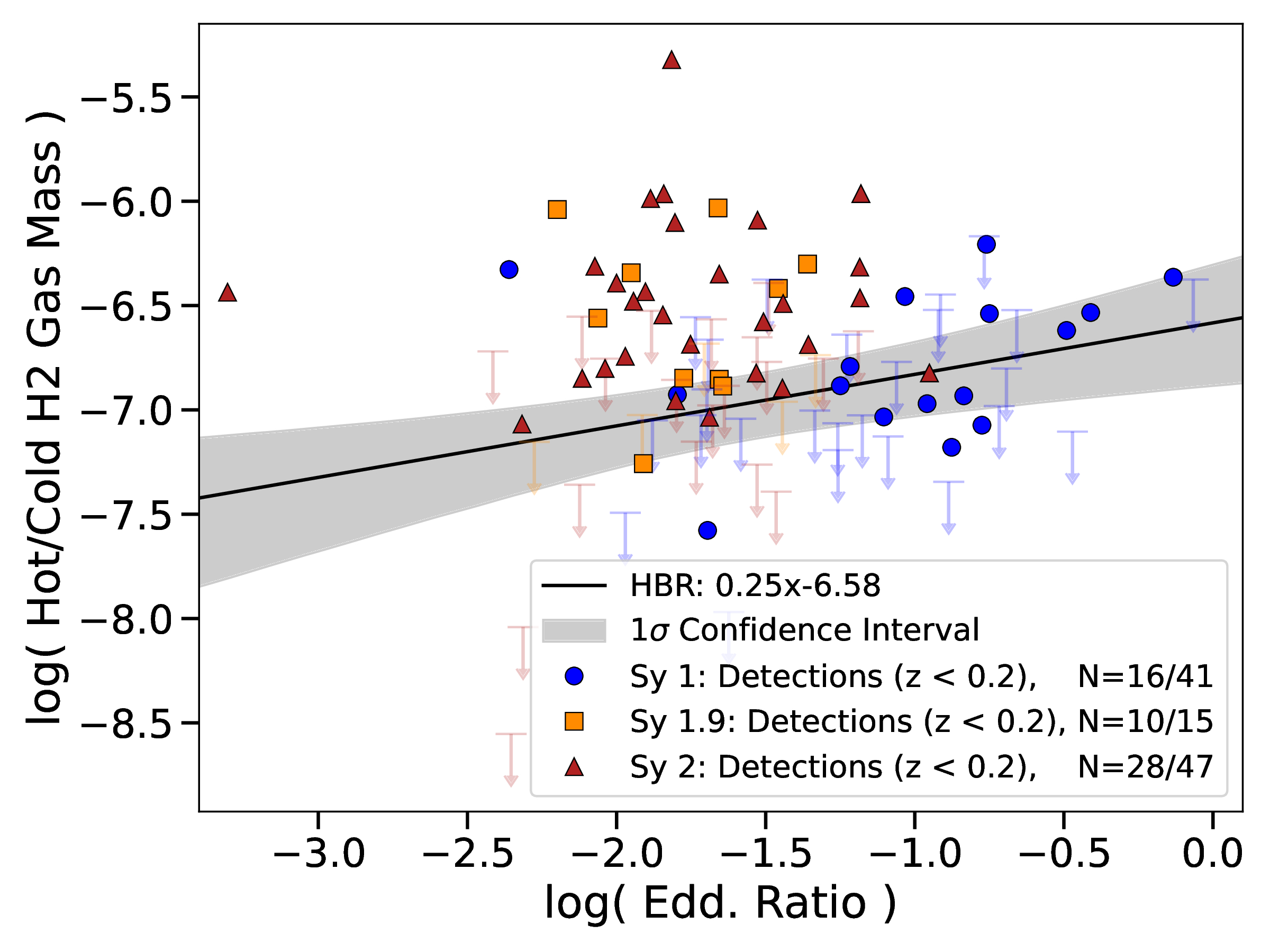}
\includegraphics[width=\columnwidth]{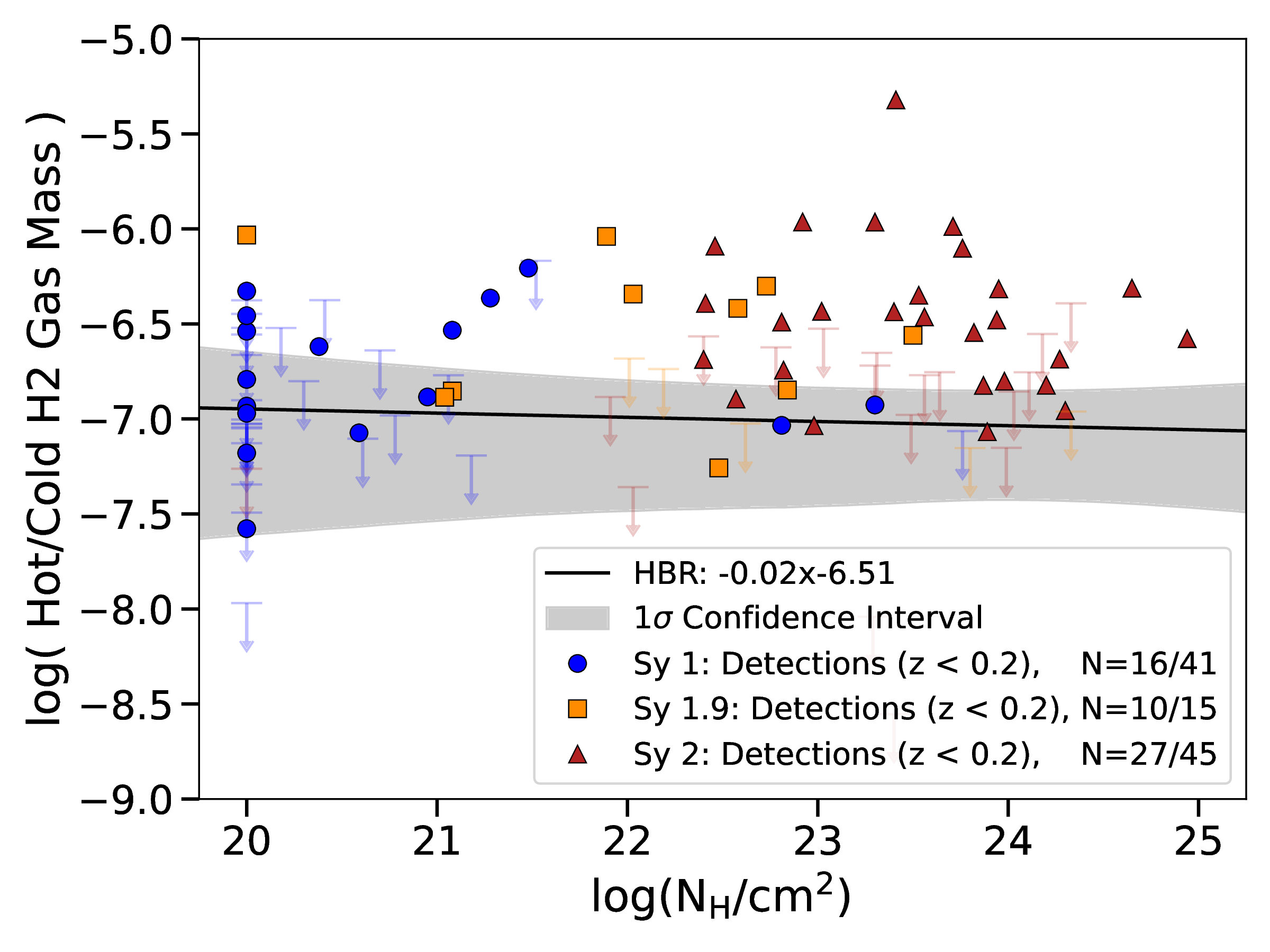}
\caption{Distributions of hot-to-cold H$_2$ gas mass ratios vs AGN characteristics for the sources with both measurements of hot and cold H$_2$ molecular gas, following the same markers as Fig.~\ref{fig:fig_h2_gas_mass_xray_eddr}. From top to bottom, the H$_2$ gas mass ratios are plotted against $L_\text{X-ray}$(14$-$150~keV), Eddington ratio, and \NH, with detection fractions 51/103 (50\%), 51/103 (50\%), and 50/99 (51\%), respectively. For $L_\text{X-ray}$(14$-$150~keV), there appears to be a moderate correlation for the relatively narrow 1.5~dex range of X-ray luminosities (p-value = 0.04). H$_2$ gas mass ratios display a weak correlation with the Eddington ratios (p-value = 0.68), and there is little correlation with \NH\ (p-value = 0.25). }
\label{fig:fig_warm_cold_h2_gas} 
\end{figure}

Understanding the hot-to-cold molecular gas ratio offers insights into the excitation mechanisms of molecular H$_2$ gas, probing the physical conditions in galaxy centers and their connection to AGN and star formation feedback. Figure~\ref{fig:fig_warm_cold_h2_gas} shows the ratio of hot-to-cold H$_2$ gas mass ratio compared to $L_\mathrm{X-ray}$(14$-$150 keV), Eddington ratio, and \NH. The hot-to-cold molecular gas ratio is computed by dividing the hot H$_2$ gas mass by cold H$_2$ gas masses provided by \citet{Koss:2021:29} for a sample of 101 BASS targets overlapping with this work. Our cold H$_2$ masses come from CO(2$-$1) observations with Atacama Pathfinder Experiment (APEX), using a Milky Way-like conversion factor from CO luminosity to H$_2$ mass ($\alpha_\text{CO}=4.3$ $M_*$(K \kms\ pc$^2$)$^{-1}$; \citealt{Koss:2021:29}). We take caution in interpreting these ratios, because the observations are from different physical regions (see discussion at the end of Section~\ref{sec:sec_discuss_h2}). Calculating statistical correlations between gas ratio and $L_\mathrm{X-ray}$(14$-$150 keV), Eddington ratio and \NH\ for the detected points, we find \rpear~=~0.27 (p-value~$=0.05$), \rpear~=~$-0.11$ (p-value~$=0.41$) and 0.12 (p-value~$=0.45$), respectively; i.e., a very marginal correlation with $L_\mathrm{X-ray}$(14$-$150 keV) and no correlations with Eddington ratio and \NH. Performing partial correlation analysis does not suggest these relations are statistically significant for $L_\mathrm{X-ray}$(14$-$150 keV) and Eddington ratio ($\rho = 0.07$, $p = 0.63$ and $\rho = 0.17$, $p = 0.22$). Including upper limits, the fitted slopes and intercepts for $L_\mathrm{X-ray} (14\text{--}150~\mathrm{keV})$, Eddington ratio, and $N_{\mathrm{H}}$ are $0.55^{+0.30}_{-0.44}$, $0.25^{+0.18}_{-0.15}$, and $0.02^{+0.13}_{-0.10}$ for the slopes, with corresponding intercepts of $-31.10^{+23.62}_{-13.24}$, $-6.58^{+0.28}_{-0.30}$, and $-6.507^{+2.17}_{-3.11}$. Considering the Seyfert classification, none exhibit a clear trend with the hot-to-cold gas ratio. Higher Eddington ratios ($>$~0.1) are dominated by Seyfert 1, while the mid-to-low Eddington ratios ($<$~0.05) are occupied by Seyfert 1.9 and Seyfert 2 sources. Higher \NH\ values (\NH~$> 10^{22}$ \cm$^{-2}$) are dominated by Seyfert 1.9 and Seyfert 2 sources, and lower \NH\ values (\NH~$< 10^{22}$ \cm$^{-2}$) are primarily Seyfert 1, consistent with Figure~\ref{fig:fig_fwhm_paa_pab_nh}.

\section{Discussion}
% Discussion
\subsection{Hidden Broad Lines}
\label{sec:sec_hidden_broad_lines}

We observe several broad emission lines in NIR, with a few sources optically identified as Seyfert 2 galaxies. These sources likely represent AGN, where the viewing angle is affected by moderate extinction levels, fully or partially obscuring the BLR. However, they also exhibit column densities exceeding log(\NH/\cmN)$= 21.9$ \citep{Koss:2017:74}. This result is not surprising, as such cases have been observed in both high and low luminosity AGN \citep[e.g.,][]{Garcet:2007:473, Oh:15:1, Kamraj:2019:255, Brok:2022:7}. 

We present cases in the bottom panel of Figure~\ref{fig:fig_fwhm_paa_pab_nh}, where some sources have broad emission in \paa\ yet have narrow optical lines. We found 20/153 (13\%) Seyfert 2 with \NH\ measured have broad \paa\ or \pab\ detection, in agreement with previous studies that include hidden broad lines \citep{Lamperti:2017:540, Brok:2022:7}. For Seyfert 1.9, we find 29/53 (55\%), and combining Seyfert 1.9 with Seyfert 2 increases the total to 49/206 (24\%). Improved counting statistics likely increased this detection rate from \citetalias{Brok:2022:7}. Our sample extends to higher redshifts, where broad \paa\ is more difficult to detect due to spectral atmospheric effects.

Our detection of numerous broad lines allows for new \Mbh\ estimates in sources that are otherwise classified as optically narrow. This phenomenon is similar to the case of Seyfert 1.9 galaxies, where dust attenuation affects different wavelengths unevenly, obscuring the bluer Balmer broad lines more than the redder Paschen broad lines. We note that variance in optical classification is not considered \citep[i.e., Changing-look AGN, e.g.,][]{Temple:2023:2938}, and therefore potential or temporary classification changes are not reflected in the analysis.

\citetalias{Brok:2022:7} suggests that the hidden broad lines in AGN does not necessarily refute the unification model. \citet{Lamperti:2017:540} found that sources with hidden broad lines are often merger systems. This suggests that the obscuration is more likely due to dust in the host galaxy rather than the nuclear torus. Furthermore, previous studies have reported lower \oiii\ to $L_{\text{X}}$ ratios in merging BAT AGN systems \citep[e.g.,][]{Koss:2010:664, Koss:2011:57, Koss:2012:220} and the correlation between higher X-ray obscuration and later merger stages \citep{Koss:2016:85, Ricci:2017:17}, supporting this view. \citetalias{Brok:2022:7} pointed to BASS examples such as 2MASX J042340.80+04080.17, which shows evidence of a merger \citep{Goncalves:99:437G}, and ESO 383-18, with dust winds that could obscure broad lines \citep{Ricci:10:47R}, reinforcing this argument.

Recent integral-field and NIR studies report that a subset of Type 2 quasars show kinematic and spectroscopic signatures consistent with a transient, heavily obscured evolutionary phase \citep[e.g.,][]{Tozzi:2024:141}. These signatures include fast ionized outflows, and centrally concentrated obscuration. Their result is complementary to orientation/covering-factor explanations, as it provides a plausible physical pathway by which broad-line regions can be temporarily hidden without excluding geometric effects in other objects.

Obscured broad lines are valuable for exploring possible variations between the broad line characteristics in optical and NIR spectra, their causes, and their impact on black hole mass estimations \citep{Mejia-Restrepo:2022:5,Temple:2023:2938,Caglar:23:60}. For Seyfert 1.9 galaxies, dust can obscure broad \ha\ emission, leading to a lower black hole mass estimate \citep{Caglar:2020:A114,Ricci:2022:8}. Even in highly dust-obscured sources, the \Mbh\ might not be accurately estimated if based on attenuated broad Paschen lines. In such cases, if Paschen lines are unavailable, \brg\ could be utilized instead, as dust attenuation less influences their wavelength. \brg\ may be a robust line to use, as it has been shown that quantities such as the virial factor $f$ may not significantly vary with redshift or obscuration, and are only mildly dependent on the specific emission line used \citep{Ricci:2022:8}. However, it should be noted that \brg\ is relatively weak, making it less detectable than the brightest NIR lines in hidden broad line sources. 

% Discussion
\subsection{Coronal Lines and X-ray Emission}

Hard X-ray luminosity is a reliable probe of AGN power, as it directly relates to the energetic processes in the vicinity of the AGN \citep[e.g.,][]{Reynolds:1997:513,Ueda:2003:886,Ricci:2017:17}. We find a similar scatter when comparing CL and X-ray (14$-$195 keV) luminosities to previous BASS NIR data releases \citep{Lamperti:2017:540,Brok:2022:7}. Both studies compare many CLs, including \oiii\ ratio with \sivi\ compared to X-ray, but do not find strong evidence to explain the scatter with $L_{\text{X}}$. In this work, the Pearson correlation coefficient is marginally higher for \sivi\ (\ppear~=~0.72) than for \oiii\ (\ppear~=~0.67) in sources with both detections, consistent with previous studies (e.g., \citetalias{Brok:2022:7}). In line with these results, recent work on the high-ionization \nev\ \lam3427~\AA\ line in BASS AGN reports a robust detection rate of approximately 43\%, and scaling with X-ray luminosity characterized by a scatter of $\lesssim0.5$~dex, further reinforcing the reliability of high-ionization lines as AGN tracers even in heavily obscured systems (Reiss, submitted.). This similarity provides definitive evidence that the suppression of CLs is not attributable to extinction. \citetalias{Brok:2022:7} notes that metallicity is not likely a large source of scatter in CL luminosities, because the sample has a relatively uniform metallicity gradient \citep{Koss:2011:57}. Also, one must consider that the state of the gas \citep[e.g., electron gas density;][]{Rodriguez-Ardila:2011:100} may influence the CL strength, but there is conflicting evidence for this explanation \citep{Landt:2015, Rodriguez-Ardila:2017:2845, Rodriguez-Ardila:2017:906}. 

Optical CL studies highlight the strong influence of local conditions on line visibility. For instance, photoionization models show that dust suppresses optical CLs via metal depletion \citep{McKaig:24:130}, while a survey of Type 1 quasars finds that these lines are rare ($\sim$4.5\% detection) and often linked to ionized outflows \citep{Doan:25:2501}. Unlike optical CLs, NIR CLs appear less affected by dust and outflows, making them more reliable tracers of AGN activity. Multi-wavelength studies will be key to disentangling these effects and further establishing high-ionization lines as robust indicators of AGN power, even in heavily obscured systems.

% Discussion
\subsection{AGN Diagnostic}

Many have found that the previously proposed NIR diagnostic is less effective than optical for identifying AGN, not just because of contamination of the AGN region by SF galaxies, but because the distributions of diagnostic line ratios for SF galaxies and AGN overlap substantially. This strong overlap implies that the diagnostic only provides robust separation in the extreme tails of the ratio distributions \citep{Dale:2004:813, Rodriguez-Ardila:2005, Martins:2013:1823, Riffel:2013:2002, Lamperti:2017:540}. We refine this method for our large sample to define regions of exclusively AGN, composite AGN+SF, and exclusively SF.

Figure \ref{fig:fig_diagnostic_plot} presents our revised NIR diagnostic plot for 93 BASS sources. The AGN region is defined with H$_2$/\brg\ $\gtrsim2$ or \feii/\pab\ $\gtrsim1.3$, SF region with H$_2$/\brg\ $\lesssim0.25$ and \feii/\pab\ $\lesssim0.45$, and between them is the composite AGN and SF region. 

Variation in these line ratios reflects multiple excitation mechanisms in our sources. \citet{Riffel:13:2587} discusses various models, but here we focus on the key aspects. \brg\ originates mainly from fully ionized regions, while H$_2$ is emitted from the adjacent semi-ionized zone \citep{Aleman:11:74}. In AGN, the harder ionizing spectrum extends this semi-ionized region \citep{Aleman:04:865}, leading to enhanced H$_2$ emission relative to \brg\ compared to SF regions. However, the recombination-line measurements are also affected by the stellar population, in particular the number of young stars relative to intermediate-age/older stars \citep{Larkin:98:59, Riffel:08:803}. The underlying absorption from stellar atmospheres can depress \pab/\brg, while a higher young star fraction increases the ionizing photon budget and continuum strength. 

Because the AGN and SF distributions overlap, the diagnostic cannot completely separate the two populations within the composite area. However, our analysis shows that whenever an object’s ratios fall outside the pure SF boundary, the diagnostic still reliably highlights it as an AGN candidate. This ensures that, despite overlap, the method retains practical utility for selecting AGN candidates for further study. It also demonstrates the diagram's utility for identifying AGN at higher line ratios, even when only upper limits are available for weaker NIR lines.  

% Discussion
\subsection{Hot Molecular Gas}
\label{sec:sec_discuss_h2}

Molecular hydrogen at 2.121\mum\ (H$_2$ 1-0 S(1) line) is a common tracer of hot molecular gas with temperatures typically in the range of 500$-$3000 K \citep{Scoville:82:136S,Riffel:14:656}. For Seyfert galaxies, hot molecular lines found in the circumnuclear regions can be excited by a variety of mechanisms: shocks from outflows, X-ray irradiation from the AGN, or local SF \citep[e.g.,][]{Riffel:21:3265}. Any observed trends in the distribution of hot molecular gas masses as a function of AGN X-ray luminosity and Eddington ratio may provide insights into AGN and galaxy interactions. We emphasize that the 2.121\mum\ line serves as a crucial indicator of the conditions and quantity of hot molecular gas in proximity to the AGN. Here, one of the largest samples of hot gas masses in unbiased Seyfert galaxies is presented, with a total of 217 detections (detection fractions of NIR molecular H$_2$ lines are shown in Table ~\ref{appendix:tab_mol_lines}). 

Figure~\ref{fig:fig_h2_gas_mass_xray_eddr} presents trends of X-ray luminosity and Eddington ratio with the hot gas mass in the circumnuclear region ($\sim0.2 - 5.5$ kpc). We find a relatively strong positive correlation between increasing gas mass and increasing $L_{\text{X}}$ in the upper panel. This trend with $L_{\text{X}}$ may indicate how AGN radiative output influences the molecular gas, potentially through heating, ionization, or outflows. \citet{Gaspar:22:230} analyzed excitation mechanisms of molecular hydrogen in the obscured AGN NGC 4945 (BATID 655), and found that the inner nuclear regions ($\sim$40~pc) are primarily excited by shocks, or more likely excited by UV fluorescence in regions radially extended from the nucleus ($\sim$200~pc). However, NGC 4945 resides in the bottom tenth percentile of Seyfert 2 in this $L_{\text{X}}$ distribution (log $L_{\text{X}}$(14$-$150~keV)~$\approx$~42.33), and may not represent the H$_2$ trend seen with higher $L_{\text{X}}$.

Examining the lower panel of Figure~\ref{fig:fig_h2_gas_mass_xray_eddr}, the trend with Eddington ratio is weaker than $L_{\text{X}}$. This trend can provide insight into the accretion efficiency of the AGN scales with the availability of hot molecular gas and provide a clue about the depletion or replenishment of the gas reservoir. The weak correlation implies that the relationship between accretion and hot molecular gas availability is less important, and other factors likely play a more significant role in regulating the gas reservoir. 

This moderate and weak correlation of hot molecular gas to X-ray luminosities and Eddington ratios is notable when compared to previous small-survey studies, which found no significant relation with kpc-scale molecular gas content \citep{Rosario:2018:5658, Garcia-Burillo:2021}, and compared to the correlations found between the soft and hard X-ray and the total CO luminosities \citep{Monje:2011:23, Koss:2021:29}. Part of these correlations are due to luminosity distance ($\propto D^2$), but this bias should have less influence on the volume-limited ($z<0.05$) subsample. Partial correlation analysis of the gas mass and distance, controlling for X-ray luminosity, is consistent with this explanation ($\rho = 0.382$, $p = 1.44 \times 10^{-09}$). However, \citet{Kawamuro:21:64} found enhanced hard X-ray emission correlated with the luminosity of cold CO molecular gas near nuclear regions, and suggested the presence of an X-ray-irradiated circumnuclear ISM, potentially disrupting SF. 

MIR studies show that AGN frequently produce an excess of warm molecular hydrogen emission and, in many cases, kinematic signatures indicative of shocks or outflows \citep{Lambrides:2019:1823, Minsley:2020:157, Riffel:2025:69}. These MIR rotational H$_2$ transitions and the NIR vibrational H$_2$ lines we study trace different temperature regimes and spatial scales, but they can arise from overlapping excitation mechanisms, X-ray heating, shocks from outflows or jets, and in some cases UV fluorescence, and so consistent trends across MIR and NIR H$_2$ support an AGN role in powering warm molecular gas. Recent JWST/MIRI spatially resolved work further confirms that AGN driven mechanical and radiative processes can heat multi-temperature molecular phases on sub-kpc scales \citep[e.g.,][]{Kakkad:2025:1102}. Together, these results show that our observed correlation between hard X-ray luminosity and NIR vibrational H$_2$ complements the MIR literature and points to AGN energy input (radiative or mechanical) as a plausible driver of warm molecular emission. Earlier Spitzer high-resolution work revealed turbulent and fast warm H$_2$ motions in AGN hosts \citep{Dasyra:2011:10}, and recent JWST/MIRI MRS mapping now spatially resolves hot/warm H$_2$ on sub-kpc scales in AGN, further supporting AGN-driven radiative and mechanical excitation of molecular gas.

Figure~\ref{fig:fig_warm_cold_h2_gas} shows that the hot-to-cold gas ratio distribution is independent of X-ray luminosity, Eddington ratio, and obscuration. However, $L_{\text{X}}$ does not extend to extremely low or high values in this subsample. The hot-to-cold gas ratio is useful for determining the state of molecular gas and in this case, the circumnuclear gas of the interstellar medium. The average hot-to-cold molecular gas ratio of circumnuclear gas is $\sim3\times10^{-7}$, spanning over four orders of magnitude, and as high as $5\times10^{-6}$. Most ratios measured in Seyfert and typical galaxies of the local Universe reside between $10^{-8}-10^{-5}$, and those above $10^{-7}$ are indicative of molecular material being affected by SF \citep{Dale:2004:813, Dasyra:2014, Emonts:2014}. Notably, \citet{Dale:2004:813} found that the hot-to-cold gas ratio correlates with the $f_{60\mu m}/f_{100\mu m}$ flux ratio, a tracer of dust temperature and indirectly star formation activity, reinforcing the connection between hot H$_2$ gas and SF processes. Studies of individual galaxies have found ratios of several times $10^{-5}$ within AGN-powered outflows of Luminous and Ultra-Luminous Infrared Galaxies \citep{Pereira-Santaella:2016, Emonts:2014, Ceci:24:2412, Ulivi:25:36}. 

An important consideration when interpreting hot-to-cold gas ratios is the spatial coverage of the measurements. Cold molecular gas traced by APEX CO observations is typically measured over large areas of the host galaxy and corrected for aperture effects. In contrast, hot H$_{2}$ observed with X-shooter is extracted through a narrow slit and is generally more centrally concentrated on circumnuclear (hundreds of parsecs) scales \citep[e.g.,][]{Hicks:2009:448}. However, we note that in galaxies undergoing mergers or with strong circumnuclear star formation, shocks may excite extended H$_{2}$ emission that could extend beyond the slit aperture \citep[e.g.,][]{Kakkad:2025:1102}. Thus, our analysis assumes that the dominant contribution to the hot H$_{2}$ luminosity is nuclear, and that any extended contribution missed by the slit would not significantly alter the ratios. Given that the hot-to-cold molecular gas mass ratios span 4-5 orders of magnitude and are 6-7 orders below unity, potential aperture mismatches are unlikely to qualitatively change our conclusions. This nuance further informs our Fig.~\ref{fig:fig_warm_cold_h2_gas} findings on the relation of hot molecular gas mass with $L_{\text{X}}$(14$-$150~keV) and Eddington ratios, and the composite nature of these galaxies at circumnuclear scales.

\section{Summary}

This study presents a comprehensive analysis of BASS DR3 NIR properties, hot molecular gas energetics, and AGN diagnostics in Seyfert galaxies, offering valuable insights into the circumnuclear environments of BASS AGNs. 

Key takeaways from this work include: 

\begin{itemize} 
\item Broad \paa\ or \pab\ emissions are observed in 44/224 (20\%) of Seyfert 2 for which we have spectral coverage, and 39/75 (52\%) of Seyfert 1.9, underscoring the importance of redder emission lines for AGN characterization in obscured sources. 
\item We find a strong correlation between CL strengths and hard X-ray luminosities, with marginally less scatter than observed for \oiii. The relatively tight correlation, particularly given that the majority of BAT AGN hosts are massive galaxies with shallow or constant metallicity gradients \citep[e.g.,][]{Koss:2011:57}, suggests that metallicity variations are unlikely to be the dominant source of scatter. Instead, these results imply that gas density and excitation mechanisms likely play a primary role in driving CL behavior. 
\item Our refined NIR diagnostic diagram effectively identifies AGN, even when some key lines are weak or yield only upper limits, with 50/89 (56.2\%) of Seyfert galaxies falling into the AGN region, and the rest as 'possible AGN' which lie in the composite AGN+SF region. 
\item With 239 detections of hot H$_2$ molecular gas, the largest compiled sample with hot-to-cold molecular gas mass ratio measurements for individual X-ray detected sources. We find a positive correlation between hot gas mass and X-ray luminosity (\rpear~$\approx 0.70$, \ppear~$\approx 1.7 \times 10^{-19}$), and the correlation persists after controlling for luminosity ($\rho \approx 0.18,\ p \approx 7.4 \times 10^{-3}$). This suggests that AGN radiative output may influence molecular gas dynamics through heating, ionization, or outflows.  
\end{itemize}

These results advance our understanding of AGN environments, emphasizing the role of NIR observations in probing obscured regions, refining black hole mass measurements, and exploring the interplay between AGN activity and molecular gas. Looking ahead, further high-resolution or high-sensitivity NIR observations promise to reveal the hidden structure of obscured BLRs, refine \mbh\ estimates, and unravel the interplay between AGN feedback and molecular gas dynamics, paving the way for a deeper understanding of galaxy evolution. This BASS NIR catalog serves as a valuable reference for studying CL properties in local AGN, and offers an opportunity for comparison with the increasing collection of high-redshift spectra from JWST.

\begin{acknowledgments}
J.G. and M.K. acknowledge support from NASA through ADAP award 80NSSC22K1126. C.R. acknowledges support from Fondecyt Regular grant 1230345, ANID BASAL project FB210003 and the China-Chile joint research fund. K.O. acknowledges support from the Korea Astronomy and Space Science Institute under the R\&D program (Project No. 2025-1-831-01), supervised by the Korea AeroSpace Administration, and the National Research Foundation of Korea (NRF) grant funded by the Korea government (MSIT) (RS-2025-00553982). Y.D. acknowledges financial support from a Fondecyt postdoctoral fellowship (3230310). I.M.C. acknowledges support from ANID programme FONDECYT Postdoctorado (3230653).
\end{acknowledgments}
 
\facilities{Du Pont (Boller \& Chivens spectrograph), Keck:I (LRIS), Magellan:Clay, Hale (DBSP), NuSTAR, Swift (XRT and BAT), VLT:Kueyen (X-Shooter), VLT:Antu (FORS2), SOAR (Goodman)}

\software{Astropy \citep{astropy:2013,astropy:2018,astropy:2022}, ESO Reflex \citep{Freudling:2013:A96}, IRAF \citep{Observatories:1999:ascl:9911.002}, Matplotlib \citep{Hunter:2007:90}, Numpy \citep{vanderWalt:2011:22}, Pandas \citep{Reback:2020} }

\appendix
\setcounter{table}{0}
\renewcommand{\thetable}{A\arabic{table}}
\setcounter{figure}{0}
\renewcommand{\thefigure}{A\arabic{figure}}

\section{Observation Information}
\label{app:obs_info}
Table~\ref{appendix:tab_info} provides an overview of the observations included in this study. This excerpt serves as a reference for readers, while the full table is available in the online journal. The table details the observational setup and additional observation properties.

\begin{deluxetable*}{lccccccccc}
\tablecaption{Sample summary of VLT/X-shooter observations and survey details from the BASS project.\label{appendix:tab_info}}
\tablewidth{0pt}
\tablehead{
  \colhead{Swift-BAT ID} & \colhead{Counterpart Object} & \colhead{Redshift} & \colhead{Date} & \colhead{Exp. Time} & \colhead{Airmass} & \colhead{Seeing} & \colhead{Spectral Resolution} & \colhead{Program ID} \\
  \colhead{}            & \colhead{}                 & \colhead{}      & \colhead{(dd.mm.yyyy)} & \colhead{(s)}  & \colhead{}    & \colhead{($''$)} & \colhead{(NIR Arm mean)}               & \colhead{}
}
\startdata
1   & 2MASXJ00004876-0709117   & 0.037 & 17.07.2021 & 2000 & 1.074 & 0.47 & 5573 & 105.20DA.001 \\
7   & SDSSJ000911.57-003654.7   & 0.073 & 11.10.2021 & 2000 & 1.096 & 1.13 & 5400 & 108.229H.001 \\
28  & NGC235A                  & 0.022 & 01.08.2021 & 1000 & 1.043 & 0.97 & 5573 & 105.20DA.001 \\
44  & 2MASXJ01003490-4752033    & 0.048 & 15.07.2021 & 1000 & 1.402 & 1.9  & 5573 & 105.20DA.001 \\
49  & MCG-7-3-7                & 0.030 & 10.07.2021 & 1000 & 1.314 & 1.02 & 5573 & 105.20DA.001 \\
55  & 2MASXJ01073963-1139117    & 0.047 & 17.07.2021 & 1000 & 1.06  & 0.92 & 5573 & 105.20DA.001 \\
58  & NGC424                   & 0.011 & 29.07.2021 & 1000 & 1.048 & 1.71 & 5573 & 105.20DA.001 \\
94  & CSRG165                  & 0.030 & 21.10.2021 & 1000 & 1.111 & 1.25 & 5573 & 108.229H.001 \\
95  & ESO354-4                 & 0.034 & 21.10.2021 & 1000 & 1.143 & 0.85 & 5573 & 108.229H.001 \\
102 & NGC788                   & 0.014 & 16.07.2021 & 500  & 1.104 & 1.27 & 5573 & 105.20DA.001 
\enddata
\tablenotetext{}{(This table will be available in its entirety in machine-readable form.) }
\end{deluxetable*}

\section{Spectral Measurements}

Table~\ref{appendix:tab_flux} presents the flux measurements obtained from the spectral fits performed in this study. The full table is accessible in the online journal.

\section{Obscuration and Hidden Broad Lines}
\label{appendix:hidden_lines}

\begin{figure*}
\centering
\includegraphics[width=\textwidth]{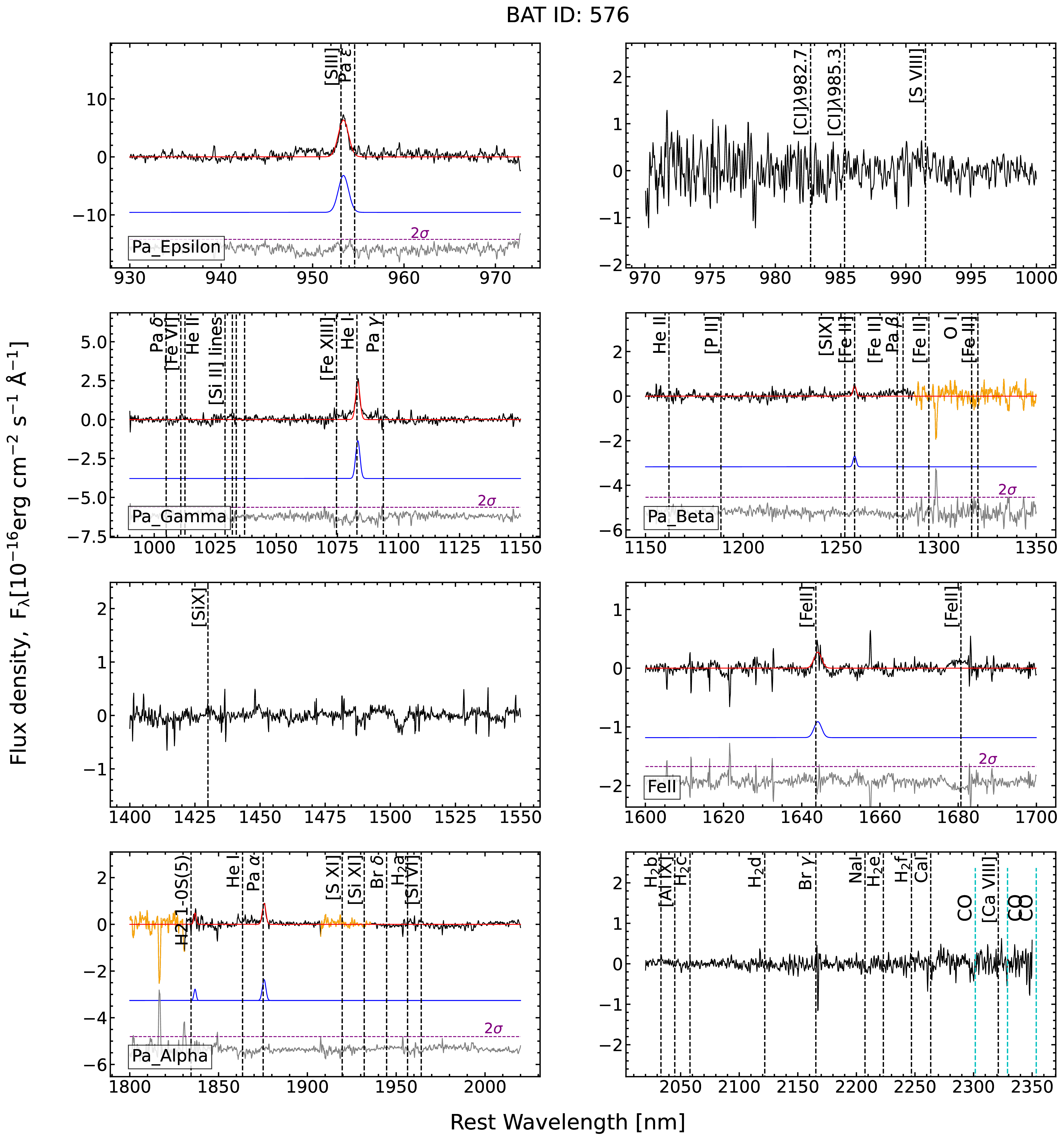}
\caption{Example of simultaneous emission-line fits for galaxy PG1149-110 (BAT ID 576), observed with VLT/X-shooter, following the same labels as Fig.~\ref{fig:fig_bat_795}. This source has no broad component in \paa, and no detection in \pab, despite being a Seyfert 1 source. There is weak emission in the wings of \hei, indicating this source may have weak broad emission, or that the data quality does not allow detection of the broad line.}
\label{appendix:fig_bat_576}
\end{figure*}

\begin{figure*}
\centering
\includegraphics[width=\textwidth]{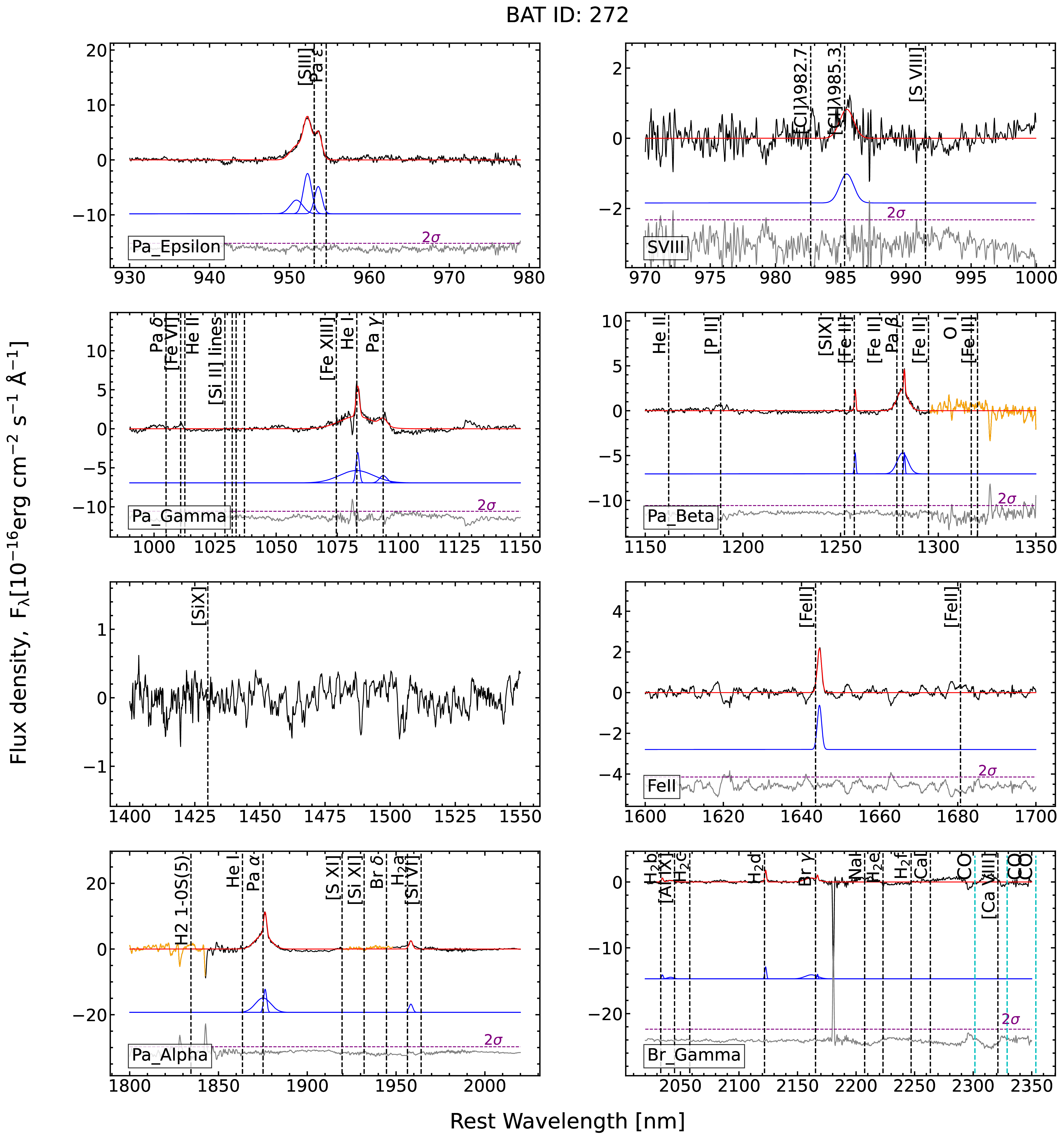}
\caption{Example of simultaneous emission-line fits for galaxy IRAS05189-2524 (BAT ID 272), observed with VLT/X-shooter, following the same labels as Fig.~\ref{fig:fig_bat_795}. This source has several broad components, including \pab\ and \paa, despite being a Seyfert 2 source. This opposing broad-emission classification could indicate an obscuring medium extinction, the bluer optical broad emission, but the NIR is less affected by the extinction. }
\label{appendix:fig_bat_272}
\end{figure*}

Figures~\ref{appendix:fig_bat_576} and \ref{appendix:fig_bat_272} presents outliers in the general distribution of \NH\ vs \paa\ FWHM with respect to Seyfert classification in the bottom panel of Fig.~\ref{fig:fig_fwhm_paa_pab_nh}.

\section{Molecular H$_2$ Detection Statistics and Ratios}

\begin{figure*}
\centering
\includegraphics[width=\columnwidth]{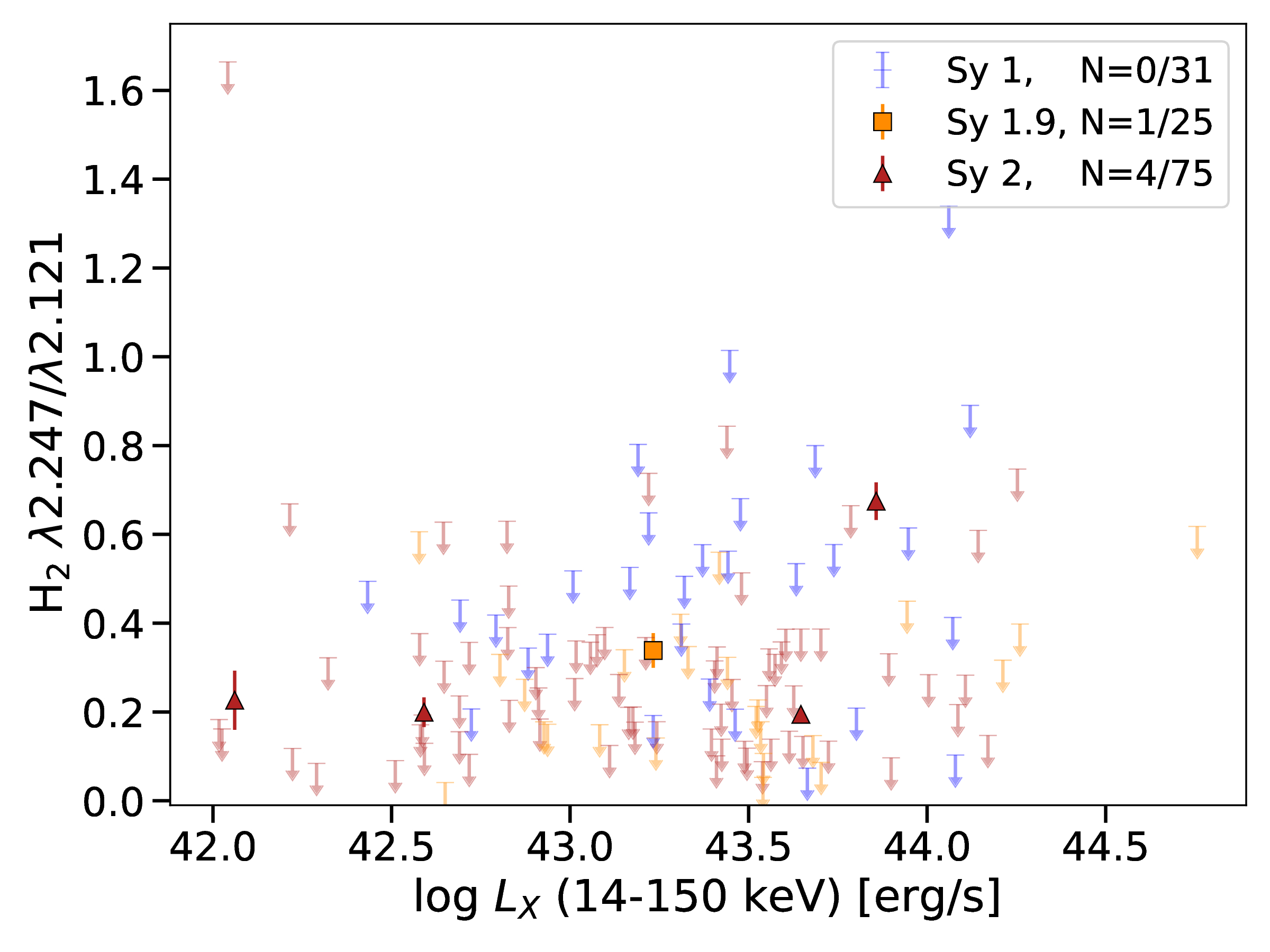} 
\includegraphics[width=\columnwidth]{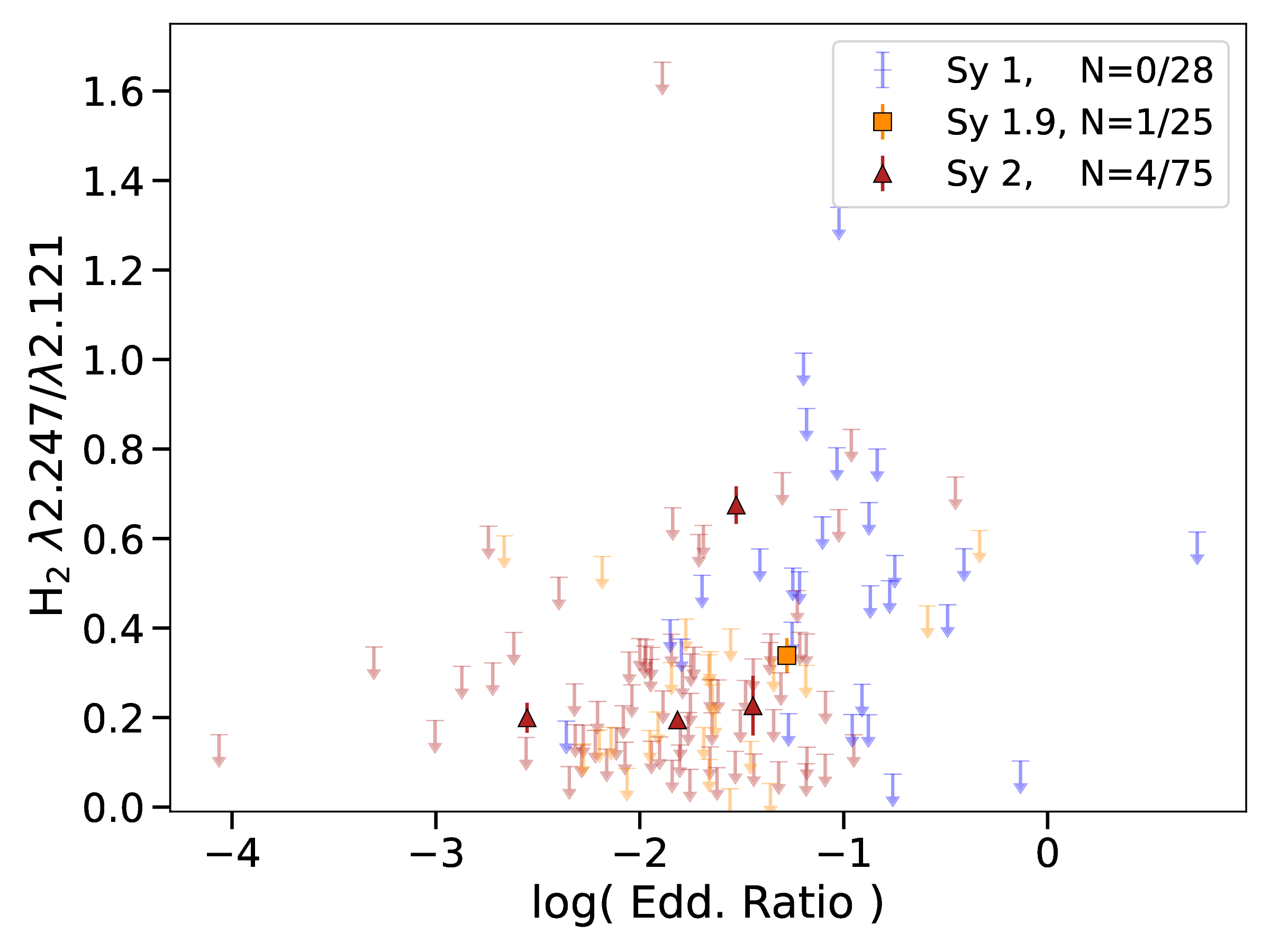} 
\includegraphics[width=\columnwidth]{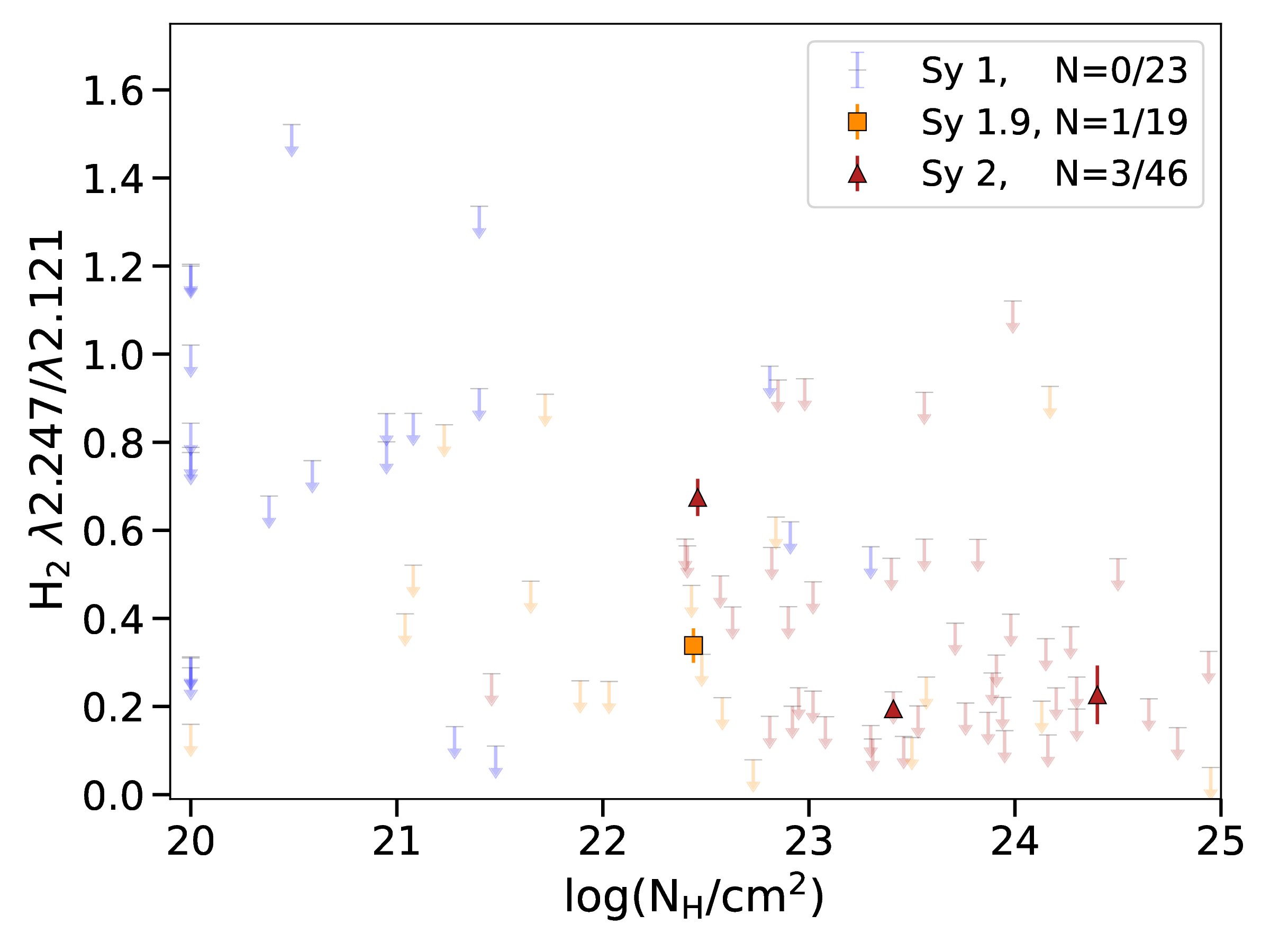}
\caption{Distributions of H$_2$ ratios from top to bottom vs $L_\text{X-ray}$(14$-$150~keV), Eddington ratio, and \NH. For X-ray luminosity, there appears to be a possibly weak correlation for the relatively narrow 1.5~dex range. Eddington ratio and \NH\ do not correlate with the gas ratio. }
\label{appendix:fig_h2_ratios}
\end{figure*}

Table~\ref{appendix:tab_mol_lines} reports a summary of detection statistics for various H$_2$ molecular emissions in the NIR, across varied Seyfert classifications. Figure~\ref{appendix:fig_h2_ratios} shows H$_2$ ($\lambda2247/\lambda2121$) versus $L_{\text{X}}$, Eddington ratio, and column density. Targets with simultaneous detection at wavelengths $\lambda2247$~nm and $\lambda2121$~nm are too scarce in this sample to make definitive statements about their trends with AGN properties. H$_2$ ($\lambda2247/\lambda2121$) flux ratio is caused by thermal ($\lesssim0.2$) and non-thermal ($\gtrsim0.5$) conditions, values based on models summarized in \citet{Mouri:1994:777}. Most detections fall near the boundary of thermal excitation models.

\startlongtable
\begin{deluxetable*}{lcccccccc}
\tablewidth{\textwidth}
\tablecaption{Cataloged emission lines measured for BAT ID 272. \label{appendix:tab_flux}}
\tablewidth{\textwidth}
\tablehead{
\colhead{Line} & \colhead{Position} &\colhead{Position Error} & \colhead{Amplitude} & \colhead{Flux} &  \colhead{Flux Error} & \colhead{FWHM} & \colhead{FWHM Error} & \colhead{S/N\tablenotemark{a}} \\
\colhead{} & \colhead{(nm)} & \colhead{(nm)} & \colhead{(erg s$^{-1}$ cm$^{-2}$ \AA$^{-1}$)} & \colhead{(erg s$^{-1}$ cm$^{-2}$)} & \colhead{(erg s$^{-1}$ cm$^{-2}$)} & \colhead{(km s$^{-1}$)} & \colhead{(km s$^{-1}$)} & \colhead{} 
}
\startdata
\sfiii\ & 952.3 & 0.007 & 7.33e$-$16 & 94.4e$-$16 & 0.9e$-$16 & 381 & 0.003 & 14.00 \\
\sfiii\ blue & 950.9 & 0.5 & 2.49e$-$16 & 4.9e$-$15 & 4.0e$-$15 & 582 & 0.3 & 4.76 \\
\sfiii\ broad & 953.1 & $-$ & $-$4.07e$-$17 & $-$2.92e$-$16 & $-$ & $-$ & $-$ & $-$ \\
\pae\ & 953.6 & 0.05 & 4.97e$-$16 & 5.7e$-$15 & 0.9e$-$15 & 337 & 33 & 9.49 \\
\pae\ broad & 954.6 & $-$ & $-$4.07e$-$17 & $-$2.92e$-$16 & $-$ & $-$ & $-$ & $-$ \\
\ci\  985 & 985.5 & 0.03 & 8.26e$-$17 & 11.0e$-$16 & 0.4e$-$16 & 381 & 0.01 & 2.68 \\
\ci\ 983 & 982.7 & $-$ & $-$3.32e$-$17 & $-$9.24e$-$17 & $-$ & $-$ & $-$ & $-$ \\
\ci\ 985 broad & 985.3 & $-$ & $-$3.44e$-$17 & $-$2.51e$-$16 & $-$ & $-$ & $-$ & $-$ \\
\ci\ 983 broad & 982.7 & $-$ & $-$3.32e$-$17 & $-$2.42e$-$16 & $-$ & $-$ & $-$ & $-$ \\
\sfviii\ & 991.5 & $-$ & $-$2.81e$-$17 & $-$7.86e$-$17 & $-$ & $-$ & $-$ & $-$ \\
\sfviii\ broad & 991.5 & $-$ & $-$2.81e$-$17 & $-$2.06e$-$16 & $-$ & $-$ & $-$ & $-$ \\
\pad & 1004.9 & $-$ & $-$3.60e$-$17 & $-$1.01e$-$16 & $-$ & $-$ & $-$ & $-$ \\
\heii\ & 1012.6 & $-$ & $-$3.55e$-$17 & $-$1.00e$-$16 & $-$ & $-$ & $-$ & $-$ \\
\siii\ 1029 & 1029.0 & $-$ & $-$1.90e$-$17 & $-$5.42e$-$17 & $-$ & $-$ & $-$ & $-$ \\
\siii\ 1032 & 1032.0 & $-$ & $-$1.96e$-$17 & $-$5.60e$-$17 & $-$ & $-$ & $-$ & $-$ \\
\siii\ 1034 & 1033.6 & $-$ & $-$2.01e$-$17 & $-$5.73e$-$17 & $-$ & $-$ & $-$ & $-$ \\
\siii\ 1037 & 1037.0 & $-$ & $-$2.18e$-$17 & $-$6.24e$-$17 & $-$ & $-$ & $-$ & $-$ \\
\hei\ & 1083.4 & 0.03 & 3.91e$-$16 & 6.8e$-$15 & 2.6e$-$15 & 452 & 52 & 14.63 \\
\fexiii\ & 1074.6 & $-$ & $-$3.48e$-$17 & $-$1.01e$-$16 & $-$ & $-$ & $-$ & $-$ \\
\fevi\ & 1010.9 & $-$ & $-$3.57e$-$17 & $-$1.01e$-$16 & $-$ & $-$ & $-$ & $-$ \\
\pag\ & 1093.8 & 0.4 & 9.32e$-$17 & 3.6e$-$15 & 1.2e$-$15 & 1001 & 369 & 3.49 \\
\pag\ broad & 1093.8 & $-$ & $-$5.18e$-$17 & $-$3.54e$-$16 & $-$ & $-$ & $-$ & $-$ \\
\hei\ broad & 1082.9 & 0.6 & 1.58e$-$16 & 27.6e$-$15 & 1.2e$-$15 & 4531 & 29 & 5.92 \\
\pab\ & 1282.6 & 0.007 & 2.37e$-$16 & 2.01e$-$15 & 0.03e$-$15 & 186 & 4 & 8.91 \\
\sfix\ & 1252.0 & $-$ & $-$1.48e$-$17 & $-$6.26e$-$17 & $-$ & $-$ & $-$ & $-$ \\
\heii\ & 1162.0 & $-$ & $-$1.72e$-$17 & $-$7.01e$-$17 & $-$ & $-$ & $-$ & $-$ \\
\feii\ 1320 & 1320.1 & $-$ & $-$6.38e$-$17 & $-$2.77e$-$16 & $-$ & $-$ & $-$ & $-$ \\
\feii\ 1257 & 1257.4 & 0.008 & 2.35e$-$16 & 2.33e$-$15 & 0.04e$-$15 & 221 & 4 & 8.85 \\
\feii\ 1279 & 1278.8 & $-$ & $-$2.23e$-$17 & $-$9.53e$-$17 & $-$ & $-$ & $-$ & $-$ \\
\pii\ & 1188.6 & $-$ & $-$2.67e$-$17 & $-$1.10e$-$16 & $-$ & $-$ & $-$ & $-$ \\
\feii\ 1295 & 1295.0 & $-$ & $-$3.96e$-$17 & $-$1.70e$-$16 & $-$ & $-$ & $-$ & $-$ \\
\oi\ & 1316.9 & $-$ & $-$6.17e$-$17 & $-$2.67e$-$16 & $-$ & $-$ & $-$ & $-$ \\
\pab\ broad & 1281.7 & 0.03 & 2.34e$-$16 & 16.2e$-$15 & 0.2e$-$15 & 1519 & 5 & 8.79 \\
\sfix\ & 1430.0 & $-$ & $-$2.10e$-$17 & $-$7.04e$-$17 & $-$ & $-$ & $-$ & $-$ \\
\feii\ 1644 & 1644.6 & 0.005 & 2.18e$-$16 & 3.01e$-$15 & 0.03e$-$15 & 237 & 3 & 12.58 \\
\feii\ 1680 & 1680.7 & $-$ & $-$2.30e$-$17 & $-$1.95e$-$16 & $-$ & $-$ & $-$ & $-$ \\
\brg\ & 2167.1 & 0.03 & 6.99e$-$17 & 9.3e$-$16 & 0.4e$-$16 & 172 & 5 & 3.31 \\
H$_2$ 1$-$0S(1) & 2122.8 & 0.01 & 1.81e$-$16 & 3.33e$-$15 & 0.10e$-$15 & 244 & 7 & 8.57 \\
H$_2$ 1$-$0S(2) & 2034.6 & 0.05 & 6.03e$-$17 & 1.13e$-$15 & 0.06e$-$15 & 260 & 31 & 2.86 \\
\hei\ & 2058.0 & $-$ & $-$1.31e$-$17 & $-$7.07e$-$17 & $-$ & $-$ & $-$ & $-$ \\
H$_2$ 1$-$0S(0) & 2223.0 & $-$ & $-$4.65e$-$17 & $-$2.61e$-$16 & $-$ & $-$ & $-$ & $-$ \\
H$_2$ 2$-$1S(1) & 2247.0 & $-$ & $-$2.63e$-$17 & $-$1.49e$-$16 & $-$ & $-$ & $-$ & $-$ \\
\nai\ & 2207.3 & $-$ & $-$3.50e$-$17 & $-$1.96e$-$16 & $-$ & $-$ & $-$ & $-$ \\
\cai\ & 2263.4 & $-$ & $-$2.39e$-$17 & $-$1.36e$-$16 & $-$ & $-$ & $-$ & $-$ \\
\alix\ & 2041.6 & $-$ & 2.47e$-$17 & 1.62e$-$15 & $-$ & 907 & $-$ & 1.17 \\
\caviii\ & 2321.1 & $-$ & $-$7.66e$-$17 & $-$4.40e$-$16 & $-$ & $-$ & $-$ & $-$ \\
\brg\ broad & 2161.8 & $-$ & 6.16e$-$17 & 8.1e$-$15 & 0.1e$-$15 & 1720 & 6 & 2.92 \\
\paa\ & 1876.3 & 0.004 & 7.06e$-$16 & 1.51e$-$14 & 0.02e$-$14 & 322 & 2 & 15.64 \\
H$_2$ 1$-$0S(5) & 1834.5 & $-$ & $-$1.43e$-$16 & $-$5.45e$-$16 & $-$ & $-$ & $-$ & $-$ \\
\hei\ & 1863.5 & $-$ & $-$6.01e$-$17 & $-$2.31e$-$16 & $-$ & $-$ & $-$ & $-$ \\
\sfxi\ & 1919.6 & $-$ & $-$5.10e$-$17 & $-$1.98e$-$16 & $-$ & $-$ & $-$ & $-$ \\
\sixi\ & 1932.0 & $-$ & $-$5.32e$-$17 & $-$2.08e$-$16 & $-$ & $-$ & $-$ & $-$ \\
\brd\ & 1944.6 & $-$ & $-$4.00e$-$17 & $-$1.57e$-$16 & $-$ & $-$ & $-$ & $-$ \\
H$_2$ a & 1958.3 & 0.01 & 2.53e$-$16 & 6.69e$-$15 & 0.06e$-$15 & 380 & 0.003 & 5.60 \\
\sivi\ & 1964.1 & $-$ & $-$5.83e$-$17 & $-$2.29e$-$16 & $-$ & $-$ & $-$ & $-$ \\
\paa\ broad & 1875.1 & 0.01 & 4.39e$-$16 & 4.75e$-$14 & 0.03e$-$14 & 1625 & 13 & 9.73 \\
\brd\ broad & 1944.6 & $-$ & $-$4.00e$-$17 & $-$4.10e$-$16 & $-$ & $-$ & $-$ & $-$ \\
\enddata
\tablecomments{Negative amplitude values correspond to the RMS of the spectrum in the line emission region. Negative flux values indicate 1$\sigma$ upper limits. Non-detections have recorded positions at the expected rest-frame location. $^a$S/N corresponds to the amplitude of the line. \tablenotetext{}{(This table will be available in its entirety in machine-readable form.)} }
\end{deluxetable*}

\begin{deluxetable*}{lcccccc}
\tablecaption{Summary of detection statistics for NIR H$_2$ molecular lines fit for each Seyfert type.\label{appendix:tab_mol_lines}}
\tablewidth{0pt}
\tablehead{
  \colhead{H$_2$ Molecular Line} & \multicolumn{2}{c}{Seyfert 1} & \multicolumn{2}{c}{Seyfert 1.9} & \multicolumn{2}{c}{Seyfert 2} \\
  \colhead{} & \colhead{N$_{\mathrm{Det.}}$} & \colhead{Fraction} & \colhead{N$_{\mathrm{Det.}}$} & \colhead{Fraction} & \colhead{N$_{\mathrm{Det.}}$} & \colhead{Fraction}
}
\startdata
$1-0\mathrm{S}(5)~\lambda1834$\,nm & 17 & 11.2($\pm$2.7)\%   & 11 & 14.9($\pm$4.5)\%   & 24 & 11.0($\pm$2.2)\% \\
$1-0\mathrm{S}(3)~\lambda1956$\,nm & 40 & 26.7($\pm$4.2)\%   & 24 & 34.8($\pm$7.1)\%   & 90 & 43.7($\pm$4.6)\% \\
$1-0\mathrm{S}(2)~\lambda2033$\,nm & 9  & 6.3($\pm$2.1)\%    & 14 & 21.9($\pm$5.8)\%   & 48 & 26.2($\pm$3.8)\% \\
$1-0\mathrm{S}(1)~\lambda2121$\,nm & 33 & 24.4($\pm$4.3)\%   & 27 & 44.3($\pm$8.5)\%   & 82 & 52.2($\pm$5.8)\% \\
$1-0\mathrm{S}(0)~\lambda2223$\,nm & 7  & 5.6($\pm$2.1)\%    & 5  & 9.3($\pm$4.1)\%    & 15 & 9.9($\pm$2.5)\% \\
$2-1\mathrm{S}(1)~\lambda2247$\,nm & 3  & 2.5($\pm$1.4)\%    & 3  & 6.0($\pm$3.5)\%    & 9  & 6.0($\pm$2.0)\% \\
\enddata
\end{deluxetable*}

\bibliography{bib}{}
\bibliographystyle{aasjournal}

\end{document}